\documentclass[sigconf,screen]{acmart}

\usepackage{multirow}
\usepackage{graphicx}
\usepackage{graphics}
\usepackage{adjustbox}
\usepackage{caption}
\usepackage{subcaption}
\usepackage{booktabs,dcolumn}
\usepackage{xcolor}
\usepackage{hyperref,tablefootnote,footnotehyper}
\usepackage[bottom]{footmisc}
\usepackage{tabularray}

\AtBeginDocument{%
  \providecommand\BibTeX{{%
    \normalfont B\kern-0.5em{\scshape i\kern-0.25em b}\kern-0.8em\TeX}}}

\setcopyright{acmlicensed}
\copyrightyear{2024}
\acmYear{2024}
\acmDOI{XXXXXXX.XXXXXXX}

\acmConference[OzCHI 2024]{November 30--December 4, 2024}{Brisbane, Australia}

\acmBooktitle{OzCHI 2024, November 30--December 4, 2024, Brisbane, Australia} 
\acmPrice{15.00}
\acmISBN{978-1-4503-XXXX-X/18/06}

\begin{document}

\title[Goldilocks Principle of `Optimal' Reflection Time]{\textit{Not Too Long, Not Too Short}: Goldilocks Principle of `Optimal' Reflection Time on Online Deliberation Platforms}

\author{ShunYi Yeo}
\email{yeoshunyi.sutd@gmail.com}
\affiliation{%
  \institution{Singapore University of Technology and Design}
  \city{Singapore}
  \country{Singapore}
}

\author{Simon Tangi Perrault}
\email{perrault.simon@gmail.com}
\affiliation{%
  \institution{Singapore University of Technology and Design}
  \city{Singapore}
  \country{Singapore}
}

\renewcommand{\shortauthors}{ShunYi Yeo and Simon T. Perrault}

\begin{abstract}
  The deliberative potential of online platforms has been widely examined but \textbf{the impact of reflection time on the quality of deliberation} remains under-explored. This paper presents two user studies involving 100 and 72 participants respectively, to investigate the impact of reflection time on the quality of deliberation in \textbf{minute-scale deliberations}. In the first study, we identified an \textit{optimal} reflection time for composing short opinion comments. In the second study, we introduced four distinct interface-based time nudges aimed at encouraging reflection near the \textit{optimal} time. While these nudges may not improve the quality of deliberation, they effectively prolonged reflection periods. Additionally, we observed mixed effects on users' experience, influenced by the nature of the time nudges. Our findings suggest that reflection time is crucial, particularly for users who typically deliberate below the \textit{optimal} reflection threshold. 
\end{abstract}






\begin{teaserfigure}
  \includegraphics[width=\textwidth]{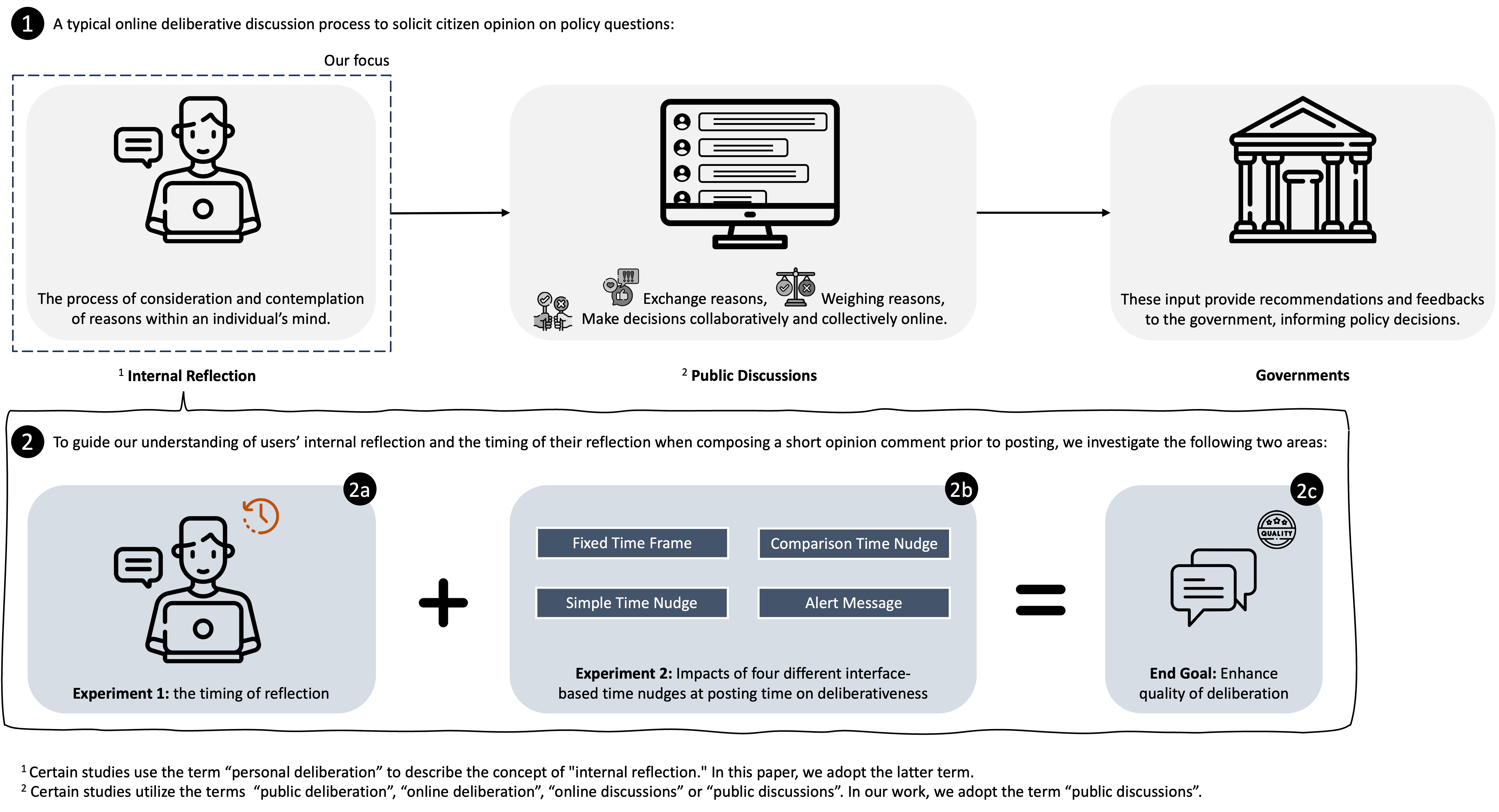}
  \caption{\textbf{1:} A typical online deliberative discussion process involves individuals deliberating independently before collectively discussing and deciding on issues, often providing inputs to the government. As such, the quality of ensuing discussions downstream is intricately linked to the quality of internal reflection upstream; \textbf{2:} In conventional online deliberation platforms, users share their opinions on a post. To enhance the public discussion quality downstream, we explore the relationship between internal reflection and reflection timing when crafting short opinion comments in minute-scale deliberations. Our investigation focuses on two primary areas: \textbf{2a:} the influence of different reflection time lengths on deliberative quality and \textbf{2b:} the effectiveness of four distinct interface-based time nudges: Fixed Time Frame, Simple Time Nudge, Comparison Time Nudge and Alert Message - in supporting the implementation of reflection time; \textbf{2c:} Ultimately, we seek to augment the quality of deliberation on online deliberation platforms.}
  \label{fig: teaser}
\end{teaserfigure}

\maketitle

\section{Introduction}

Advancements in information and communication technologies have enabled the emergence of online forums dedicated to deliberative democracy~\cite{Bertot2010}. Encompassing various formats (e.g., online discourse and reviews, discussion forums like Reddit and social media like Facebook)~\cite{davies2013online}, these forums, sometimes called online deliberation platforms~\cite{Bertot2010, davies2013online}, serve as virtual democratic public spheres that foster political participation~\cite{Keane2000}. On these platforms, large and heterogeneous groups of people contribute their knowledge and opinions~\cite{Jacobs2009, fearon1998deliberation}. Asynchronous exchanges facilitated through discussion threads further empower citizens' participation to collaboratively engage in thoughtful discussions on societal matters~\cite{Jacobs2009, Habermas1984} and informing policy decisions~\cite{Dekker2015, Gastil2014}. This process is commonly known as online deliberative discussions~\cite{davies2013online, saldivar2019civic} (see Figure~\ref{fig: teaser}-1). In this way, online deliberation platforms facilitate the process by enabling well-informed and reasoned discussions about public-interest issues, with the potential involvement of all stakeholders~\cite{zhang2014perceived}.

In some philosophical discussions of deliberation, it is assumed that informed, rational and open-minded discussions would naturally occur on these online deliberation platforms without further assistance~\cite{Habermas1984, Farina2014, kim2015factful}. However, many of these platforms are not designed with deliberation in mind~\cite{friess2015systematic, manin2011comment}, lacking in features that support high quality discussions~\cite{Davies2009}. It is thus not uncommon to see posts that are incendiary or irrelevant~\cite{voggeser2018self}, as well as contributions lacking substantial information~\cite{zhang2005online}. Such common occurrences disrupt the internal reflection process and undermine the quality of subsequent public discussions, leading to chaotic and disjointed discourse~\cite{voggeser2018self, Bishop1980, Neuman1986}. Fundamentally, the essence of public discussions is intricately tied to the process of internal reflection, wherein individuals formulate reflective opinions that contribute to the deliberative process~\cite{zhang2021nudge, hansen2013nudge}. This is further reinforced by Goodin et al.~\cite{Goodin2000, Goodin2003} who postulated that internal reflection is fundamental in the deliberation process: \textit{`... deliberation of the more internal-reflective sort precede formal group deliberations of the more discursive sort'}. Recognizing the primacy of internal reflection~\cite{bohman2000public, Goodin2003, Farina2014}, scholarly attention has emphasized that fostering reflection is crucial for deliberative democracy to flourish~\cite{chambers2003deliberative, Dryzek2002, Goodin2000, muradova2021seeing, 10.1145/3613904.3642530}. Central to this discussion is the recognition that providing adequate reflection time is essential to enhance the quality of deliberation~\cite{moon2013reflection, fleck2010reflecting, li2010stage}. As Moon succinctly argues~\cite{moon2013reflection}: \textit{`reflection takes time; therefore creating or allowing time for reflection is essential'}. Yet, the examination of reflection time materialized in interface design remains largely unexplored. 

Motivated by Moon's insights, we thus seek to explore two aspects of reflection time on the quality of deliberation within the internal reflection process: (1) how long it takes and (2) how it can be supported on online deliberation platforms (see Figure~\ref{fig: teaser}-2). Our examination is facilitated by designing and implementing a simple interface design intended to encourage reflection time at minute-scale deliberations. To support our examination, we pose the following research questions: \textbf{RQ1: What is the impact of reflection time on the quality of deliberation?} and \textbf{RQ2: How do different interface-based time nudges affect the quality of deliberation?} We conducted two user studies, each addresses the aforementioned questions respectively.

To answer RQ1, we conducted our first study with 100 participants, where participants deliberate on an issue at different reflection time intervals. Study results revealed that there exists an \textit{optimal} reflection time for composing short opinion comments in minute-scale deliberations and that increasing reflection time is most favourable for users who deliberate below the optimal time. To answer RQ2, we conducted our second study with 72 participants to assess the effectiveness of different time nudges on the quality of deliberation. We designed and implemented four interface-based time nudges (i.e., Fixed Time Frame, Simple Time Nudge, Comparison Time Nudge and Alert Message), to facilitate the integration of nudges to existing online deliberation platforms. Study results revealed that all four time nudges effectively extended reflection time but they did not differentiate in their impact on deliberative quality. Moreover, the time nudges exhibited distinct traits, leading to varying levels of user experience when they craft their opinions online. We discuss later how these results provide important implications in the design of reflection time on online deliberation platforms.

The contribution of this work is twofold: 
\begin{itemize} 
\item The results of two studies investigating how the quality of deliberation is affected by the time length for reflection and four distinct interface-based time nudges in minute-scale deliberations.
\item A set of design guidelines detailing the integration of the dimension of reflection time and the application of these interface-based time nudges in online deliberation platforms and in other similar contexts.
\end{itemize}
\section{Background and Related Work}
We first begin with an overview of our main focus: reflection time within the deliberation process. Subsequently, we discuss its assessment methods and recent works on nudging to enhance deliberativeness.

\begin{figure*}[!htbp]
   \centering
   \includegraphics[width=\textwidth]{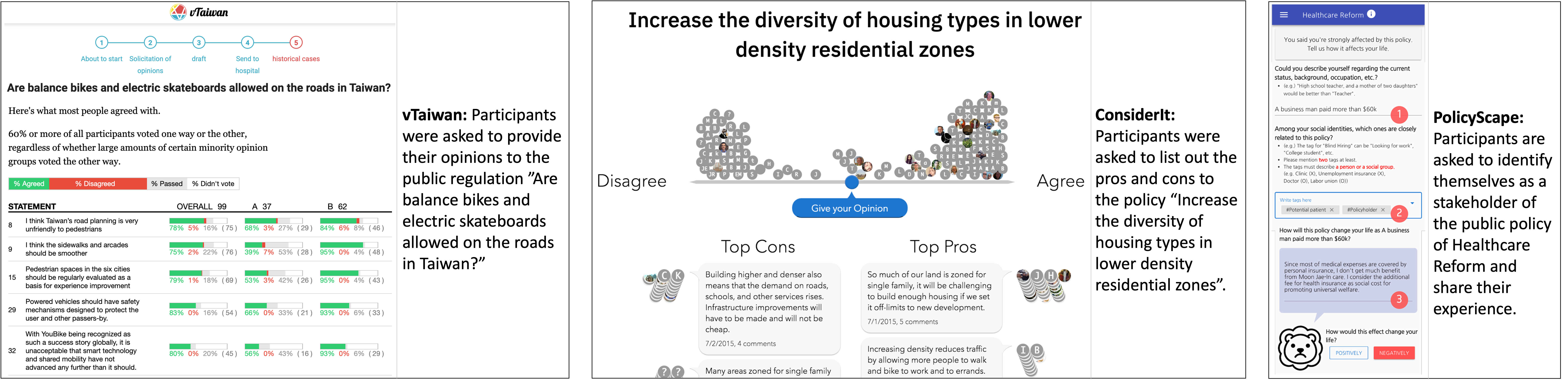}
   \caption{Examples of existing online deliberation platforms (from left): \textbf{vTaiwan}\protect\footnotemark is a platform for opinion exchange to produce public regulations that align with the expectations and needs of stakeholders. The next two examples shows an illustration of online deliberation platforms that incorporate reflection approaches: \textbf{ConsiderIt}\protect\footnotemark is a platform to support users' reflection process by guiding them to reflect on trade-offs of policies through the creation of pros and cons points; \textbf{PolicyScape}\protect\footnotemark is an online system designed to support users' reflection on public policies by helping them to explore diverse stakeholders' perspectives.}
   \label{fig: example of existing online deliberation platforms}
\end{figure*}

\subsection{The Role of Reflection Time in Deliberation}

Deliberation denotes the process of thoughtful consideration by an \textit{individual}~\cite{hobbes1946, Goodin2003, davies2013online, Barnes1984-BARCWO-2}. It involves carefully weighing the reasons for and against a given measure~\cite{Goodin2003, barnes1984complete, davies2013online, hobbes1946}, empowering \textit{individuals} to make informed and rational choices~\cite{zhang2021nudge, christiano2009debates}. Deliberation theorists emphasize reflection as a fundamental element with its significance well-supported within deliberation studies~\cite{Sartori1987, Dryzek2002, chambers2003deliberative, Goodin2000, muradova2021seeing}. As articulated by Goodin and Niemeyer~\cite{Goodin2003}: `\textit{internal reflection inherently precedes public discussion}', laying the cognitive groundwork that shapes an individual's stance and engagement in subsequent public discussions~\cite{Goodin2003, Goodin2000, chambers2003deliberative, Dryzek2002, muradova2021seeing}. 

Research into enhancing deliberation quality through reflection has been widely examined on online deliberation platforms~\cite{Dryzek2002, kriplean2012supporting, kim2019crowdsourcing, zhang2021nudge, arceneaux2017taming, price2002does, 10.1145/3613904.3642530} (see Figure~\ref{fig: example of existing online deliberation platforms}). However, ensuring effective reflection necessitates establishing appropriate conditions. Scholars~\cite{moon2013reflection, fleck2010reflecting} postulate that reflection \textit{takes time}, emphasizing that `creating or allowing time for reflection is essential'. Addressing this, designing for interactive systems should mitigate time as a barrier to reflection~\cite{li2010stage}, acculturating users to take some time for attention and reflection before commenting~\cite{Farina2014}. Collectively, these studies underscore the significance of allocating time for reflection, supporting a shared conclusion: reflection time possess the potential to influence deliberativeness, a normatively desired effect for deliberation. We thus propose \textbf{H1: More reflection time increases deliberativeness.}  

While reflection time is crucial for reflection, how reflection time influences deliberativeness remains unexplored even within the relevant CSCW and HCI fields of deliberation studies. For this study, we thus focus on examining the role of reflection time, particularly in \textbf{how long it takes} and it \textbf{how it can be supported} in the context of minute-scale deliberations, where users craft short opinion comments on online deliberation platforms. Given the prevalence of short-form communication on online deliberation platforms~\cite{takayoshi2015short, umansky2023dances, zhang2024form}, which fosters rapid interactions and quick information dissemination~\cite{zhang2022driving, jarvenpaa2010research}, focusing on minute-scale deliberations capture this fast-paced nature where users often respond swiftly to posts or comments of online discourse~\cite{jarvenpaa2010research}. Understanding how reflection operates within these short time frames is crucial for enhancing the quality of deliberation in such contexts. 

\addtocounter{footnote}{-2}
\footnotetext[\thefootnote]{\href{https://vtaiwan.tw/}{vTaiwan}}
\addtocounter{footnote}{1}
\footnotetext[\thefootnote]{\href{https://consider.it/}{ConsiderIt}. Access the complete paper at~\cite{kriplean2012supporting}.}
\addtocounter{footnote}{1}
\footnotetext[\thefootnote]{Access the complete paper at~\cite{kim2019crowdsourcing}.}

\subsection{Measuring Deliberation: Deliberativeness}
\label{sec:quality}

Deliberativeness denotes the quality of deliberation made by an individual~\cite{Trenel2004}, a fundamental aspect of public discussions~\cite{menon2020nudge}. In our study, we adopt the term `deliberativeness' to characterize the quality of an individual's opinion~\cite{price2002does, steenbergen2003measuring, stromer2007measuring, graham2003search, bohman2000public, zhang2021nudge} on online deliberation platforms. 

Deliberativeness is a composite measure with multiple dimensions~\cite{Trenel2004, zhang2021nudge}. Spatariu et al.~\cite{Spatariu2004} identified four methods to analyze deliberativeness, with three pertaining to public discussions among groups of people. The fourth method, argument structure analysis, applies even to individuals' opinions. As we focus on internal reflection, we assess written opinions, emphasizing argumentation as our measure of deliberativeness and adopting methodologies from the literature that cover this aspect: Tr\'enel~\cite{Trenel2004} proposed a coding scheme for deliberativeness with eight measurements that point to the quantity, duality and diversity of opinions. Cappella et al.~\cite{cappella2002argument} propose \textit{argument repertoire} which counts the number of non-redundant arguments for and against an issue, a common measure used in deliberative research~\cite{menon2020nudge, zhang2021nudge, cappella2002argument, kim2021starrythoughts, 10.1145/3613904.3642530}. Menon et al.~\cite{menon2020nudge} further include \textit{response word count} as they find it to be an `overall interesting indicator of the evolution of a conversation on online platforms'. Other papers also emphasize the significance of \textit{argument diversity}~\cite{anderson2016all, gao2023coaicoder, richards2018practical, 10.1145/3613904.3642530}, which measures the diversity of perspectives in an opinion, accounting for varied interpretations and viewpoints of an issue. 

For this study, we follow the practice of prior work, assessing deliberativeness with three measures: argument repertoire, argument diversity and word count. 

\subsection{Nudging to Augment Deliberativeness}
\label{sec:nudges}

Nudges involving simple interface changes have been shown to effectively change users' behaviour~\cite{thaler2008nudge}. Nudges seek to influence choices and decision process by effectively intervening and moderating an individual's usual behaviour, redirecting to shift from their automatic mode of thinking towards the value introduced by the nudge~\cite{thaler2008nudge}. The idea of nudge was eagerly adopted in HCI~\cite{wang2014field, caraban201923, harbach2014using, park2009newscube}. Caraban et al.~\cite{caraban201923} presented a systematic review of the use of nudging in HCI research, of which they found 23 distinct mechanisms of nudging that can be grouped into 6 categories. Their work discussed the factors shaping the nudge's effectiveness.  

Within the realm of deliberation, there has been some interest in using simple interface designs to enhance the deliberativeness of online discourse~\cite{xiao2015design, zhang2013structural}. For instance, Menon et al.\cite{menon2020nudge} demonstrated that interface nudges such as partitioning text fields, could lead to a significant increase in response length by up to 35\% and the generation of more arguments by up to 25\%. Likewise, Yeo et al.~\cite{10.1145/3613904.3642530} found a significant positive impact of textual based reflective nudges on deliberative quality and how different reflective nudges influence the dynamics of online deliberations. Additionally, Zhang et al.\cite{zhang2021nudge} investigated the use of reflection nudges through simple question prompts on users' perceived issue knowledge and their expression of attitudes and opinions. Their findings showed that prompting participants with questions like `What are your opinions on this issue?' enhanced the quality of participants' opinions.

Drawing from existing efforts, we utilize the concept of nudges in the dimension of time. Our goal is to enhance deliberativeness through various interface-based time nudges. We anticipate that these time nudges will enhance deliberativeness.
\section{Study 1: Reflection Time for Deliberation}
To address \textbf{RQ1} and validate \textbf{H1}, we conducted an experimental study with 100 participants to investigate the most appropriate reflection time needed to achieve the highest deliberativeness. 

\subsection{Independent Variable and Experimental Design}
\label{sec:study 1 fixed time frame}

A between-subject experiment was conducted with the independent variable being Reflection Time with five levels that are modeled at one minute intervals: \{One minute, Two minutes, Three minutes, Four minutes, Five minutes\}. Since no prior in-depth investigation on reflection times exists (prior research that comes closest is on investigating the optimal delay time for pre-moderated content on public displays~\cite{greis2014can}) and our focus on minute-scale deliberation, we chose concrete values to best capture the impact on deliberativeness.

Reflection time was modeled following Alós-Ferrer and Buckenmaier~\cite{alos2021cognitive}, as the sum of the times resulting from a chain of steps of reasoning to arrive at an observed response. Thus, reflection time is the total time taken by an iterative process of reading, reflecting, typing and reviewing. As it is also difficult to disentangle the different steps in reflection, using this framework allows us to capture the entire journey from task onset to submission. 

\begin{figure*}[!htbp]
  \centering
  \includegraphics[width=\textwidth]{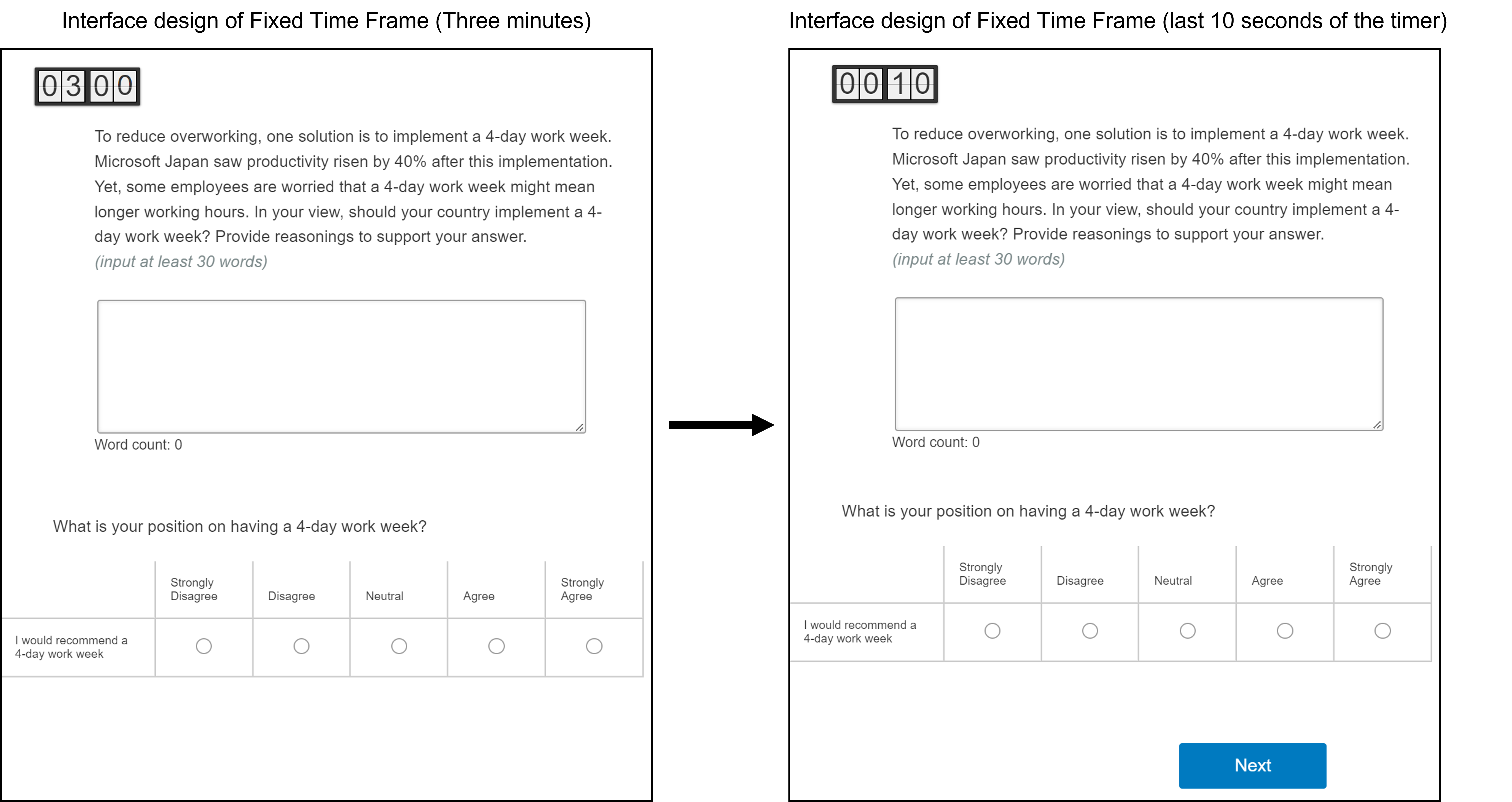}
  \caption{Key features of the interface design in study 1. Left: The timer is positioned on the top left corner and there is no submit button for participants to proceed even if they have finished typing their responses before the timer ends. Right: When the timer reaches 10 seconds, the submit button appears. When the timer ends, participants' responses will be automatically submitted. This figure shows a screenshot of the Three minutes experimental condition but is the same for all five reflection time lengths used in study 1.}
  \label{experiment1design}
\end{figure*}

The interface design, inspired by literature on hyperbolic time discounting and dual process theory, seek to encourage users to pause and reflect during the specified reflection periods. Following Wang et al.'s design considerations for timer nudge~\cite{wang2014field}, our interface includes three key features, depicted in Figure~\ref{experiment1design}. 

\begin{itemize}
    \item \emph{Delay Submit Button.} \emph{Delay Submit Button} hides the `Submit' button until a certain reflection time is reached, inspired by the nudging mechanism of hiding~\cite{caraban201923}. This ensures that participants stay on the task page to reflect for the duration specified in their condition. The `Submit' button becomes visible only 10 seconds before the timer concludes. 
    \item \emph{Timer.} To give participants a visual delay of time passing, a count down \emph{Timer} is displayed at the top left corner of the interface. The timer would start at 60, 120, 180, 240 and 300 seconds for the five experimental conditions respectively. 
    \item \emph{Auto Submission.} To constraint participants to only reflect up to the specific time interval, the \emph{Auto Submission} feature would kick in when the timer reaches 0 and automatically submit the participant's response.
\end{itemize}

All three features will be automatically reset if the participant exits and reopens the survey link, or navigate to a different page and then return to the same page.
 
\subsection{Dependent Variables}
\label{sec:dvs}

As deliberativeness is multi-dimensional, we operationalized it through three measurements: argument repertoire, argument diversity and word count as discussed in section~\ref{sec:quality}.

Argument repertoire and argument diversity were derived from a content analysis of participants’ responses by two coders (see Appendix section~\ref{sec:methodologyappendixA1}). Both coders were PhD students with respectively two and five years of experience using content analysis. Cohen’s Kappa was used to determine the agreement between the two coders’ judgments: Argument Repertoire ($\kappa = 0.808$) and Argument Diversity ($\kappa = 0.808$). Kappa scores for both metrics were above the satisfactory threshold of 0.80~\cite{10.2307/2529310,fleiss1979large, morse1997perfectly}. 

The dependent variables are coded as follows:
\begin{itemize}
    \item \emph{Argument repertoire} is the number of non-redundant arguments regarding each position of the discussion topic. Ideas produced along the two positions were combined.
    \item \emph{Argument diversity} was coded by counting the number of unique themes present in the entire response. A higher diversity count indicates more varied perspectives present in the participant’s responses~\cite{anderson2016all, gao2023coaicoder}.
    \item \emph{Word count} is a simple count of the number of words in the participants' opinion to the discussion topic.
\end{itemize}

\subsection{Power Analysis}
We conducted a power calculation for a five-group ANOVA study seeking a medium effect size according to Cohen's conventions~\cite{cohen2013statistical}, at 0.90 observed power with an alpha of 0.05, giving \textit{N} = 20 per experimental condition, hence we recruited 100 participants.

\subsection{Participants and Ethics}
A total of 100 participants was recruited through Amazon Mechanical Turk (MTurk). Refer to Appendix Table~\ref{tab:demographic-experiment1} on the breakdown of the demographic profile in each experimental condition. We ensured that the demographic profiles across the five conditions were similar so as to control for any fixed effects resulting from the differences in demographic factors. We obtained ethics approval from our local Institutional Review Board (IRB) and reimbursed participants at an hourly rate of 7.25 USD, as per IRB guidelines. 

\subsection{Task and Material}
\label{sec:common task}
For the task, we identified a discussion topic on four-day workweek. Participants had to write at least 30 words when expressing their take on the topic\footnote{We conducted a pilot of this task with 22 participants on MTurk and found that a simple text box often led to responses with low informational content, such as \textit{`Four-day workweek is a good thing in my view. - P4, Pilot'}, which made it difficult to compare the results of the experimental conditions. Following lessons learnt from other studies~\cite{menon2020nudge,bandura1982self,switzer1991judgment}, we decided to add a minimal word count requirement that seek to improve the quality of responses.}. We selected this topic as it is readily accessible and politically relevant\footnote{The topic on four-day workweek has garnered attention in various countries, including governments and in corporate discussions: \href{https://www.nytimes.com/2023/02/22/business/four-day-workweek-study.html}{The New York Times} and \href{https://www.straitstimes.com/business/world-s-largest-4-day-work-week-trial-finds-few-are-going-back?login=true}{Straits Times}.}, thereby promoting constructive and open debate. We chose to provide minimal deliberation materials, consisting mainly of the topic background to align with the context of minute-scale deliberations.

\subsection{Procedure}
\label{sec:experiment1procedure}
The study contained three parts with Qualtrics and MTurk being used for data collection.

\subsubsection{Pre-Task: Consent, Instructions manual} MTurk workers who met the worker's requirements for our task\footnote{To be eligible for participation, participants had to be: (1) 18 years old and above, (2) fluent in English, with (3) a HIT (i.e., human intelligence task) approval rating above 98\% and (4) the number of HITs approved greater than 10,000. The fairly stringent cutoffs are to ensure that participating workers have consistently delivered reliable task quality. Qualifiers were also assigned to ensure unique worker participation in the experiment.} were invited to complete our study. Participation consent was obtained before the study. Participants were informed that they had to express their opinions on a discussion topic, which was undisclosed at this point, under a specified reflection time frame. They then proceeded to read the instruction manual which outlined the key features of the interface. Afterwards, participants saw a screen informing them that the timer would start immediately after.

\subsubsection{Main Task} Detailed in section~\ref{sec:common task}. 

\subsubsection{Post-Task: Post-Task Feedback, Familiarity, Demographics} After completing the main task, participants completed a short post-task survey on the duration of the reflection time. This was assessed on a five-point Likert scale (1 = Too Short, 2 = A Little Short, 3 = Just Right, 4 = A Little Long and 5 = Too Long). 

To control for any fixed effects, participants were asked to rate their familiarity on the discussion topic from a scale of 0 to 100 with 0 being not familiar at all and 100 being very familiar. Demographic information (age, gender, country) was also collected. 

\subsection{Outliers}
\label{sec:outliers}
Before analyzing our data, we plotted a box plot (box and whisker plot) to visually show the dispersion of our data and to identify any potential outliers~\cite{schwertman2004simple}. We adopted the approach by Dawson~\cite{dawson2011significant} to discard data points that were either below $Q1 - 3 IQR$ or above $Q3 + 3 IQR$ (where Q1 is the first quartile, Q3 the third, and IQR is the interquartile range). This led us to discard two data points from the Four minute condition. Hence, our final sample size was 98. 
\section{Findings (Study 1)}

\subsection{Results}
Unless otherwise specified, ANOVA was used to identify main effects and pairwise t-tests with Bonferroni correction for post-hoc comparisons~\cite{norman2010likert} for the three measurements of deliberativeness.

\subsubsection{Word Count}
We found a statistically significant main effect of \emph{Reflection Time} on Word Count ($F_{4,93}$ = 8.21, $p<.0001$). There was a statistical difference between One minute ($M=31.45$ words) and Three ($M=71$ words), Four ($M=73.78$ words) and Five minutes ($M=76.55$ words) (all $p<.01$). 
We also found a significant difference between Two minutes ($M=47.15$ words) and Five minutes ($p=.032$). Results are summarized in Figure~\ref{fig:exp1-wordcount}.

\subsubsection{Argument Repertoire}
We found a significant main effect of \emph{Reflection Time} on Argument Repertoire ($F_{4,93} = 4.99, p<.001$). There was a statistical difference between One minute and Three, Four and Five minutes such that One minute ($M=1.85$ arguments) has lower number of arguments compared to Three ($M=3.3$ arguments, $p<.05$ ), Four ($M=3.39$ arguments, $p<.01$) and Five minutes ($M=3.35$ arguments, $p<.05$). We did not find any other pairwise differences. Results are summarized in Figure \ref{fig:exp1-argument}.

\subsubsection{Argument Diversity}
\begin{figure*}[!htbp]
\centering
\begin{subfigure}{.5\textwidth}
  \centering
  \includegraphics[scale=0.45]{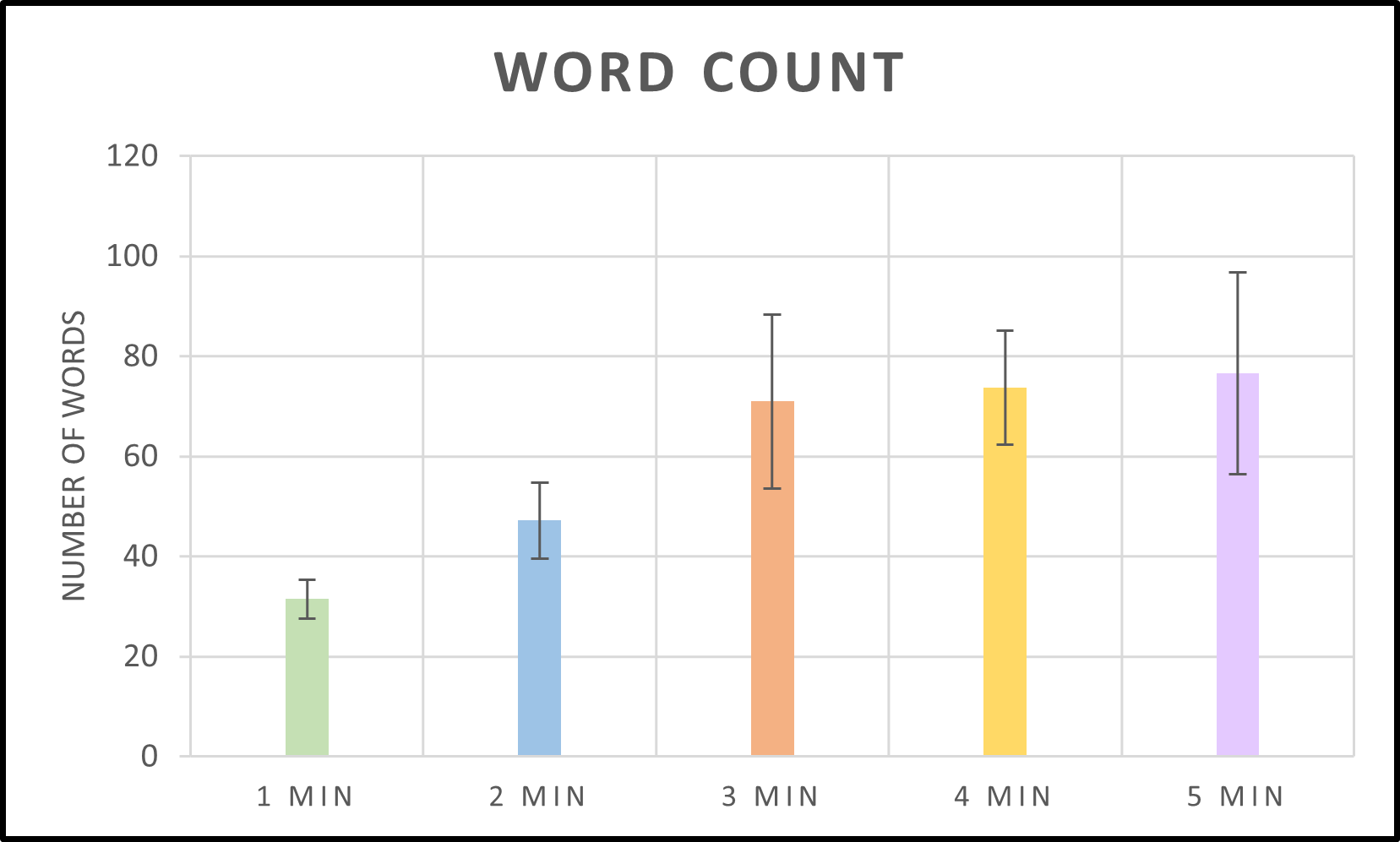}
  \caption{Word Count}
  \label{fig:exp1-wordcount}
\end{subfigure}%
\begin{subfigure}{.5\textwidth}
  \centering
  \includegraphics[scale=0.45]{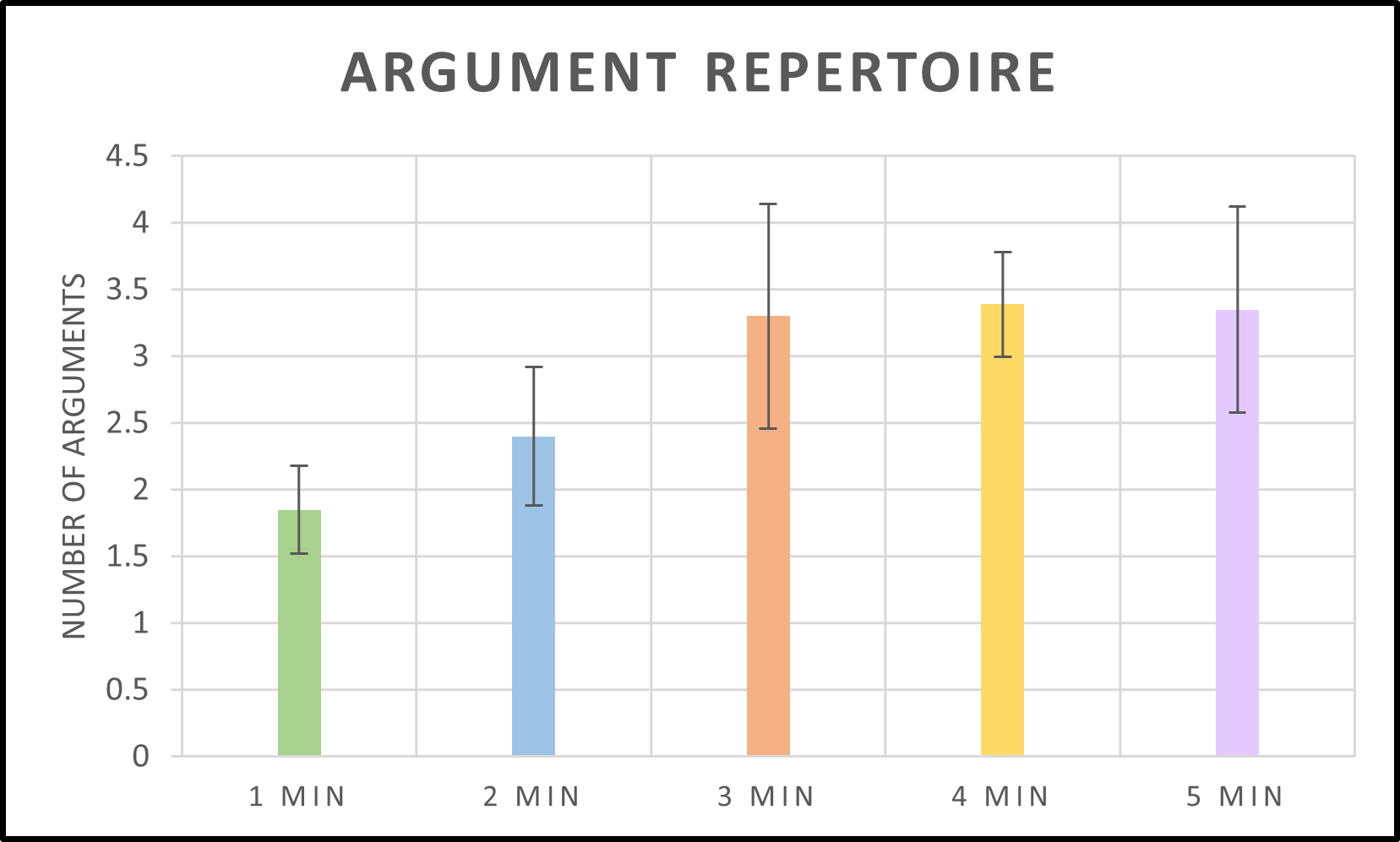}
  \caption{Argument Repertoire}
  \label{fig:exp1-argument}
\end{subfigure}
\begin{subfigure}{.5\textwidth}
  \centering
  \includegraphics[scale=0.45]{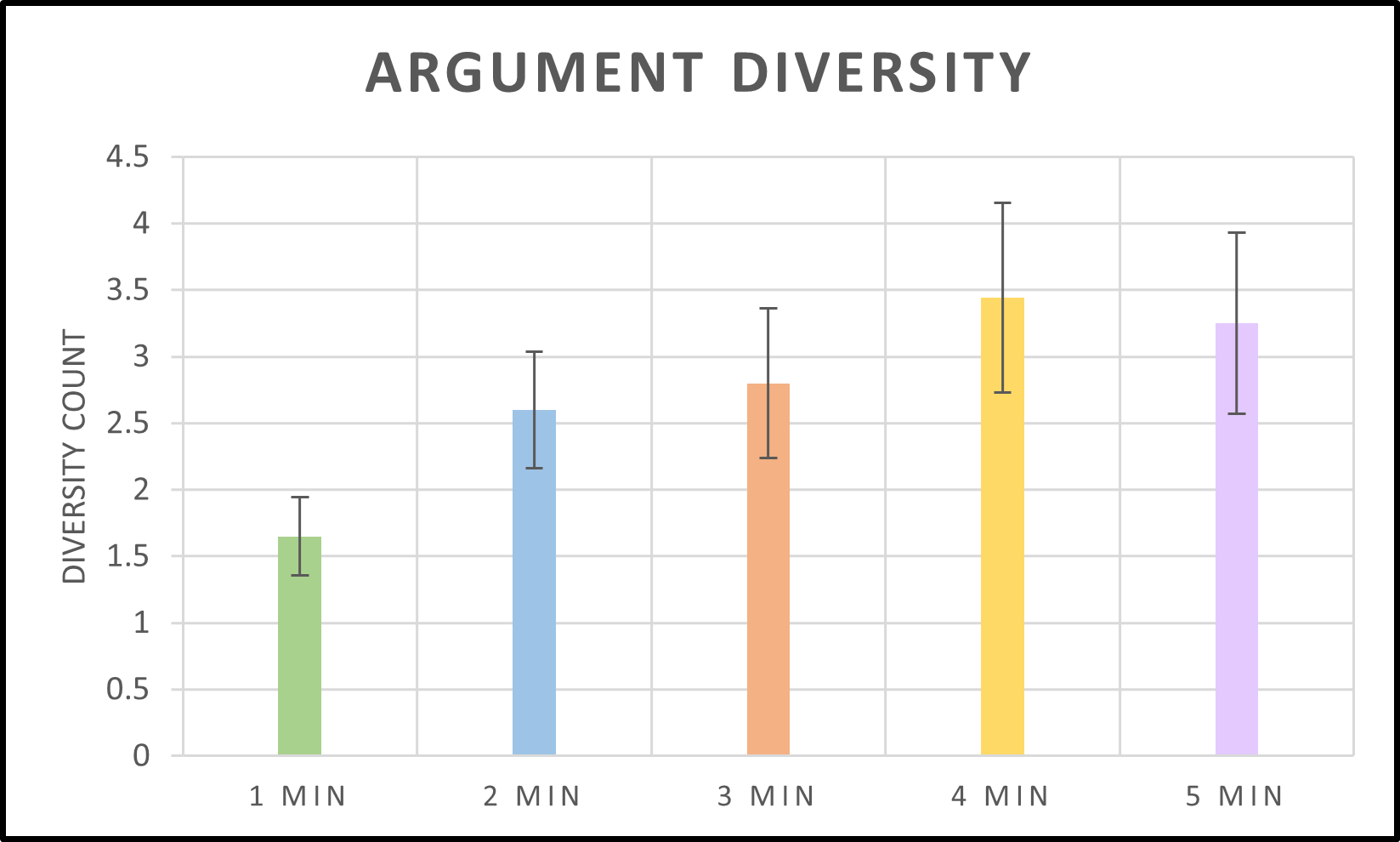}
  \caption{Argument Diversity}
  \label{fig:experiment1-diversity}
\end{subfigure} 
\caption{Word count, argument repertoire and argument diversity for each reflection time. Error bars show 95\% confidence intervals.}
\label{fig:results-experiment1}
\end{figure*}

We found a statistically significant main effect of \emph{Reflection Time} on Argument Diversity ($F_{4,93} = 6.16$, $p<.001$). 
Similarly, we found significant differences between One minute ($M=1.65$ diversity) and Three ($M=2.8$ diversity, $p<.05$), Four ($M=3.44$ diversity, $p<.01$) and Five minutes ($M=3.25$ diversity, $p <.01$). Results are summarized in Figure~\ref{fig:experiment1-diversity}.

\subsubsection{Summary of Results} Overall, deliberative quality by the three measurements is summarized in Table~\ref{tab:summary of opinion quality}. A careful look at the data reveals that Four minutes achieves the highest overall deliberativeness in the lowest amount of time among all the other reflection times. It is also worth mentioning that the typical reflection time\footnote{Before conducting study 1, we measured participants' ($N = 34$) typical reflection time on the four-day workweek task without any experimental influence. The mean time was 3.07 minutes, with a median of 2.71 minutes (range = 42.3 seconds to 6.14 minutes, $SD = 1.73$ minutes, $Q1 = 1.63$ minutes, $Q3 = 4.86$ minutes). We then conducted study 1 to find the most appropriate reflection time and if the mean typical reflection time aligns with this appropriate reflection time. It is noteworthy to note that the mean typical reflection time is below the `optimal' reflection time of Four minutes.} participants took fell short of Four minutes.  

\begin{table*}[!htbp]
 \caption{Deliberative quality across the different reflection times. $\alpha$, $\beta$, $\gamma$ and $\mu$ show significant pairwise comparisons ($p<.05$).}
 \label{tab:summary of opinion quality}
 \scalebox{0.84}{
 \begin{tabular}{@{}cccccc@{}}
 \hline
  \multirow{3}{*}{\textbf{\begin{tabular}[c]{@{}c@{}}Deliberativeness \\ \\ \end{tabular}}} & \multicolumn{5}{c}{\textbf{Reflection Time}} \\ \cline{2-6}
  & {\textbf{\begin{tabular}[c]{@{}c@{}}1 min \\ ($M \pm S.D.$)\end{tabular}}} & {\textbf{\begin{tabular}[c]{@{}c@{}}2 mins \\ ($M \pm S.D.$)\end{tabular}}} & {\textbf{\begin{tabular}[c]{@{}c@{}}3 mins \\ ($M \pm S.D.$)\end{tabular}}} & {\textbf{\begin{tabular}[c]{@{}c@{}}4 mins \\ ($M \pm S.D.$)\end{tabular}}} & {\textbf{\begin{tabular}[c]{@{}c@{}}5 mins \\ ($M \pm S.D.$)\end{tabular}}} \\ 
 \hline
 \textbf{Word Count} & 31.45 $\pm$ 8.83$^{\alpha,\beta,\gamma}$ & 47.15 $\pm$ 17.16$^{\mu}$ & 71.00 $\pm$ 39.71$^{\alpha}$ & 73.78 $\pm$ 24.60$^{\beta}$ & 76.55 $\pm$ 45.99$^{\gamma, \mu}$ \\
 \textbf{Argument Repertoire} & 1.85 $\pm$ 0.75$^{\alpha,\beta,\gamma}$ & 2.40 $\pm$ 1.19 & 3.30 $\pm$ 1.92$^{\alpha}$ & 3.39 $\pm$ 0.85$^{\beta}$ & 3.35 $\pm$ 1.76$^{\gamma}$ \\
 \textbf{Argument Diversity} & 1.65 $\pm$ 0.67$^{\alpha,\beta,\gamma}$ & 2.60 $\pm$ 0.99 & 2.80 $\pm$ 1.28$^{\alpha}$ & 3.44 $\pm$ 1.54$^{\beta}$ & 3.25 $\pm$ 1.55$^{\gamma}$ \\
 \hline
 \end{tabular}}
\end{table*}

\subsubsection{Relationship between Reflection Time and Deliberativeness}
Based on our results, we plotted three scatter plots to examine the relationship between reflection time and deliberativeness.

To determine which regression model (i.e., linear or logarithmic) provides a reasonable fit, we plotted both models in all three scatter plots (see Figure~\ref{fig:log-linear}). We then ran regression analyses for both functional forms (see Appendix Table~\ref{tab:regression analysis}). The logarithmic regression model provided a better fit across all three scatter plots, with an increase of $R^2$ for Word Count (0.872 to 0.973), Argument Repertoire (0.782 to 0.980) and Argument Diversity (0.798 to 0.984). The main reason for preferring the logarithm model is that it makes more sense theoretically as we explain now. 

\begin{figure*}[!htbp]
\centering
\begin{subfigure}{.5\textwidth}
  \centering
  \includegraphics[scale=0.45]{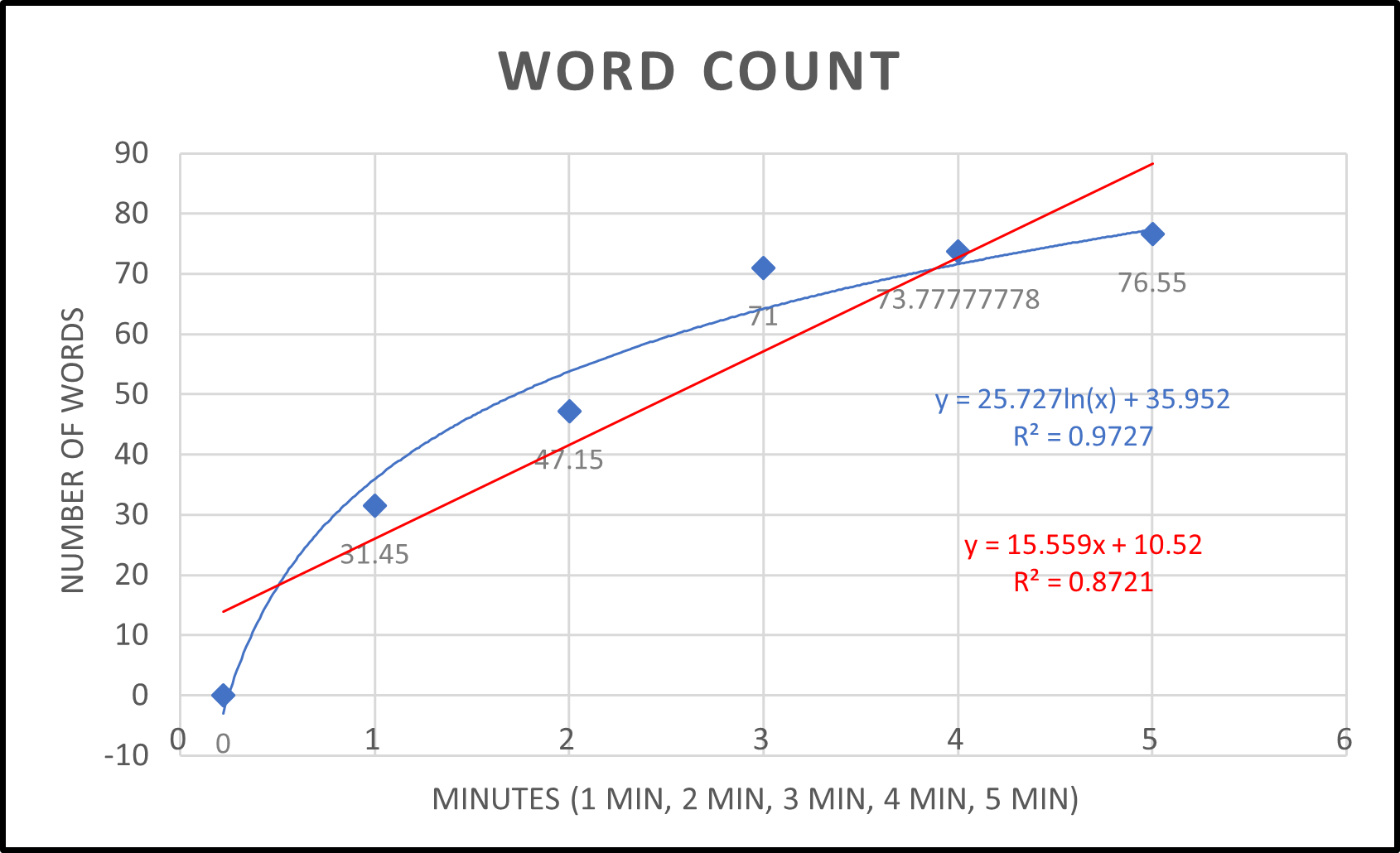}
  \caption{Scatter plot of Word Count}
  \label{fig:scatter plot ave word count}
\end{subfigure}%
\begin{subfigure}{.5\textwidth}
  \centering
  \includegraphics[scale=0.45]{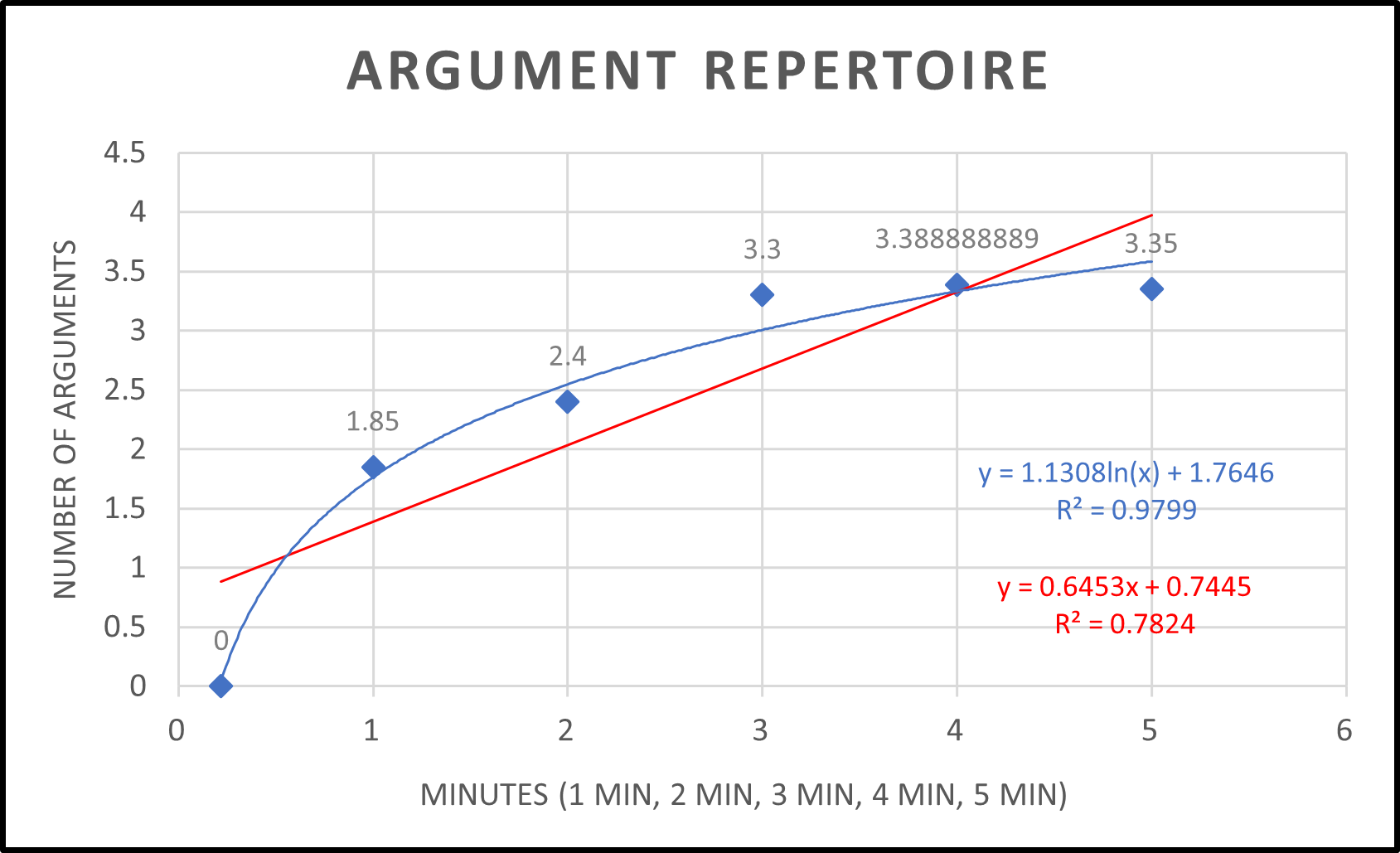}
  \caption{Scatter plot of Argument Repertoire}
  \label{fig:scatter plot argument count}
\end{subfigure} \\[1.5ex]
\begin{subfigure}{.5\textwidth}
  \centering
  \includegraphics[scale=0.45]{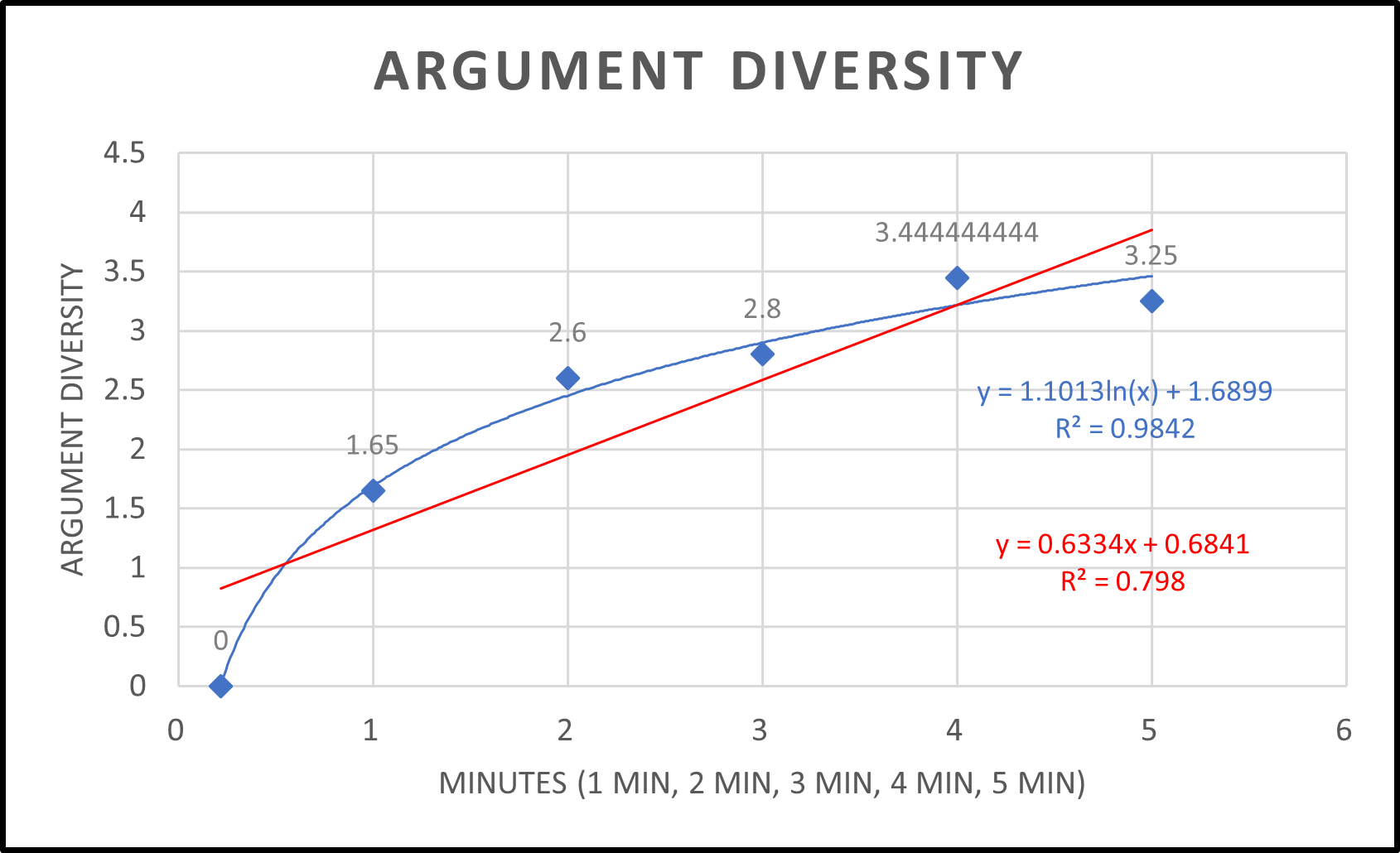}
  \caption{Scatter plot of Argument Diversity}
  \label{fig:scatter plot argument diversity}
\end{subfigure}
\caption{Logarithm function (blue) vs linear function (red) for the three measurements of deliberativeness.}
\label{fig:log-linear}
\end{figure*}

\paragraph{Plateauing Effect. } Firstly, by examining the data from all three scatter plots (see Figure~\ref{fig:log-linear}), word count, argument repertoire and argument diversity increases with reflection time but plateau after three minutes. This hinted that the relationship between reflection time and deliberative quality might be nonlinear.

\paragraph{Intercept. } Secondly, it does not make sense to have words or arguments present at time 0 minutes ($t = 0$). For instance, at $t = 0$, average word count is 10.52 as in the linear model (see Figure~\ref{fig:scatter plot ave word count}), which is practically impossible. Hence, a logarithmic model may make more sense here as it accounts for the time lag participants took before responding.

All three scatter plots for the logarithmic function (see Figure~\ref{fig:log-linear}) showed that word count, argument repertoire and argument diversity remained at 0 until reflection time reaches 0.22 minutes (approximately 13.2 seconds). This reflects the average time participants took to respond to the task, aligning with our recorded data.

\paragraph{Summary. } From this logarithmic law, and by observing our data, we can conclude that \textbf{increasing reflection time exhibit diminishing returns in terms of the increase for deliberative quality}. 

\subsubsection{Post-Task Feedback on the Duration for Reflection}
\label{sec:subjective feedback}
\begin{figure*}[!htbp]
  \centering
  \includegraphics[width=0.5\textwidth]{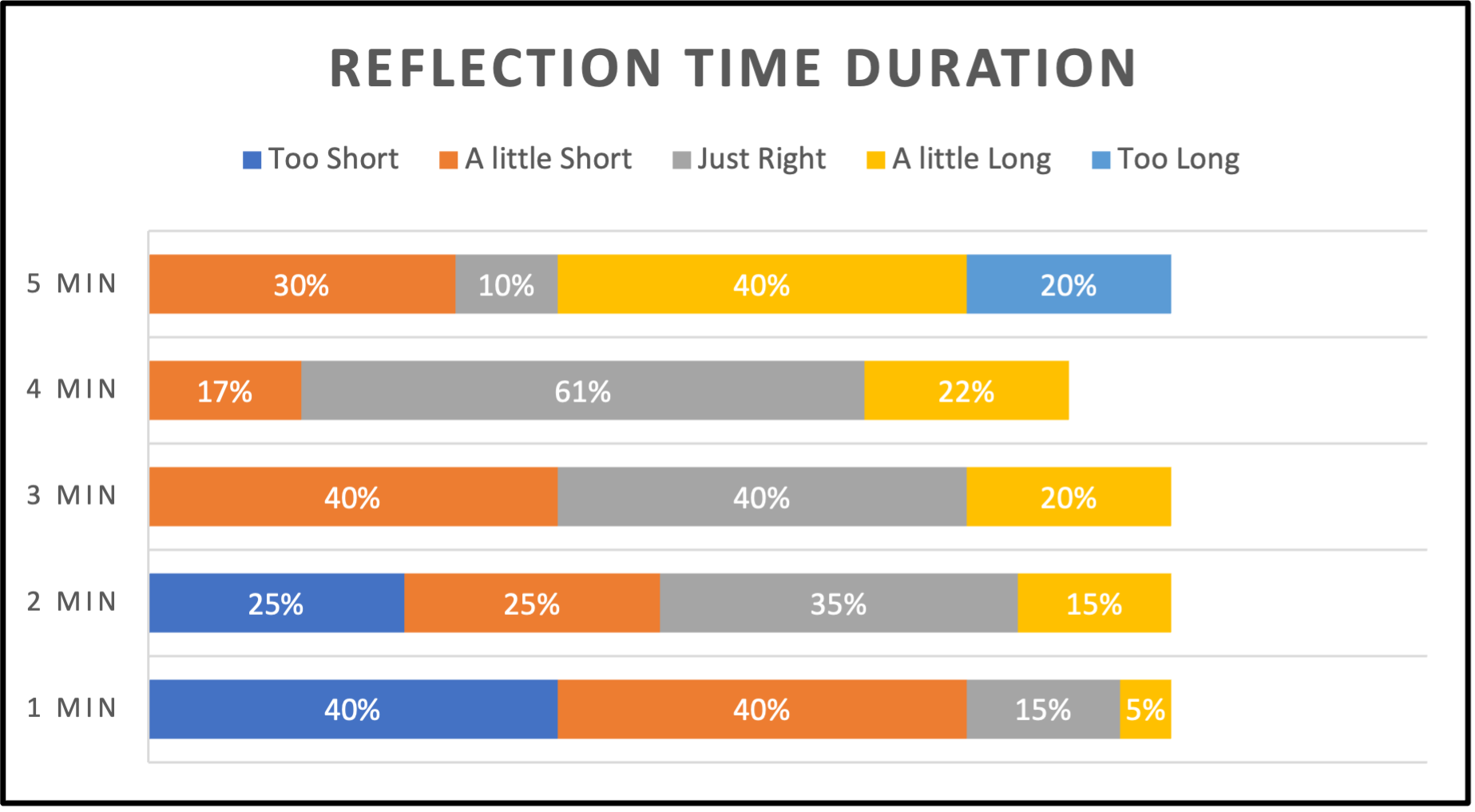}
  \caption{Post-task feedback on the length of reflection time in each experimental condition. Note: the Four Minutes condition is missing two discarded data points but the total still amounts to 100\%.}
  \label{time-duration}
\end{figure*}

One minute was widely considered inadequate, with 80\% of participants indicating it was \textit{Too Short} (40\%) or \textit{A little Short} (40\%). At Two minutes, the proportion finding it too short decreased to 50\% (25\% \textit{Too Short}, 25\% \textit{A little Short}), while 35\%, considered it \textit{Just Right}. At Three minutes, none found it \textit{Too Short}, though 40\% still felt it was \textit{A little Short}, and 40\% thought it was \textit{Just Right}. At Four minutes, 17\% found it \textit{A little Short}, and 61\% considered it \textit{Just Right}. At Five minutes, 60\% of participants found it long (20\% \textit{Too Long}, 40\% \textit{A little Long}), while 10\% found it \textit{Just Right} - a significant drop from Four minutes. Surprisingly, 30\% felt that Five minutes was \textit{A little Short}, up from 17\% at Four minutes. The results are summarized in Figure~\ref{time-duration}.

\subsection{Discussion}
In this section, we discussed our findings to answer \textbf{RQ1} and to validate \textbf{H1}. We also discussed how our results corroborate and enrich previous studies.

\subsubsection{Logarithmic Relationship between Reflection Time and Deliberativeness}
Our findings indicate a significant positive effect of reflection time on deliberativeness. This is inline with Alós-Ferrer and Buckenmaier~\cite{alos2021cognitive} work, which showed that higher observed depth of reasoning is linked to longer reflection times. Our results would also mean that the element of time in reflection is important in the deliberation process. This aligns with prior research that allowing for time in reflection is crucial and thus should be accounted for~\cite{fleck2010reflecting,lindley2009desiring,li2010stage,farina2012rulemaking,moon2013reflection}.  

However, our study goes beyond to uncover the impact of increasing reflection time. We found that increasing reflection time does not necessarily increase deliberative quality indefinitely. There seems to exists a \textit{limit} to how much time can enhance deliberativeness. Our results showed that all three measurements of deliberativeness obey the logarithmic law, indicating a non-linear relationship between reflection time and deliberativeness. Specifically, all three measurements provide converging evidence that as reflection time increases beyond a certain limit, deliberative quality plateaus or even drops. This suggests that increasing reflection time beyond the limit diminishes the rate of return for deliberativeness in minute-scale deliberations. Thus, \textbf{H1 is partially supported}.

\subsubsection{Goldilocks Principle of `Optimal' Time}
The logarithmic relationship suggests that there exist a possibility of an \textit{`optimal time'} for reflection in minute-scale deliberations. The optimal time is the \textit{limit} to how time can improve deliberative quality. It can also be interpreted as the point where deliberative quality is maximized most efficiently. From study 1, we deduce that the optimal reflection time is \emph{Four minutes} because it has the best deliberative quality across the five experimental conditions in terms of argument repertoire and argument diversity. Moreover, participants' subjective feedback for \emph{Four minutes} is the most favourable (i.e., \emph{Four minutes} garnered the highest percentage of 61\% for \textit{Just Right}). 

The existence of an optimal reflection time in deliberation has a twofold effect when designing for minute-scale deliberations. First is the case of spending too little time which is below the optimal duration, leading to lower deliberative quality. In this case, increasing reflection time arbitrarily close to the optimum, may significantly enhance deliberativeness, signifying that designers should work towards enabling users to reflect at the optimal time. Second is the case of spending too much time that is well beyond the optimum. In this case, it may not necessarily improve deliberativeness any further and might even introduce trade-offs. We define this phenomenon as the Goldilocks Principle\footnote{The Goldilocks principle is named after a children's story of "The Three Bears". It is used to illustrate the concept of having "just the right amount".} of `Optimal' Reflection Time, too short is unsatisfactory but too long can be undesirable. 

\subsubsection{Below Optimal Time: Time-Induced Reflection}
In the first case, we coined a term `time-induced reflection' - a way of using time to induce participants to reflect more whenever they reflect below the optimal time. Study 1 illustrated that reflecting below the optimal time of Four minutes (e.g., One and Two minutes) results in lower deliberativeness. Consequently, extending reflection time nearer to the optimum (e.g., Three - Four minutes) significantly enhance deliberativeness. Therefore, designers should aim not only to provide time for reflection but, more importantly, work towards encouraging users to reflect arbitrarily close or at the optimum. Time-induced reflection can be easily achieved through nudging techniques~\cite{thaler2008nudge} such as prompting users to take more time to reflect or review before posting. In this aspect, time-induced reflection acts as an anchor influencing individuals' effort~\cite{switzer1991judgment,menon2020nudge}. Most task performers acting alone go about their usual behaviours~\cite{hinsz1997using}, and coupled with the lack of incentives, they do not push themselves towards producing better quality of work~\cite{menon2020nudge}. Introducing an anchor to encourage more reflection shifts their previous performance towards the value introduced by the anchor~\cite{menon2020nudge,cervone1986anchoring}. This is especially beneficial for two groups of users: (1) impulsive individuals and (2) underthinking individuals. Impulsive individuals often exhibit time-urgent personality, having a predisposition to rushing~\cite{rendon2012pilot} and completing tasks hastily, all of which tended to correlate more with negative affect~\cite{emmons1986influence}, resulting in poorer performance~\cite{moritz2014judgmental}. Time-induced reflection can effectively slow them down, compelling them to invest more time to consciously process their thoughts and be more thoughtful in their responses, consequentially moving them towards reflecting at the optimum. This intervention breaks their habitual behaviour and tendency of just progressing through, ultimately producing more reflective responses and enhancing deliberative quality. For underthinking individuals, time-induced reflection signals a cognitive awareness by reinforcing their engagement back to the task. It allows users to revisit and rethink their responses, ultimately enhancing deliberative quality.

\subsubsection{Above Optimal Time: Time-Subduced Reflection}
On the other hand, increasing reflection time beyond the optimum may not necessarily be beneficial. Our results showed that as time goes beyond the optimum, deliberative quality falls in terms of argument repertoire and argument diversity. We termed this phenomenon `time-subduced reflection' - the point at which the drawbacks of extending reflection time beyond optimality outweigh the benefits. In other words, while reflection time supports the deliberation process~\cite{moritz2014judgmental,moon2013reflection,lindley2009desiring,li2010stage,fleck2010reflecting}, nudging individuals to take more time when they have already done so does not create additional value and meaning to their current thought-process. This could even demoralize and demotivate people, making them more impatient (see section~\ref{sec:subjective feedback}), or overwhelms them, especially so when incentives are lacking. It is also harmful as longer reflection time without introducing new thinking prompts to support users may result in a decision paralysis~\cite{moritz2014judgmental,cartwright1941relation,moyer1976mental}. This can happen to individuals who are constantly dwelling with the same thoughts, driving them into the same state of thinking. This causes unproductive thinking and may even result in worse performance~\cite{moritz2014judgmental} as individuals exhibit more randomness in their response~\cite{rustichini2009neuroeconomics}. This may explain why deliberative quality in the Five minutes condition deteriorates after the optimal time. In some cases, it may even lower their confidence levels~\cite{hamblin2022negative,lischetzke2003attention} and cause unfavourable feelings of self-doubts when participants could not add further dimensions to their thinking process~\cite{hamblin2022negative}. This might explain why some participants find Five minutes limited (i.e., 30\% indicated \textit{A little Short}). All in all, additional reflection induced by increasing time beyond optimality adversely impacts deliberative quality.

Therefore, when designing for time on minute-scale deliberations, designers should be wary of the Goldilocks principle of `optimal' reflection time.
\section{Study 2: Time Nudges to Encourage Time-Induced Reflection}
From study 1, we identified a Goldilocks principle of `optimal' reflection time. However, a sizeable number of participants still considered One and Two minutes to be ideal (see Figure~\ref{time-duration}). Leveraging the concept of nudge (see section~\ref{sec:nudges}), we design time nudges to encourage users to deliberate longer. Our design considerations largely follows the nudging mechanisms by Caraban et al.~\cite{caraban201923}. In study 2, we thus address \textbf{RQ2} by examining the effects of various interface-based time nudges on deliberativeness and users' perceptions of the nudge to determine which time nudge best promotes time-induced reflection.

\subsection{Independent Variable and Experimental Design}
A between-subject experiment was conducted with the independent variable being Time Nudges with four levels: \{Fixed Time Frame, Simple Time Nudge, Comparison Time Nudge and Alert Message\}. The time nudges are described below:

\begin{figure*}[!htbp]
\centering
\begin{subfigure}{.5\textwidth}
  \centering
  \includegraphics[scale=0.3]{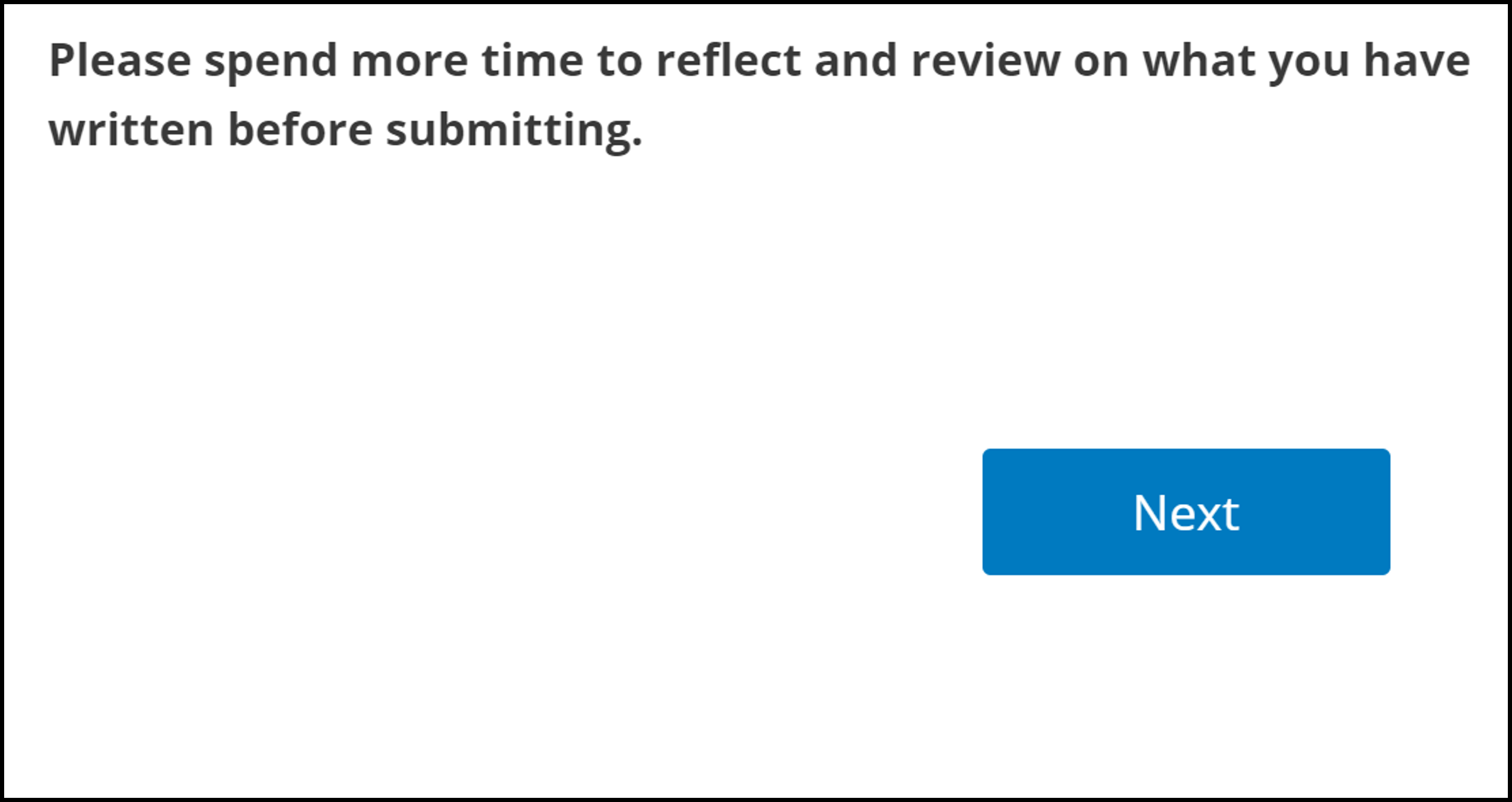}
  \caption{Simple Time Nudge}
  \label{fig:simple time nudge}
\end{subfigure}%
\begin{subfigure}{.5\textwidth}
  \centering
  \includegraphics[scale=0.3]{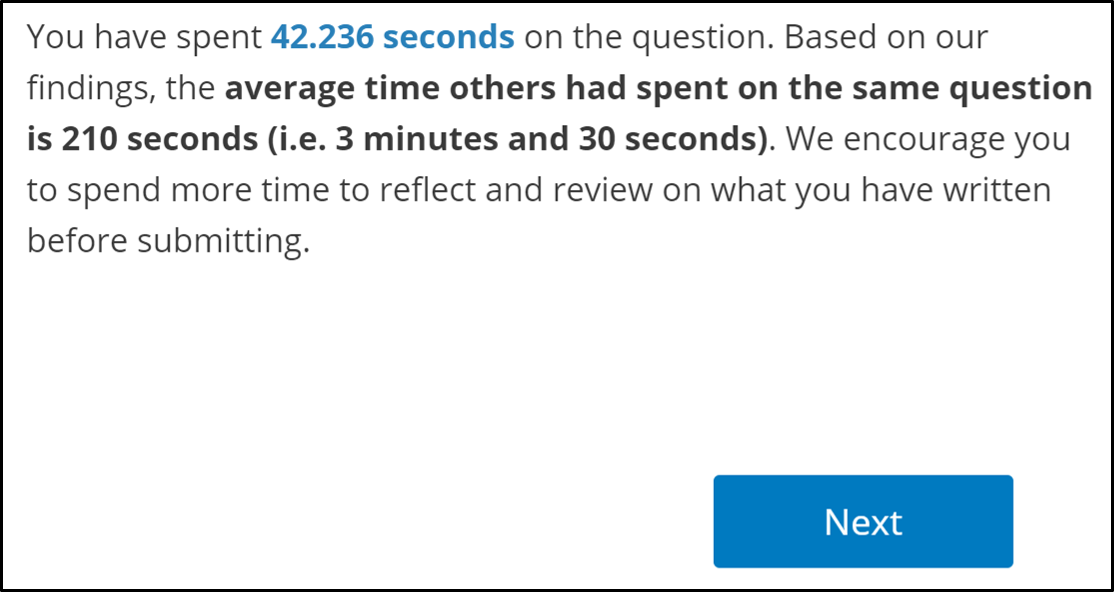}
  \caption{Comparison Time Nudge}
  \label{fig:comparison time nudge}
\end{subfigure} \\[1.5ex]
\begin{subfigure}{.4\textwidth}
  \centering
  \includegraphics[scale=0.28]{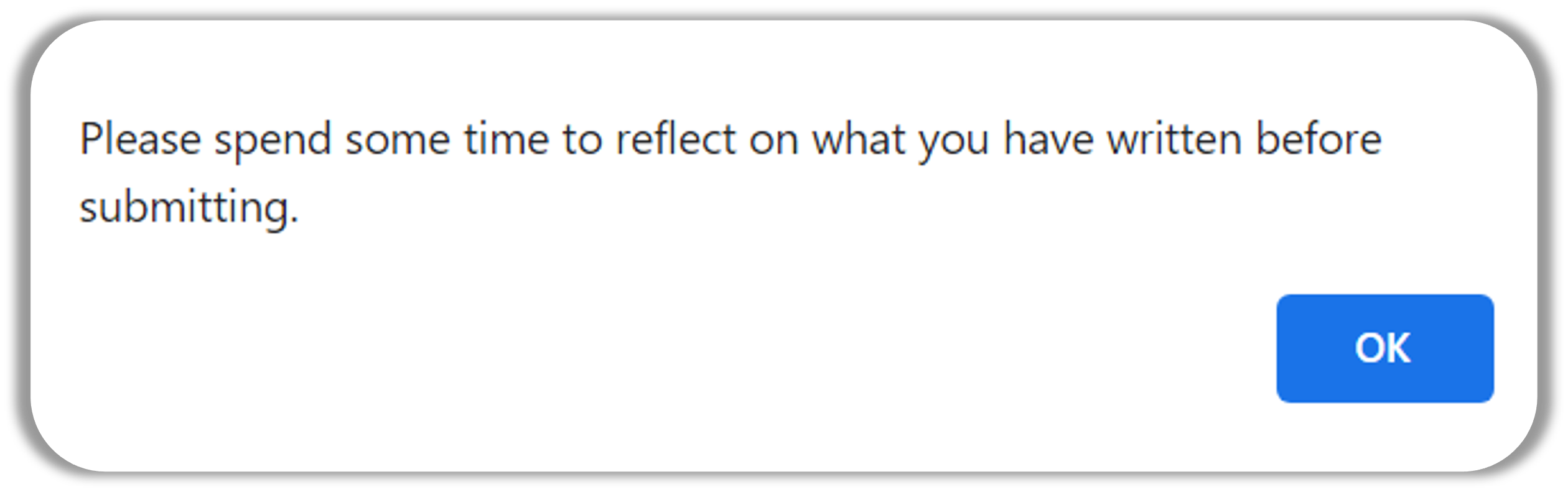}
  \caption{Alert Message. Users will have to click on the Alert Message before they can continue with anything they are doing on the screen.}
  \label{fig:alert message}
\end{subfigure} \hspace{5mm}%
\begin{subfigure}{.5\textwidth}
  \centering
  \includegraphics[scale=0.25]{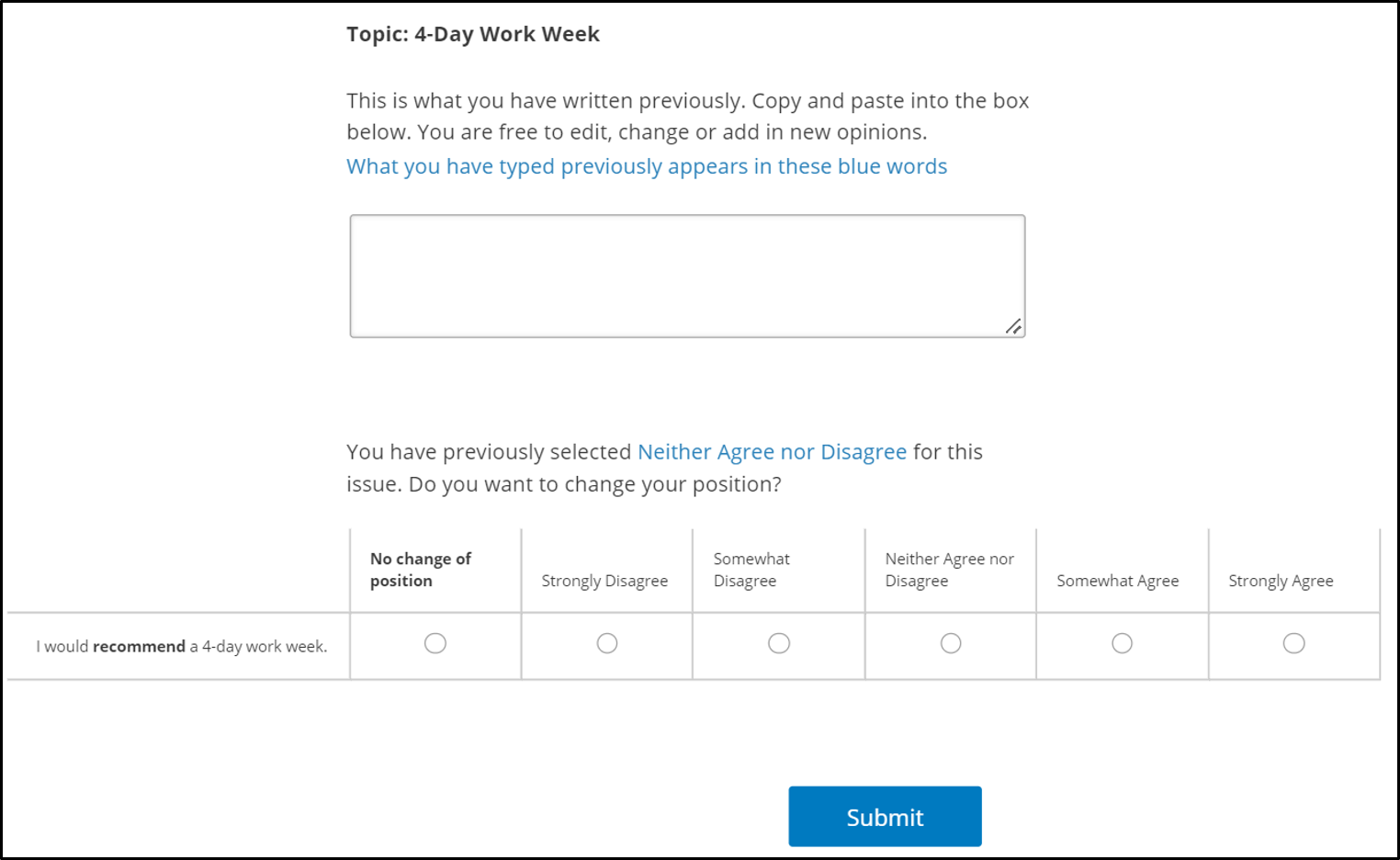}
  \caption{Interface design after users click on the time nudges for Simple Time Nudge and Comparison Time Nudge. Words shown in the colour blue are the responses participants had typed previously.}
  \label{after simple/comparison time nudge}
\end{subfigure}
\caption{Interface designs for the Time Nudges excluding Fixed Time Frame which was shown in Figure~\ref{experiment1design}. }
\label{fig:simple-comparison-alert nudges}
\end{figure*}

\subsubsection{Fixed Time Frame}
Fixed Time Frame follows the design principles by Wang et al.~\cite{wang2014field}, retaining the same interface design as in study 1 (see Figure~\ref{experiment1design}). We applied the nudge concept to encourage participants to spend a specific reflection time on the discussion topic. This was set to be 3.5 minutes (210 seconds), slightly below the optimal Four minutes (240 seconds) based on feedback from study 1, where 22\% of participants found Four minutes a little long (see Figure~\ref{time-duration}).

\subsubsection{Simple Time Nudge}
Simple Time Nudge (see Figure~\ref{fig:simple time nudge}) follows the nudging mechanism of reminding~\cite{caraban201923}. The nudge appears after users have furnished their responses on the discussion topic. It prompts users to pause and review their responses before posting. Users can review their responses, shown in blue, and make edits before submitting their final opinions (see Figure~\ref{after simple/comparison time nudge}). 

\subsubsection{Comparison Time Nudge}
Comparison Time Nudge follows the nudging mechanism of social influence through social comparisons~\cite{caraban201923} to nudge users towards a desirable behaviour by increasing self-awareness~\cite{erickson2000social} and tapping into \textit{herd instinct bias}~\cite{festinger1954theory, colusso2016designing, gouveia2016exploring, caraban201923}. The nudge works similar to \emph{Simple Time Nudge} but displays a prompt revealing the user's reflection time compared to the average time spent by others (see Figure~\ref{fig:comparison time nudge}). The average time is set at 3.5 minutes, serving as a goal-based anchor~\cite{zhang2021nudge} to nudge users towards this reflection duration. Like the \emph{Simple Time Nudge}, users can then review and edit their responses before posting (see Figure~\ref{after simple/comparison time nudge}).

\subsubsection{Alert Message}
\label{sec: study2 alert message design}
Alert Message follows the nudging mechanism of reinforcement~\cite{caraban201923, zhu2017exploring}. It is a pop-up nudge that appears after 20 seconds, reminding users to take more time to reflect before submitting their responses (see Figure~\ref{fig:alert message}). 

\subsection{Dependent Variables}
\label{sec:dvs in study 2}
To discern the impact of time nudges on deliberativeness and users' perceptions of the nudges, the study has three dependent variables: deliberativeness, reflection time and users' perceptions (users' experience, thinking process and perceived helpfulness). 

\subsubsection{Deliberativeness}
Deliberativeness was measured in the same manner as in study 1 of word count, argument repertoire and argument diversity (see section~\ref{sec:dvs}). 

\subsubsection{Reflection Time}
To assess the impact of time nudges on reflection time, we did not restrict reflection duration, except in the \emph{Fixed Time Frame} condition. Participants could and were encouraged to take as much time as needed. Reflection time was measured as the total time users spent reflecting in the presence of the time nudge. 

\subsubsection{Users' Perceptions}
\label{sec:dv users'perceptions}
Users' perceptions were measured using subjective measures (i.e., self-reported scores in questionnaires). These subjective measures encompassed eight question items on a 1-5 Likert scale (1 = strongly disagree to 5 = strongly agree), split into three categories: \textit{Users' Experience}, \textit{Thinking Process} and \textit{Perceived Helpfulness} (see Table~\ref{tab:users' perceptions eight questions}).

\begin{itemize}
    \item \emph{Users' experience} measures user interaction and engagement with time nudges, adapted from prior studies that studies haste~\cite{rendon2012pilot}.
    \item \emph{Thinking process} evaluates the perceived impact of time nudges on users' cognitive processes.
    \item \emph{Perceived helpfulness} assesses the overall perceived utility of time nudges in aiding participants.
\end{itemize}

\begin{table*}[!htbp]
 \caption{The eight question items assessing users' perceptions on their experience, thinking process and perceived helpfulness of the time nudges. Participants have to rate each question on a five-point Likert scale (1 = strongly disagree to 5 = strongly agree).}
 \label{tab:users' perceptions eight questions}
\centering
\begin{tblr}{
  width = \linewidth,
  colspec = {Q[213]Q[727]},
  cell{2}{1} = {r=2}{},
  cell{4}{1} = {r=4}{},
  cell{8}{1} = {r=2}{},
  vlines,
  hline{1-2,4,8,10} = {-}{},
  hline{3,5-7,9} = {2}{},
}
\textbf{Categories of Users' Perceptions} & \textbf{Question Items}\\
\textbf{Users' Experience } & I felt in a hurry (\textbf{time pressure}) to complete the task.\\
 & The timer/prompt (i.e., asking me to spend more time) is \textbf{distractive}.\\
\textbf{Thinking Process } & This process has helped me to \textbf{understand} my thinking process better.\\
 & This process has helped me to be more \textbf{mindful} of my own thoughts.\\
 & This process has helped me to be more \textbf{aware} of my own thoughts.\\
 & This process has helped me to be more \textbf{confident} of my own opinions.\\
\textbf{Perceived Helpfulness } & This process allows me to \textbf{reflect} on my opinions before submitting.~\\
 & I believe this process is a \textbf{good way} to encourage people to take more time to review on what they have written before submitting.
\end{tblr}
\end{table*}

\subsection{Power Analysis}
We conducted a power calculation for a four-group ANOVA study seeking a medium effect size according to Cohen’s conventions~\cite{cohen2013statistical}, at 0.80 observed power with an alpha of 0.05, giving \textit{N} = 18 per experimental condition, hence we recruited 72 participants.

\subsection{Participants and Ethics}
A total of 72 participants were recruited through MTurk. Refer to Appendix Table~\ref{tab:demographic-experiment2} for the complete demographic profiles. We excluded participants from study 1 as the task was the same. Similar to study 1, we ensured that the demographic profiles across the four conditions were similar to control for any fixed effects resulting from the differences in demographic factors. Likewise, we got ethics approval from our local IRB and and reimbursed participants at an hourly rate of 7.25 USD.

\subsection{Procedure and Task}
The procedure for this study closely follows that of study 1 outlined in section~\ref{sec:experiment1procedure}. In the pre-task, participants went through the instructions for the study. 

In the main task, participants type and post a response to the topic of four-day workweek under the influence of the time nudge. We maintain consistency in the topic across the two studies to draw robust comparisons between the different time nudges and their impact on deliberativeness within the same thematic domain~\cite{cahit2015internal}.

In the post-task, participants completed a short survey comprising of the eight question items as well as to provide feedback on the time nudge they engaged with. Lastly, data on demographic and familiarity of the topic were collected.
\section{Findings (Study 2)}

\subsection{Quantitative Results}
\label{sec: quanti study 2}
Similar to study 1, unless otherwise specified, ANOVA was used to identify main effects and pairwise t-tests with Bonferroni correction for post-hoc comparisons~\cite{norman2010likert} for deliberativeness, reflection time and users' perceptions (users' experience, thinking process, perceived helpfulness). 

\subsubsection{Deliberativeness}
No statistically significant effect of \emph{Time Nudges} on word count, argument repertoire and argument diversity were found (all $p>.05$). This implies that there is no significant differences in the deliberative quality among the four time nudges. The results are summarized in Table~\ref{tab:quality-experiment2}. The values we measured for study 2 are overall aligned with the values observed in study 1 even when taking the reflection time per condition into account.

\label{sec:study2 opinion quality}
\begin{table*}[!htbp]
\caption{Deliberative quality of word count, argument repertoire and argument diversity across the four time nudges.}
\label{tab:quality-experiment2}
\begin{tabular}{@{}ccccc@{}}
\toprule
\multirow{3}{*}{\textbf{Deliberativeness}} & \multicolumn{4}{c}{\textbf{Time Nudges}} \\ \cline{2-5}
& \textbf{\begin{tabular}[c]{@{}c@{}}Fixed Time \\ Frame\end{tabular}} & \textbf{\begin{tabular}[c]{@{}c@{}}Simple Time \\ Nudge\end{tabular}} & \textbf{\begin{tabular}[c]{@{}c@{}}Comparison\\ Time Nudge\end{tabular}} & \textbf{\begin{tabular}[c]{@{}c@{}}Alert\\ Message\end{tabular}} \\
\midrule
Word Count & 66.2 & 61.7 & 66.7 & 57.1 \\
Argument Repertoire & 3.33 & 3.39 & 3.00 & 3.06 \\
Argument Diversity & 2.67 & 2.94 & 3.39 & 3.28\\
\bottomrule
\end{tabular}
\end{table*}

\subsubsection{Reflection Time}
\label{sec:study2 deliberation time}
\begin{figure*}[!htbp]
  \centering
  \includegraphics[scale=0.4]{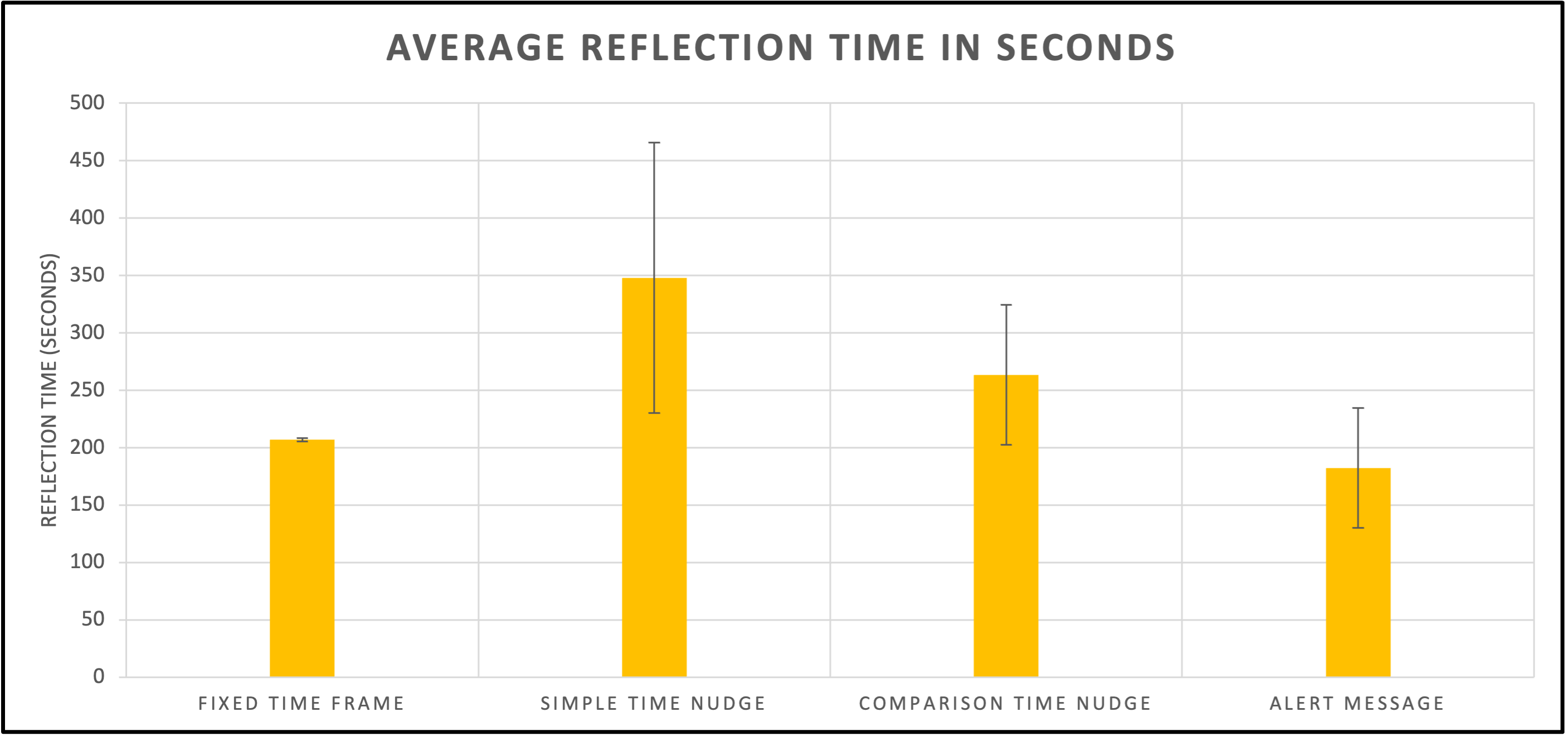}
  \caption{Reflection Time in seconds for each time nudge. The Fixed Time Frame condition was the only one where participants had a fixed time to reflect. Error bars show 95\% confidence intervals. }
  \label{fig:error bar and average deliberation times}
\end{figure*}

We found a significant main effect of \emph{Time Nudges} on Reflection Time ($F_{3,68}$ = 4.08, $p<.01$). There was a statistical difference between \emph{Fixed Time Frame} ($M = 3.5$ minutes) and \emph{Simple Time Nudge} ($M = 5.8$ minutes, $p<.05$) which was explainable as users in \emph{Fixed Time Frame} are bounded by the time limit of 3.5 minutes. We also found a significant difference between \emph{Simple Time Nudge} and \emph{Alert Message} ($M = 3.0$ minutes, $p<.05$). We did not find any other pairwise differences. The results are summarized in Figure \ref{fig:error bar and average deliberation times}.

\subsubsection{Users' Experience}
\label{sec:time manipulation}
Table~\ref{tab:summary of time manipulation} shows the average scores, main effects and pairwise differences. Overall, participants felt that the \emph{Fixed Time Frame} gave them a significantly higher time pressure compared to the other three time nudges. In terms of distraction, \emph{Comparison Time Nudge} was deemed more distractive compared to \emph{Simple Time Nudge}.

\begin{table*}[!htbp]
 \caption{Users' Experience. $\alpha$, $\beta$ and $\gamma$ show significant pairwise comparisons ($p<.05$) for each question item.}
 \label{tab:summary of time manipulation}
 \scalebox{0.85}{\begin{tabular}{@{}ccccccc@{}}
 \hline
  \multirow{4}{*}{\textbf{Users' Experience}} & \multirow{4}{*}{\textbf{$F_{3,68}$}} & \multirow{4}{*}{\textbf{$p$}} & \multicolumn{4}{c}{\textbf{Time Nudges}} \\ \cline{4-7}
  & & &\textbf{\begin{tabular}[c]{@{}c@{}} Fixed \\ Time Frame\\ ($M \pm S.D.$)\end{tabular}} & \textbf{\begin{tabular}[c]{@{}c@{}} Simple \\ Time Nudge\\ ($M \pm S.D.$)\end{tabular}} & \textbf{\begin{tabular}[c]{@{}c@{}} Comparison \\ Time Nudge\\ ($M \pm S.D.$)\end{tabular}} & \textbf{\begin{tabular}[c]{@{}c@{}} Alert \\ Message\\ ($M \pm S.D.$)\end{tabular}} \\   
 \hline
 \begin{tabular}[c]{@{}c@{}} I felt in a hurry \textbf{(time} \\ \textbf{pressure)} to complete the task.\end{tabular} & 7.44 & <.001 & 2.78 $\pm$ 1.35$^{\alpha,\beta,\gamma}$ & 1.56 $\pm$ 0.62$^{\alpha}$ & 1.72 $\pm$ 1.27$^{\beta}$ & 1.28 $\pm$ 0.57$^{\gamma}$ \\ [0.3cm]
 \begin{tabular}[c]{@{}c@{}} The prompt (i.e., asking me to\\ spend more time) is \textbf{distractive}.\end{tabular} & 3.24 & <.05 & 2.28 $\pm$ 1.36 & 2.22 $\pm$ 1.11$^{\alpha}$ & 3.39 $\pm$ 1.14$^{\alpha}$ & 2.56 $\pm$ 1.42 \\ 
 \hline
 \end{tabular}}
\end{table*}

\subsubsection{Thinking Process}
\label{sec:thinking process}
Results for \textit{Thinking Process} are summarized in Table \ref{tab:summary of thinking process}. We found a main effect of \emph{Time Nudges} for every question item. Users perceived \emph{Simple Time Nudge} ($M = 4.22$) to help them better understand their thoughts compared to \emph{Comparison Time Nudge} and \emph{Alert Message}.
We observed the same trend for the other three questions where \emph{Simple Time Nudge} scores were significantly higher compared to other scores for \emph{Comparison Time Nudge} (all questions) and \emph{Alert Message} (all questions except \textbf{awareness}). In summary, \emph{Simple Time Nudge} is perceived to better support users' thinking process when compared to \emph{Comparison Time Nudge} and \emph{Alert Message}.

\begin{table*}[!htbp]
 \caption{Thinking Process. $\alpha$, $\beta$ and $\gamma$ show significant pairwise comparisons ($p<.05$) for each question item.}
 \label{tab:summary of thinking process}
 \scalebox{0.85}{\begin{tabular}{@{}ccccccc@{}}
 \hline
  \multirow{4}{*}{\textbf{Thinking Process}} & \multirow{4}{*}{\textbf{$F_{3,68}$}} & \multirow{4}{*}{\textbf{$p$}} & \multicolumn{4}{c}{\textbf{Time Nudges}} \\ \cline{4-7}
  & & &\textbf{\begin{tabular}[c]{@{}c@{}} Fixed \\ Time Frame\\ ($M \pm S.D.$)\end{tabular}} & \textbf{\begin{tabular}[c]{@{}c@{}} Simple \\ Time Nudge\\ ($M \pm S.D.$)\end{tabular}} & \textbf{\begin{tabular}[c]{@{}c@{}} Comparison \\ Time Nudge\\ ($M \pm S.D.$)\end{tabular}} & \textbf{\begin{tabular}[c]{@{}c@{}} Alert \\ Message\\ ($M \pm S.D.$)\end{tabular}} \\   
 \hline
 \begin{tabular}[c]{@{}c@{}} Going through this process \\  has helped me to \textbf{understand} \\  my thinking process better.\end{tabular} & 4.43 & < .01 & 3.78 $\pm$ 0.88 & 4.22 $\pm$ 0.55$^{\alpha,\beta}$ & 3.11 $\pm$ 1.41$^{\alpha}$ & 3.11 $\pm$ 1.32$^{\beta}$ \\ [0.5cm]
 \begin{tabular}[c]{@{}c@{}} Going through this process has \\ helped me to be more \textbf{mindful} \\ of my own thoughts.\end{tabular} & 3.61 & <.05 & 3.72 $\pm$ 0.96 & 4.39 $\pm$ 0.61$^{\alpha,\beta}$ & 3.22 $\pm$ 1.56$^{\alpha}$ & 3.28 $\pm$ 1.45$^{\beta}$ \\ [0.5cm]
\begin{tabular}[c]{@{}c@{}} Going through this process has \\ helped me to be more \textbf{aware} \\ of my own thoughts.\end{tabular} & 3.88 & <.05 & 3.72 $\pm$ 0.96 & 4.33 $\pm$ 0.59$^{\alpha}$ & 3.06 $\pm$ 1.47$^{\alpha}$ & 3.28 $\pm$ 1.56 \\ [0.5cm]
\begin{tabular}[c]{@{}c@{}} Going through this process has \\ helped me to be more \textbf{confident} \\ of my own opinions.\end{tabular} & 4.69 & <.01 & 3.61 $\pm$ 0.85 & 4.39 $\pm$ 0.98$^{\alpha,\beta}$ & 3.28 $\pm$ 1.45$^{\alpha}$ & 3.00 $\pm$ 1.33$^{\beta}$ \\ 
 \hline
 \end{tabular}}
\end{table*}

\subsubsection{Perceived Helpfulness}
\label{sec:perceived helpfulness}
Results for \textit{Perceived Helpfulness} is summarized in Table \ref{tab:summary of perceived helpfulness}. Pairwise t-test with Benjamini-Hochberg correction was conducted only for the first question item. We found a significant main effect of \emph{Time Nudges} for both question items. Participants perceive \emph{Simple Time Nudge} ($M = 4.61$) to help them reflect better compared to the other three nudges. Participants also perceived \emph{Simple Time Nudge} to be a better way in encouraging people to take more time to review before submitting when compared to \emph{Fixed Time Frame}.   

\begin{table*}[!htbp]
 \caption{Perceived Helpfulness. $\alpha$, $\beta$ and $\gamma$ show significant pairwise comparisons ($p<.05$) for each question item.}
 \label{tab:summary of perceived helpfulness}
\scalebox{0.85}{\begin{tabular}{@{}ccccccc@{}}
 \hline
  \multirow{4}{*}{\textbf{Perceived Helpfulness}} & \multirow{4}{*}{\textbf{$F_{3,68}$}} & \multirow{4}{*}{\textbf{$p$}} & \multicolumn{4}{c}{\textbf{Time Nudges}} \\ \cline{4-7}
  & & &\textbf{\begin{tabular}[c]{@{}c@{}} Fixed \\ Time Frame\\ ($M \pm S.D.$)\end{tabular}} & \textbf{\begin{tabular}[c]{@{}c@{}} Simple \\ Time Nudge\\ ($M \pm S.D.$)\end{tabular}} & \textbf{\begin{tabular}[c]{@{}c@{}} Comparison \\ Time Nudge\\ ($M \pm S.D.$)\end{tabular}} & \textbf{\begin{tabular}[c]{@{}c@{}} Alert \\ Message\\ ($M \pm S.D.$)\end{tabular}} \\   
 \hline
 \begin{tabular}[c]{@{}c@{}} This process allows me to \textbf{reflect} \\ on my opinions before submitting.\end{tabular} & 2.91 & <.05 & 3.72 $\pm$ 1.13$^{\alpha}$ & 4.61 $\pm$ 0.50$^{\alpha,\beta,\gamma}$ & 3.83 $\pm$ 1.15$^{\beta}$ & 3.83 $\pm$ 1.15$^{\gamma}$ \\ [0.3cm]
 \begin{tabular}[c]{@{}c@{}} I believe this process is a \textbf{good} \\ \textbf{way} to encourage people to take \\ more time to review on what \\ they have written before submitting.\end{tabular} & 2.92 & <.05 & 3.44 $\pm$ 1.04$^{\alpha}$ & 4.33 $\pm$ 0.59$^{\alpha}$ & 3.72 $\pm$ 0.96 & 3.94 $\pm$ 1.06 \\ 
 \hline
 \end{tabular}}
\end{table*}

\subsubsection{Summary of Quantitative Results}
Overall, there is no significant difference in terms of deliberativeness across the four time nudges, which was explainable because all the  reflection times are arbitrary close or beyond the optimal time of Four minutes (see section~\ref{sec:study2 deliberation time}) - \emph{Fixed Time Frame} ($M = 3.5$ minutes), \emph{Simple Time Nudge} ($M = 5.8$ minutes), \emph{Comparison Time Nudge} ($M = 4.4$ minutes) and \emph{Alert Message} ($M = 3.0$ minutes). This is inline with study 1 results where we observed deliberativeness to have a plateauing effect after the Three minutes mark.

Although there is no significant difference in terms of deliberativeness across the four time nudges, they successfully convince users to spend more time to reflect. However, users' experience for each time nudge differs greatly. Significant effects of the \emph{Time Nudges} were found to be subjectively helpful to users in terms of their thinking process and perceived helpfulness. One condition that stands out from the others seems to be \emph{Simple Time Nudge}, as participants felt less pressure, did not find it distractive, found it helpful in their thinking process and generally found it useful. 

\subsection{Qualitative Results}
\label{sec:subjective preferences}
In this section, we present our findings derived from the post-task feedback. We first detail the benefits of the time nudges followed by their drawbacks.

\subsubsection{Common Benefit: Self-evaluate, Rethink and Reflect}
Overall, users reported two key benefits in all of the time nudges. First, the time nudges allow users to self-evaluate, rethink and reflect. Participants appreciated the additional time provided by the time nudges as it offered a brief pause for reflection (Alert Message, P12), allow them to rethink their previous responses (Fixed Time Frame, P12), consider what they want to communicate in their writing and make adjustments and additions accordingly (Fixed Time Frame, P12; Simple Time Nudge, P4), and even enhanced the organization of their thoughts (Alert Message, P12). All of these are succinctly summarized by P17 from Fixed Time Frame condition: \textit{`I think the timer [...] makes you want to go back and adjust what you wrote especially for those people who completed their thoughts very quickly. It allows a bit of time to finish, reflect and add more if needed.'}

\subsubsection{Common Benefit: Review, Re-read and Checks}
Secondly, users reported that the time nudges allow them to review, re-read and to check for mistakes. They appreciated the opportunity for revisions and error avoidance from posting anything they might regret typing (Fixed Time Frame, P18) and to correct grammatical or thought errors caused by momentary distractions (Comparison Time Nudge, P17). Others appreciated the opportunity for double checking what they have written to see if they could fix up their thoughts or strengthen them (Simple Time Nudge, P10). These benefits are succinctly put forth by P18 from Comparison Time Nudge condition: \textit{`I like how the reminders made me check for silly typos. I always make them no matter how slow my original formulation/typing is. And a reminder to check and make sure you have missed anything can be useful for jogging the brain and refining initial thoughts.'} The above two benefits were present in the top key benefits for all of the time nudges.

\subsubsection{Fixed Time Frame: More Focused}
\label{sec:focused}
Unique benefits pertaining to the specific time nudges are described below. For Fixed Time Frame, users reported that it allows them to be more focused at the task at hand. \textit{`It allows me to concentrate as best [as I could] so that I can answer the question within the time limit.'} (P2).

\subsubsection{Simple Time Nudge: Confirmation}
For Simple Time Nudge, specific benefits include serving as a confirmation for users to be sure of their opinions. \textit{`I like how it actually makes me want to check over what I had written and to really take in the ideas and thoughts I have and make sure they're the ones that I want to [say].'} (P3). Another participant echoed the same: \textit{`[...] having the opportunity to review, reflect and confirm that this is really your opinion. This is good so people don't make a hasty opinion.'} (P16)

\subsubsection{Simple Time Nudge \& Alert Message: Prevent Impulsive Comments}
Both Simple Time Nudge and Alert Message prevent impulsivity. Participants were quoted saying that the time nudges help them realize that: \textit{`we can't be anxious and should pay more attention; not to be hasty, and be sure that our writing reflects our true opinion.'} (Simple Time Nudge, P14). As such, it encourages thoughtful contemplation on the issue rather than providing spontaneous reactions (Alert Message, P13). This could also help prevent negative online phenomena, as participants noted that both time nudges provided an opportunity to cool down before posting a hot-headed comment in a rage (Simple Time Nudge, P12) and may compel individuals to change their minds before posting toxic comments (Alert Message, P18).

\subsubsection{Alert Message: Encourage Sincere Opinion}
Moreover, Alert Message was reported to encourage users to provide a more sincere opinion. \textit{`It helps me give the best answers I can provide knowing that I had the time to do so.'} (P3). Another participant also echoed that it prompts them to give a more truthful response: \textit{`I think what I like most is you do find yourself thinking deeper about what you really feel and think, and I think you give a more truthful, well-thought out response.'} (P7).

\subsubsection{Common Drawbacks: Time-consuming/Extra steps}
A key dislike present in all time nudges is that they can be time-consuming. In Fixed Time Frame, a user responded: \textit{`I did not like that it takes a very long time for the submit  button to appear, and I had to spend the extra time either on filler or rambling.'} (P9). For Simple Time Nudge, users noted: \textit{`I like least that I need to spend additional time to reflect on my communication which isn't always what I do.'} (P4). Overall participants complained about the extra steps required to validate their messages in every condition.

\subsubsection{Fixed Time Frame: Time Pressure \& Being Forced}
Unique drawback for Fixed Time Frame is that it induces time pressure on users. \textit{`I hate the way it causes the likelihood of panicking and hurriedness.'} (P16), and \textit{`I don't think working against a running clock is the best way. Some people who are very indecisive and anxious would need to write their thoughts down.'} (P17). Additionally, users feel forced when adhering to the strict time rules. \textit{`I hate how it arbitrarily stops you from advancing. People don't like being forced to do things and I am one of them.'} (P18).

\subsubsection{Comparison Time Nudge: Self-doubt and Discouraging \& Induce Pressure}
\label{sec:self-doubt}
Specific drawback of the Comparison Time Nudge is that it induces self-doubt and discourages users which has a spillover effect of inducing pressure. \textit{`It sounds kind of condescending. Like they're telling you your work is probably wrong or subpar just because you took less time.'} (P4). Another user also noted that: \textit{`I dislike being compared to others, especially since I had spent a lot more time than others, but was still being told to slow down. It felt more like an admonishment overall, rather than an opportunity or encouragement.'} (P18). Another user felt discouraged: \textit{`This prompt suggests that I did not spend enough time thinking about the topic although I feel that I had. This is discouraging.'} (P6). One user even felt self-conscious and felt the need to change his/her way of doing: \textit{`I felt like I didn't write enough. I felt self-conscious on what I was writing and I feel the need to change.'} (P11). Another lost confidence in his/her responses: \textit{`The prompt took away some of the confidence I had that my answer was `correct'.'} (P14). As such, these negative implications pressurized users: \textit{`It makes me feel pressured knowing that I had spent lesser time than others.'} (P15).

\subsubsection{Alert Message: Distractive \& Annoying}
\label{sec:distractive}
Unique drawback for Alert Message is being distractive, leading to some users indicating that the time nudge is annoying. \textit{`I was actually kind of annoyed because the reminder came up when I only typed 3-4 words. It interfered with my thought pattern[s] because I was halfway typing a sentence. It stopped my train of thought, right when I was at the mid-sentence. It stopped me and I was kind of put out. It might help someone rephrase or re-frame their written text, but mostly it make me kind of mad.'} (P2). Another was quoted saying: \textit{`[...] That pop-up box was really annoying.'} (P17).

\paragraph{\textup{Key benefits and drawbacks for the respective time nudges are summarized in Table~\ref{tab:likes-dislikes}. For a comprehensive review of the feedback associated with the four time nudges, please refer to Appendix Figures~\ref{fig:likes fixedtimeframe} -~\ref{fig:dislikes alertmessage}.}}  

\begin{table*}[!htbp]
\caption{Key Benefits and Drawbacks for each Time Nudge. Note that this table presents only the most significant benefits and drawbacks. Benefits and drawbacks are ranked with the top being the most key for the respective time nudge. For benefits and drawbacks that are ranked on the same footing, they are in the colour \textcolor{cyan}{blue} (benefits) and \textcolor{magenta}{pink} (drawbacks). Note that Simple Time Nudge only has one key drawback.}
\label{tab:likes-dislikes}
\scalebox{0.9}{
\begin{tabular}{@{}lll@{}}
\toprule
\textbf{Time Nudge Interface} & \textbf{Key Benefits} & \textbf{Key Drawbacks}\\
\midrule
\multirow{3}{*}{Fixed Time Frame} & Self-evaluate, Rethink \& Reflect & Time Pressure \\ & Review, Re-read \& Check & Time Consuming \\ & More Focused & Being Forced\\ 
\midrule
\multirow{4}{*}{Simple Time Nudge} & \textcolor{cyan}{Self-evaluate, Rethink \& Reflect} & Time Consuming \\ & \textcolor{cyan}{Review, Re-read \& Check} \\ & Confirmation \\ & Prevent Impulsive Comments \\
\midrule
\multirow{3}{*}{Comparison Time Nudge} & Self-evaluate, Rethink \& Reflect & Self-doubt, Discouraging \\ & Review, Re-read \& Check & \textcolor{magenta}{Time Consuming} \\ & & \textcolor{magenta}{Pressure} \\ 
\midrule
\multirow{4}{*}{Alert Message} & Self-evaluate, Rethink \& Reflect & Distractive \\ & Prevent Impulsive Comments & \textcolor{magenta}{Time Consuming} \\ & \textcolor{cyan}{Review, Re-read \& Check} & \textcolor{magenta}{Annoying} \\ & \textcolor{cyan}{Encourage Sincere Opinions} \\
\bottomrule
\end{tabular}}
\end{table*}    
\subsection{Discussion}
In this section, we discuss our findings to answer \textbf{RQ2}. We synthesized both quantitative data (see section~\ref{sec: quanti study 2}) and qualitative data (see section~\ref{sec:subjective preferences}) as well as findings from our initial study to provide a comprehensive assessment of the time nudges on deliberativeness.

\subsubsection{Time Length is Key}
Overall, our results from studies 1 and 2 showed that the length of time to reflect is key.

\paragraph{Longer Time is Beneficial...}
Results from study 1 showed increase deliberativeness with longer reflection time length, especially between One/Two minutes conditions compared to Three/Four-Five minutes condition. This suggests that, for minute-scale deliberations, reflection time is key to increasing deliberativeness of online opinions.

\paragraph{...But has Diminishing Returns.}
We did however notice that the increase in deliberativeness was plateauing after the Three minutes mark. While this specific value may be topic-dependent, forcing users to wait too long may not be beneficial, as they would simply wait it out, or simply drop and not post opinions. Our modelling of deliberativeness over time shows a clear logarithmic relationship between the two quantities.

\subsubsection{Time Nudges Trigger Time-Induced Reflection}
Study 2 showed that each time nudge effectively extended participants' reflection time, approaching or surpassing the optimal reflection time determined in study 1.

\paragraph{Time Nudges may Not Improve Deliberativeness...} 
From our findings, we found that the different time nudges may not improve overall deliberativeness (see section \ref{sec:study2 opinion quality}).

\paragraph{...But Trigger Time-Induced Reflection}
However, each time nudge managed to convince participants to spend more time and reach around the optimal reflection time determined in study 1 (see section~\ref{sec:study2 deliberation time}). Moreover, the key benefit these time nudges served was to allow users to self-evaluate, rethink and reflect better (see section \ref{sec:subjective preferences}). This would suggest that the time nudges at some point triggered time-induced reflection - using time to help participants to reflect more - but we could not sieve out which time nudge had done better in terms of deliberative quality. However, users' experience with each time nudge greatly differs as we discussed below.

\subsubsection{Fixed Time Frame Induces Stress}
Fixed Time Frame was intended to stop users from progressing through only after reflecting for a certain time duration. However, both quantitative and qualitative results showed that it induces a psychological stress of time pressure. While this stress could be an asset for some such as making users to be more focused at the task (see section~\ref{sec:focused}), it may not be well-received by the majority. As seen from Table \ref{tab:likes-dislikes}, users indicated time pressure to be their top drawback. Thus, our findings suggest that the \textbf{usefulness of Fixed Time Frame is very much dependent on the users' tolerance to stress}. As the majority do not sit well with it, \textit{`working against a running clock'} might not be the best way to support reflection. In this way, Fixed Time Frame could be seen as a \textbf{double-edge sword}. It does allows users to reflect and review before posting but the key drawback is having to endure the psychological stress. This time nudge was an interesting baseline that allowed us to understand the interplay between time, user experience and deliberativeness.

\subsubsection{Simple Time Nudge Subjectively Better Overall}
Our quantitative results showed a significant effect of Simple Time Nudge on \textit{Thinking Process} and \textit{Perceived Helpfulness}. This was also visibly seen from users' feedback wherein the top three key benefits were the ability to reflect, review, provide assurance and confirmation as well as preventing impulsivity.

A closer look at the reflection time induced by Simple Time Nudge amounts to 5.8 minutes (see section \ref{sec:study2 deliberation time}) which is beyond the optimal time deduced in study 1. Yet, no significant difference was reported in terms of deliberativeness which is inline with study 1 results. In this case, Simple Time Nudge may have increase deliberativeness by means of helping users to reflect better, depicted in both our quantitative and qualitative results. But \textbf{any additional reflection time beyond optimality helps users psychologically by means of making them `feel better'} as they are able to look back at their previous response before submission. Providing confirmation and assurance was reflected as one of the key benefit of Simple Time Nudge. However, \textbf{this assurance comes at a cost as users reported that this process is time consuming} and not all users have the available time to do so (see Table~\ref{tab:likes-dislikes} which depicts time consuming as the sole key drawback). Hence, additional reflection beyond optimality induced by the time nudge only gives users a \textbf{peace of mind} and does not seek to improve their actual deliberativeness.   

\subsubsection{Comparison Time Nudge - Don't Compare Users' Performance for Reflection}
We design Comparison Time Nudge using a goal-based time anchor to influence users to move towards reflecting arbitrarily close to the optimal reflection time. We do this by comparing the actual reflection time users took versus the time they should work towards, in the hopes of increasing their deliberative quality. Instead, we found opposite effects. Users found that the comparison is discouraging them, with some saying that it is an \textit{`admonishment'}, leading them to feel pressurized. This is in contrast with previous studies that a judgemental and goal-based anchor bring change to individuals' performance~\cite{menon2020nudge, cervone1986anchoring}. Furthermore, as individuals react negatively towards Comparison Time Nudge - having feelings of self-consciousness and lowering of their self-confidence (see section~\ref{sec:self-doubt}), naturally they perceive the nudge to be more distractive than Simple Time Nudge (see section~\ref{sec:time manipulation}). Distractive here could have implied that the time nudge was more disturbing to users as it cause them to have unfavourable feelings. This indicates that \textbf{reflection is a personal thing and therefore, should not be compared with others}.

\subsubsection{Alert Message for Sincere Opinions but Needs Tweaking}
Users' feedback that the Alert Message was distracting, causing them to lose their train of thoughts (see section~\ref{sec:distractive}). This may be because we set the Alert Message to appear at 20 seconds (see section~\ref{sec: study2 alert message design}) which might be too early. To mitigate this, the Alert Message could be delayed at a longer interval (e.g., 40 seconds) but still appear before users submit their opinions. This adjustment could help users provide more thoughtful and non-hasty opinions, as users found the Alert Message to be a good reminder (see Table~\ref{tab:likes-dislikes}). However, it is challenging to predict exactly when users will submit their opinions, making the timing of nudge inherently uncertain~\cite{caraban201923}. 
\section{Implications For Design}
\subsection{Design Implications}
To our knowledge, this work is the first to explore the impact of reflection time on deliberativeness in minute-scale deliberation. Based on the outcomes of studies 1 and 2, we have derived guidelines to inform designers and researchers about the influence of reflection time in this context.

\subsubsection{Consider the Optimal Timing for Reflection (Study 1)}
We found that there exists a Goldilocks principle of `optimal' time in reflection. Specifically, there is a diminishing marginal rate of return for deliberative quality after the optimal time. Hence, it would be meaningless to just ask users to \textit{`take more time to reflect'}. Instead, we propose encouraging users to reflect close to or at the optimal time. This is because deliberative quality significantly improves for users who reflect below optimal time. As it might be difficult to find the exact optimality point, based on our study, we propose a range of three minutes to four minutes for reflection in the context of minute-scale deliberation.

Subsequently, we design different time nudges in study 2 to nudge users who reflect below optimality to take more time to reflect, ultimately moving them towards optimal time. In the case where users have reflected till the optimal time length, we recommend applying \textbf{non-time nudges} to further support their reflection process. For example, the use of thinking prompts~\cite{10.1145/3613904.3642530} could be helpful and complement the timing approach.

\subsubsection{Time Factors to Consider to Enhance Reflection (Study 2)} 
We found that each time nudge uniquely affected users' experiences. Based on our results, we created a set of four design guidelines (G1-G4).

\paragraph{\textbf{G1:} Avoid Permanent Timers on the Interface.}
For Fixed Time Frame, restricting users to reflect against a countdown timer significantly increases their stress, which is not ideal. To mitigate this, we could remove the auto submission feature and have the countdown timer to be used only at the start to allow users to read and think. This would take away the effect of time pressure as users can submit their responses as and when they are done. Putting the timer at the start would also stimulate reflection by getting users to get into the momentum. In cases where working against a running clock proves to be effective, a possible mitigation could be using the concept of time blindness \cite{birth2017time} by hiding the timer and only allowing the submit button to appear after a certain reflection time is reached. All this while, letting users know that the submit button will appear after a certain reflection time (e.g., 3.5 minutes). This would mean that although users innately know that there is a timer, they are unable to sense the passing of time, thus reducing their stress. 

\paragraph{\textbf{G2:} Confirmation and Validation is Essential.}
Confirmation and validation proves to be essential in the reflection process, evident in the case of Simple Time Nudge. Though it may not necessarily translate to an increase in deliberativeness, taking in one's own thoughts before posting is perceived to be helpful in users' thinking process. Therefore, having an opportunity for users to check back before posting may help in their psychological thinking.  

\paragraph{\textbf{G3:} Reflection is Personal - Don't Compare Performance.}
From Comparison Time Nudge, we learnt that reflection is personal and thus should not be compared with others. Hence, we propose that designers should not design time nudges that seek to compare oneself with others.

\paragraph{\textbf{G4:} Careful use of Time Nudge in a Non-Distractive Way.}
Finally, time nudges could act as a helpful and non-distractive reminder if it appears at a later timing but just before users post their opinions, as in the case of Alert Message. However, this could be tricky as it can be hard to find the best time interval for the nudge to appear~\cite{caraban201923}. From our results, we suggest that the nudge could occur between two to three minutes, at the half past mark time interval just before the optimal time ends for the targeted behaviour to take place. 

\subsection{Implications to Other Areas}
While the task performed in this study was specific and done in a minute-scale deliberation context, we believe that our results could pave the way towards leveraging reflection time in other contexts.

\subsubsection{Online Discussions} 
Online discussions enable users to engage anonymously or pseudonymously, which can lead to toxic online disinhibition~\cite{voggeser2018self, joinson2003understanding, joinson2007disinhibition, suler2004online, lapidot2012effects}. Trolling phenomena such as flame wars occurs on impulse~\cite{voggeser2018self} and derail constructive societal and political discourse~\cite{voggeser2018self}. Moreover, poor design of online discussions platforms can further exacerbate these issues by encouraging ill-intentioned behavior.

Integrating a reflection duration on platforms like Reddit could reduce the prevalence of such phenomena by preventing impulsivity and regrets associated with users' postings~\cite{wang2011regretted}, evident from our findings. The idea of giving time for users to reflect before posting has been implemented on subreddit /r/discussion\_patiente/\footnote{\href{https://www.reddit.com/r/discussion\_patiente/}{subreddit Patient Discussion}} (i.e., `patient discussion' in French) where commenters need to wait 24 hours to fully reflect before posting. Our results suggest that even a short delay of three to four minutes can significantly enhance deliberativeness, without the need for an excessively long wait.

Moreover, specialized communities such as subreddit/r/AskHistorians/\footnote{\href{https://www.reddit.com/r/AskHistorians/}{subreddit Ask Historians}}, which enforce strict community rules and emphasize quality answers (i.e., `\textit{Answers must be in-depth and comprehensive, or they will be removed}'), could benefit from integrating a time nudge. This could enhance response quality by encouraging users to invest more time in thoughtful reflection before posting, potentially reducing the need for moderators over time.

\subsubsection{Online Review Systems} 
Online reviewing platforms such as Yelp or Rotten Tomatoes could benefit from adding a reflection delay before users post reviews. For instance, allowing users to post only after a two minutes reflection period. This can be easily implemented by displaying the submit button after two minutes. Such a delay could foster higher levels of empathy and responsible online behaviour. For instance, users unhappy with a restaurant service can cool down before posting, preventing impulsive comments — a key benefit reflected in our results. Additionally, it could reduce review-bombing\footnote{\href{https://www.hollywoodreporter.com/tv/tv-news/lord-of-the-rings-the-rings-of-power-amazon-review-bombed-1235211190/}{Hollywood Review-Bombing}}, where products or services are unfairly judged based on biases rather than merits. As immediate posting is not crucial, leveraging time nudges could lead to more nuanced and balanced reviews.

\subsubsection{Beyond Human-Generated Content Platforms}
Time nudges designed for minute-scale deliberation can be adapted in other contexts such as decision-making processes of non-human agents. For instance, integrating reflection periods in autonomous systems to allow the system to consult its decision-making algorithms can enhance decision quality. 

In hybrid human-AI collaborative environments, integrating a reflection period can improve the quality of joint decision-making and optimize the interaction between human and AI. For instance, after the AI presents information, a prompt can appear, suggesting the human to take a two-minute reflection period to reflect on the AI's suggestions and consider their own knowledge. This would ensure that decisions made are not hasty and allow for a more thorough deliberation process, combining the strengths of both human judgment and artificial intelligence.
\section{Limitations}
There are limitations in this work. Firstly, the tasks in both studies were specific, aiming to achieve depth and clarity in understanding the influences of reflection time within a well-defined context~\cite{slack2001establishing}. While this establishes internal validity, it may limit generalizability and ecological validity. Assessing ecological validity relies on first establishing internal validity~\cite{cahit2015internal, campbell2015experimental, cook1979quasi, slack2001establishing}, thus our focused approach allows for a thorough examination of reflection time on deliberativeness. Future research could extend the application of reflection time on other contentious topics with varying complexity and nature. 

Similarly, both studies investigates reflection times in the context of minute-scale deliberation. This focus allows us to capture the fast-paced nature and rapid interaction on online deliberation platforms~\cite{takayoshi2015short, umansky2023dances, zhang2024form, zhang2022driving, jarvenpaa2010research, jarvenpaa2010research}. Future work could explore different deliberation contexts, including those with longer deliberation time frames. 

Lastly, in study 1, we identified an optimal reflection time of four minutes using a public issue of four-day workweek. However, this optimal reflection time may vary with the complexity of the issue/theme discussed and the exact nature of the task in realistic contexts. Therefore, we do not claim the generalization of the optimal timing of four minutes. 
\section{Conclusion}
In this work, we investigated how reflection time impacts deliberativeness in minute-scale deliberations. We provided the missing link between the two by conducting two studies. In the first study, our results suggest that there exists an optimal reflection time on the quality of minute-scale deliberation when writing a short opinion comment, and that increasing reflection time is most favourable when users reflect below the optimum. In the second study, we provide design considerations for enacting the dimension of time where it is warranted to account for better deliberative quality in minute-scale deliberations, through direct applications of nudging techniques. Our results expand current work on minute-scale deliberation and is a first step towards a more complete account of how time supports reflection in the deliberation process as well as how researchers and designers can better incorporate time in minute-scale deliberations. 

\bibliographystyle{ACM-Reference-Format}
\bibliography{reference.bib}


\begin{thebibliography}{108}


\ifx \showCODEN    \undefined \def \showCODEN     #1{\unskip}     \fi
\ifx \showDOI      \undefined \def \showDOI       #1{#1}\fi
\ifx \showISBNx    \undefined \def \showISBNx     #1{\unskip}     \fi
\ifx \showISBNxiii \undefined \def \showISBNxiii  #1{\unskip}     \fi
\ifx \showISSN     \undefined \def \showISSN      #1{\unskip}     \fi
\ifx \showLCCN     \undefined \def \showLCCN      #1{\unskip}     \fi
\ifx \shownote     \undefined \def \shownote      #1{#1}          \fi
\ifx \showarticletitle \undefined \def \showarticletitle #1{#1}   \fi
\ifx \showURL      \undefined \def \showURL       {\relax}        \fi
\providecommand\bibfield[2]{#2}
\providecommand\bibinfo[2]{#2}
\providecommand\natexlab[1]{#1}
\providecommand\showeprint[2][]{arXiv:#2}

\bibitem[Al{\'o}s-Ferrer and Buckenmaier(2021)]%
        {alos2021cognitive}
\bibfield{author}{\bibinfo{person}{Carlos Al{\'o}s-Ferrer} {and} \bibinfo{person}{Johannes Buckenmaier}.} \bibinfo{year}{2021}\natexlab{}.
\newblock \showarticletitle{Cognitive sophistication and deliberation times}.
\newblock \bibinfo{journal}{\emph{Experimental Economics}} \bibinfo{volume}{24}, \bibinfo{number}{2} (\bibinfo{year}{2021}), \bibinfo{pages}{558--592}.
\newblock


\bibitem[Anderson et~al\mbox{.}(2016)]%
        {anderson2016all}
\bibfield{author}{\bibinfo{person}{Ross~C Anderson}, \bibinfo{person}{Meg Guerreiro}, {and} \bibinfo{person}{Joanna Smith}.} \bibinfo{year}{2016}\natexlab{}.
\newblock \showarticletitle{Are all biases bad? Collaborative grounded theory in developmental evaluation of education policy}.
\newblock \bibinfo{journal}{\emph{Journal of Multidisciplinary Evaluation}} \bibinfo{volume}{12}, \bibinfo{number}{27} (\bibinfo{year}{2016}), \bibinfo{pages}{44--57}.
\newblock


\bibitem[Arceneaux and Vander~Wielen(2017)]%
        {arceneaux2017taming}
\bibfield{author}{\bibinfo{person}{Kevin Arceneaux} {and} \bibinfo{person}{Ryan~J Vander~Wielen}.} \bibinfo{year}{2017}\natexlab{}.
\newblock \bibinfo{booktitle}{\emph{Taming intuition: How reflection minimizes partisan reasoning and promotes democratic accountability}}.
\newblock \bibinfo{publisher}{Cambridge University Press}.
\newblock


\bibitem[Bandura(1982)]%
        {bandura1982self}
\bibfield{author}{\bibinfo{person}{Albert Bandura}.} \bibinfo{year}{1982}\natexlab{}.
\newblock \showarticletitle{Self-efficacy mechanism in human agency.}
\newblock \bibinfo{journal}{\emph{American psychologist}} \bibinfo{volume}{37}, \bibinfo{number}{2} (\bibinfo{year}{1982}), \bibinfo{pages}{122}.
\newblock


\bibitem[Barnes(1984a)]%
        {barnes1984complete}
\bibfield{author}{\bibinfo{person}{Jonathan Barnes}.} \bibinfo{year}{1984}\natexlab{a}.
\newblock \bibinfo{booktitle}{\emph{Complete works of Aristotle, volume 1: The revised Oxford translation}}. Vol.~\bibinfo{volume}{1}.
\newblock \bibinfo{publisher}{Princeton University Press}.
\newblock


\bibitem[Barnes(1984b)]%
        {Barnes1984-BARCWO-2}
\bibfield{editor}{\bibinfo{person}{Jonathan Barnes}} (Ed.). \bibinfo{year}{1984}\natexlab{b}.
\newblock \bibinfo{booktitle}{\emph{Complete Works of Aristotle, Volume 2: The Revised Oxford Translation}}.
\newblock \bibinfo{publisher}{Princeton University Press}.
\newblock


\bibitem[Bertot et~al\mbox{.}(2010)]%
        {Bertot2010}
\bibfield{author}{\bibinfo{person}{John~Carlo Bertot}, \bibinfo{person}{Paul~T. Jaeger}, \bibinfo{person}{Sean Munson}, {and} \bibinfo{person}{Tom Glaisyer}.} \bibinfo{year}{2010}\natexlab{}.
\newblock \showarticletitle{Social Media Technology and Government Transparency}.
\newblock \bibinfo{journal}{\emph{Computer}} \bibinfo{volume}{43}, \bibinfo{number}{11} (\bibinfo{year}{2010}), \bibinfo{pages}{53--59}.
\newblock
\urldef\tempurl%
\url{https://doi.org/10.1109/MC.2010.325}
\showDOI{\tempurl}


\bibitem[Birth(2017)]%
        {birth2017time}
\bibfield{author}{\bibinfo{person}{Kevin Birth}.} \bibinfo{year}{2017}\natexlab{}.
\newblock \bibinfo{booktitle}{\emph{Time Blind}}.
\newblock \bibinfo{publisher}{Springer}.
\newblock


\bibitem[Bishop et~al\mbox{.}(1980)]%
        {Bishop1980}
\bibfield{author}{\bibinfo{person}{George~F. Bishop}, \bibinfo{person}{Robert~W. Oldendick}, \bibinfo{person}{Alfred~J. Tuchfarber}, {and} \bibinfo{person}{Stephen~E. Bennett}.} \bibinfo{year}{1980}\natexlab{}.
\newblock \showarticletitle{Pseudo-opinions on public affairs.}
\newblock \bibinfo{journal}{\emph{Public Opinion Quarterly}}  \bibinfo{volume}{44} (\bibinfo{year}{1980}), \bibinfo{pages}{198--209}.
\newblock
\urldef\tempurl%
\url{https://doi.org/10.1086/268584}
\showDOI{\tempurl}


\bibitem[Bohman(2000)]%
        {bohman2000public}
\bibfield{author}{\bibinfo{person}{James Bohman}.} \bibinfo{year}{2000}\natexlab{}.
\newblock \bibinfo{booktitle}{\emph{Public deliberation: Pluralism, complexity, and democracy}}.
\newblock \bibinfo{publisher}{MIT press}.
\newblock


\bibitem[Cahit(2015)]%
        {cahit2015internal}
\bibfield{author}{\bibinfo{person}{Kaya Cahit}.} \bibinfo{year}{2015}\natexlab{}.
\newblock \showarticletitle{Internal validity: A must in research designs}.
\newblock \bibinfo{journal}{\emph{Educational Research and Reviews}} \bibinfo{volume}{10}, \bibinfo{number}{2} (\bibinfo{year}{2015}), \bibinfo{pages}{111--118}.
\newblock


\bibitem[Campbell and Stanley(2015)]%
        {campbell2015experimental}
\bibfield{author}{\bibinfo{person}{Donald~T Campbell} {and} \bibinfo{person}{Julian~C Stanley}.} \bibinfo{year}{2015}\natexlab{}.
\newblock \bibinfo{booktitle}{\emph{Experimental and quasi-experimental designs for research}}.
\newblock \bibinfo{publisher}{Ravenio books}.
\newblock


\bibitem[Cappella et~al\mbox{.}(2002)]%
        {cappella2002argument}
\bibfield{author}{\bibinfo{person}{Joseph~N Cappella}, \bibinfo{person}{Vincent Price}, {and} \bibinfo{person}{Lilach Nir}.} \bibinfo{year}{2002}\natexlab{}.
\newblock \showarticletitle{Argument repertoire as a reliable and valid measure of opinion quality: Electronic dialogue during campaign 2000}.
\newblock \bibinfo{journal}{\emph{Political Communication}} \bibinfo{volume}{19}, \bibinfo{number}{1} (\bibinfo{year}{2002}), \bibinfo{pages}{73--93}.
\newblock


\bibitem[Caraban et~al\mbox{.}(2019)]%
        {caraban201923}
\bibfield{author}{\bibinfo{person}{Ana Caraban}, \bibinfo{person}{Evangelos Karapanos}, \bibinfo{person}{Daniel Gon{\c{c}}alves}, {and} \bibinfo{person}{Pedro Campos}.} \bibinfo{year}{2019}\natexlab{}.
\newblock \showarticletitle{23 ways to nudge: A review of technology-mediated nudging in human-computer interaction}. In \bibinfo{booktitle}{\emph{Proceedings of the 2019 CHI Conference on Human Factors in Computing Systems}}. \bibinfo{pages}{1--15}.
\newblock


\bibitem[Cartwright(1941)]%
        {cartwright1941relation}
\bibfield{author}{\bibinfo{person}{Dorwin Cartwright}.} \bibinfo{year}{1941}\natexlab{}.
\newblock \showarticletitle{Relation of decision-time to the categories of response}.
\newblock \bibinfo{journal}{\emph{The American Journal of Psychology}} \bibinfo{volume}{54}, \bibinfo{number}{2} (\bibinfo{year}{1941}), \bibinfo{pages}{174--196}.
\newblock


\bibitem[Cervone and Peake(1986)]%
        {cervone1986anchoring}
\bibfield{author}{\bibinfo{person}{Daniel Cervone} {and} \bibinfo{person}{Philip~K Peake}.} \bibinfo{year}{1986}\natexlab{}.
\newblock \showarticletitle{Anchoring, efficacy, and action: The influence of judgmental heuristics on self-efficacy judgments and behavior.}
\newblock \bibinfo{journal}{\emph{Journal of Personality and social Psychology}} \bibinfo{volume}{50}, \bibinfo{number}{3} (\bibinfo{year}{1986}), \bibinfo{pages}{492}.
\newblock


\bibitem[Chambers(2003)]%
        {chambers2003deliberative}
\bibfield{author}{\bibinfo{person}{Simone Chambers}.} \bibinfo{year}{2003}\natexlab{}.
\newblock \showarticletitle{Deliberative democratic theory}.
\newblock \bibinfo{journal}{\emph{Annual review of political science}} \bibinfo{volume}{6}, \bibinfo{number}{1} (\bibinfo{year}{2003}), \bibinfo{pages}{307--326}.
\newblock


\bibitem[Charmaz(2014)]%
        {charmaz2014constructing}
\bibfield{author}{\bibinfo{person}{Kathy Charmaz}.} \bibinfo{year}{2014}\natexlab{}.
\newblock \bibinfo{booktitle}{\emph{Constructing grounded theory}}.
\newblock \bibinfo{publisher}{sage}.
\newblock


\bibitem[Chen et~al\mbox{.}(2016)]%
        {chen2016challenges}
\bibfield{author}{\bibinfo{person}{Nan-chen Chen}, \bibinfo{person}{Rafal Kocielnik}, \bibinfo{person}{Margaret Drouhard}, \bibinfo{person}{Vanessa Pe{\~n}a-Araya}, \bibinfo{person}{Jina Suh}, \bibinfo{person}{Keting Cen}, \bibinfo{person}{Xiangyi Zheng}, {and} \bibinfo{person}{Cecilia~R Aragon}.} \bibinfo{year}{2016}\natexlab{}.
\newblock \showarticletitle{Challenges of applying machine learning to qualitative coding}. In \bibinfo{booktitle}{\emph{ACM SIGCHI Workshop on Human-Centered Machine Learning}}.
\newblock


\bibitem[Christiano and Christman(2009)]%
        {christiano2009debates}
\bibfield{author}{\bibinfo{person}{Thomas Christiano} {and} \bibinfo{person}{John Christman}.} \bibinfo{year}{2009}\natexlab{}.
\newblock \bibinfo{booktitle}{\emph{Debates in Political Philosophy}}.
\newblock \bibinfo{publisher}{Wiley Online Library}.
\newblock


\bibitem[Cohen(2013)]%
        {cohen2013statistical}
\bibfield{author}{\bibinfo{person}{Jacob Cohen}.} \bibinfo{year}{2013}\natexlab{}.
\newblock \bibinfo{booktitle}{\emph{Statistical power analysis for the behavioral sciences}}.
\newblock \bibinfo{publisher}{Routledge}.
\newblock


\bibitem[Colusso et~al\mbox{.}(2016)]%
        {colusso2016designing}
\bibfield{author}{\bibinfo{person}{Lucas Colusso}, \bibinfo{person}{Gary Hsieh}, {and} \bibinfo{person}{Sean~A Munson}.} \bibinfo{year}{2016}\natexlab{}.
\newblock \showarticletitle{Designing closeness to increase gamers' performance}. In \bibinfo{booktitle}{\emph{Proceedings of the 2016 CHI Conference on Human Factors in Computing Systems}}. \bibinfo{pages}{3020--3024}.
\newblock


\bibitem[Cook et~al\mbox{.}(1979)]%
        {cook1979quasi}
\bibfield{author}{\bibinfo{person}{Thomas~D Cook}, \bibinfo{person}{Donald~Thomas Campbell}, {and} \bibinfo{person}{Arles Day}.} \bibinfo{year}{1979}\natexlab{}.
\newblock \bibinfo{booktitle}{\emph{Quasi-experimentation: Design \& analysis issues for field settings}}. Vol.~\bibinfo{volume}{351}.
\newblock \bibinfo{publisher}{Houghton Mifflin Boston}.
\newblock


\bibitem[Corbin and Strauss(2014)]%
        {corbin2014basics}
\bibfield{author}{\bibinfo{person}{Juliet Corbin} {and} \bibinfo{person}{Anselm Strauss}.} \bibinfo{year}{2014}\natexlab{}.
\newblock \bibinfo{booktitle}{\emph{Basics of qualitative research: Techniques and procedures for developing grounded theory}}.
\newblock \bibinfo{publisher}{Sage publications}.
\newblock


\bibitem[Davies and Chandler(2013)]%
        {davies2013online}
\bibfield{author}{\bibinfo{person}{Todd Davies} {and} \bibinfo{person}{Reid Chandler}.} \bibinfo{year}{2013}\natexlab{}.
\newblock \showarticletitle{Online deliberation design: Choices, criteria, and evidence}.
\newblock \bibinfo{journal}{\emph{arXiv preprint arXiv:1302.5177}} (\bibinfo{year}{2013}).
\newblock


\bibitem[Davies and Gangadharan(2009)]%
        {Davies2009}
\bibfield{author}{\bibinfo{person}{Todd Davies} {and} \bibinfo{person}{Seeta~Peña Gangadharan}.} \bibinfo{year}{2009}\natexlab{}.
\newblock \bibinfo{booktitle}{\emph{Online Deliberation: Design, Research, and Practice}}.
\newblock \bibinfo{publisher}{CSLI Publications}.
\newblock


\bibitem[Dawson(2011)]%
        {dawson2011significant}
\bibfield{author}{\bibinfo{person}{Robert Dawson}.} \bibinfo{year}{2011}\natexlab{}.
\newblock \showarticletitle{How significant is a boxplot outlier?}
\newblock \bibinfo{journal}{\emph{Journal of Statistics Education}} \bibinfo{volume}{19}, \bibinfo{number}{2} (\bibinfo{year}{2011}).
\newblock


\bibitem[Dekker and Bekkers(2015)]%
        {Dekker2015}
\bibfield{author}{\bibinfo{person}{Rianne Dekker} {and} \bibinfo{person}{Victor Bekkers}.} \bibinfo{year}{2015}\natexlab{}.
\newblock \showarticletitle{The contingency of governments' responsiveness to the virtual public sphere: A systematic literature review and meta-synthesis}.
\newblock \bibinfo{journal}{\emph{Government Information Quarterly}} \bibinfo{volume}{32}, \bibinfo{number}{4} (\bibinfo{year}{2015}), \bibinfo{pages}{496--505}.
\newblock
\showISSN{0740-624X}
\urldef\tempurl%
\url{https://doi.org/10.1016/j.giq.2015.09.007}
\showDOI{\tempurl}


\bibitem[Dryzek(2002)]%
        {Dryzek2002}
\bibfield{author}{\bibinfo{person}{John~S. Dryzek}.} \bibinfo{year}{2002}\natexlab{}.
\newblock \bibinfo{booktitle}{\emph{{Deliberative Democracy and Beyond: Liberals, Critics, Contestations}}}.
\newblock \bibinfo{publisher}{Oxford University Press}.
\newblock
\showISBNx{9780199250431}
\urldef\tempurl%
\url{https://doi.org/10.1093/019925043X.001.0001}
\showDOI{\tempurl}


\bibitem[Emmons and Diener(1986)]%
        {emmons1986influence}
\bibfield{author}{\bibinfo{person}{Robert~A Emmons} {and} \bibinfo{person}{Ed Diener}.} \bibinfo{year}{1986}\natexlab{}.
\newblock \showarticletitle{Influence of impulsivity and sociability on subjective well-being.}
\newblock \bibinfo{journal}{\emph{Journal of Personality and social psychology}} \bibinfo{volume}{50}, \bibinfo{number}{6} (\bibinfo{year}{1986}), \bibinfo{pages}{1211}.
\newblock


\bibitem[Erickson and Kellogg(2000)]%
        {erickson2000social}
\bibfield{author}{\bibinfo{person}{Thomas Erickson} {and} \bibinfo{person}{Wendy~A Kellogg}.} \bibinfo{year}{2000}\natexlab{}.
\newblock \showarticletitle{Social translucence: an approach to designing systems that support social processes}.
\newblock \bibinfo{journal}{\emph{ACM transactions on computer-human interaction (TOCHI)}} \bibinfo{volume}{7}, \bibinfo{number}{1} (\bibinfo{year}{2000}), \bibinfo{pages}{59--83}.
\newblock


\bibitem[Farina et~al\mbox{.}(2014)]%
        {Farina2014}
\bibfield{author}{\bibinfo{person}{Cynthia~R. Farina}, \bibinfo{person}{Dmitry Epstein}, \bibinfo{person}{Josiah Heidt}, {and} \bibinfo{person}{Mary~J. Newhart}.} \bibinfo{year}{2014}\natexlab{}.
\newblock \showarticletitle{Designing an Online Civic Engagement Platform: Balancing "More" vs. "Better" Participation in Complex Public Policymaking}.
\newblock \bibinfo{journal}{\emph{Int. J. E-Polit.}} \bibinfo{volume}{5}, \bibinfo{number}{1} (\bibinfo{date}{jan} \bibinfo{year}{2014}), \bibinfo{pages}{16–40}.
\newblock
\showISSN{1947-9131}
\urldef\tempurl%
\url{https://doi.org/10.4018/ijep.2014010102}
\showDOI{\tempurl}


\bibitem[Farina et~al\mbox{.}(2012)]%
        {farina2012rulemaking}
\bibfield{author}{\bibinfo{person}{Cynthia~R Farina}, \bibinfo{person}{Mary Newhart}, {and} \bibinfo{person}{Josiah Heidt}.} \bibinfo{year}{2012}\natexlab{}.
\newblock \showarticletitle{Rulemaking vs. democracy: Judging and nudging public participation that counts}.
\newblock \bibinfo{journal}{\emph{Mich. J. Envtl. \& Admin. L.}}  \bibinfo{volume}{2} (\bibinfo{year}{2012}), \bibinfo{pages}{123}.
\newblock


\bibitem[Fearon(1998)]%
        {fearon1998deliberation}
\bibfield{author}{\bibinfo{person}{James~D Fearon}.} \bibinfo{year}{1998}\natexlab{}.
\newblock \showarticletitle{Deliberation as}.
\newblock \bibinfo{journal}{\emph{Deliberative democracy}}  \bibinfo{volume}{1} (\bibinfo{year}{1998}), \bibinfo{pages}{44}.
\newblock


\bibitem[Festinger(1954)]%
        {festinger1954theory}
\bibfield{author}{\bibinfo{person}{Leon Festinger}.} \bibinfo{year}{1954}\natexlab{}.
\newblock \showarticletitle{A theory of social comparison processes}.
\newblock \bibinfo{journal}{\emph{Human relations}} \bibinfo{volume}{7}, \bibinfo{number}{2} (\bibinfo{year}{1954}), \bibinfo{pages}{117--140}.
\newblock


\bibitem[Fleck and Fitzpatrick(2010)]%
        {fleck2010reflecting}
\bibfield{author}{\bibinfo{person}{Rowanne Fleck} {and} \bibinfo{person}{Geraldine Fitzpatrick}.} \bibinfo{year}{2010}\natexlab{}.
\newblock \showarticletitle{Reflecting on reflection: framing a design landscape}. In \bibinfo{booktitle}{\emph{Proceedings of the 22nd Conference of the Computer-Human Interaction Special Interest Group of Australia on Computer-Human Interaction}}. \bibinfo{pages}{216--223}.
\newblock


\bibitem[Fleiss et~al\mbox{.}(1979)]%
        {fleiss1979large}
\bibfield{author}{\bibinfo{person}{Joseph~L Fleiss}, \bibinfo{person}{John~C Nee}, {and} \bibinfo{person}{J~Richard Landis}.} \bibinfo{year}{1979}\natexlab{}.
\newblock \showarticletitle{Large sample variance of kappa in the case of different sets of raters.}
\newblock \bibinfo{journal}{\emph{Psychological bulletin}} \bibinfo{volume}{86}, \bibinfo{number}{5} (\bibinfo{year}{1979}), \bibinfo{pages}{974}.
\newblock


\bibitem[Friess and Eilders(2015)]%
        {friess2015systematic}
\bibfield{author}{\bibinfo{person}{Dennis Friess} {and} \bibinfo{person}{Christiane Eilders}.} \bibinfo{year}{2015}\natexlab{}.
\newblock \showarticletitle{A systematic review of online deliberation research}.
\newblock \bibinfo{journal}{\emph{Policy \& Internet}} \bibinfo{volume}{7}, \bibinfo{number}{3} (\bibinfo{year}{2015}), \bibinfo{pages}{319--339}.
\newblock


\bibitem[Gao et~al\mbox{.}(2023)]%
        {gao2023coaicoder}
\bibfield{author}{\bibinfo{person}{Jie Gao}, \bibinfo{person}{Kenny Tsu~Wei Choo}, \bibinfo{person}{Junming Cao}, \bibinfo{person}{Roy Ka-Wei Lee}, {and} \bibinfo{person}{Simon Perrault}.} \bibinfo{year}{2023}\natexlab{}.
\newblock \showarticletitle{CoAIcoder: Examining the Effectiveness of AI-assisted Human-to-Human Collaboration in Qualitative Analysis}.
\newblock \bibinfo{journal}{\emph{ACM Transactions on Computer-Human Interaction}} (\bibinfo{year}{2023}).
\newblock


\bibitem[Gastil et~al\mbox{.}(2014)]%
        {Gastil2014}
\bibfield{author}{\bibinfo{person}{John Gastil}, \bibinfo{person}{Robert Richards}, {and} \bibinfo{person}{Katherine Knobloch}.} \bibinfo{year}{2014}\natexlab{}.
\newblock \showarticletitle{Vicarious Deliberation: How the Oregon Citizens' Initiative Review Influenced Deliberation in Mass Elections}.
\newblock \bibinfo{journal}{\emph{International Journal of Communication}} \bibinfo{volume}{8}, \bibinfo{number}{0} (\bibinfo{year}{2014}).
\newblock
\showISSN{1932-8036}
\urldef\tempurl%
\url{https://ijoc.org/index.php/ijoc/article/view/2235}
\showURL{%
\tempurl}


\bibitem[Goodin(2000)]%
        {Goodin2000}
\bibfield{author}{\bibinfo{person}{Robert~E. Goodin}.} \bibinfo{year}{2000}\natexlab{}.
\newblock \showarticletitle{Democratic Deliberation within}.
\newblock \bibinfo{journal}{\emph{Philosophy \& Public Affairs}} \bibinfo{volume}{29}, \bibinfo{number}{1} (\bibinfo{year}{2000}), \bibinfo{pages}{81--109}.
\newblock
\showISSN{00483915, 10884963}
\urldef\tempurl%
\url{http://www.jstor.org/stable/2672865}
\showURL{%
\tempurl}


\bibitem[Goodin and Niemeyer(2003)]%
        {Goodin2003}
\bibfield{author}{\bibinfo{person}{Robert~E. Goodin} {and} \bibinfo{person}{Simon~J. Niemeyer}.} \bibinfo{year}{2003}\natexlab{}.
\newblock \showarticletitle{When Does Deliberation Begin? Internal Reflection versus Public Discussion in Deliberative Democracy}.
\newblock \bibinfo{journal}{\emph{Political Studies}} \bibinfo{volume}{51}, \bibinfo{number}{4} (\bibinfo{year}{2003}), \bibinfo{pages}{627--649}.
\newblock
\urldef\tempurl%
\url{https://doi.org/10.1111/j.0032-3217.2003.00450.x}
\showDOI{\tempurl}
\showeprint{https://doi.org/10.1111/j.0032-3217.2003.00450.x}


\bibitem[Gouveia et~al\mbox{.}(2016)]%
        {gouveia2016exploring}
\bibfield{author}{\bibinfo{person}{R{\'u}ben Gouveia}, \bibinfo{person}{F{\'a}bio Pereira}, \bibinfo{person}{Evangelos Karapanos}, \bibinfo{person}{Sean~A Munson}, {and} \bibinfo{person}{Marc Hassenzahl}.} \bibinfo{year}{2016}\natexlab{}.
\newblock \showarticletitle{Exploring the design space of glanceable feedback for physical activity trackers}. In \bibinfo{booktitle}{\emph{Proceedings of the 2016 ACM international joint conference on pervasive and ubiquitous computing}}. \bibinfo{pages}{144--155}.
\newblock


\bibitem[Graham and Witschge(2003)]%
        {graham2003search}
\bibfield{author}{\bibinfo{person}{Todd Graham} {and} \bibinfo{person}{Tamara Witschge}.} \bibinfo{year}{2003}\natexlab{}.
\newblock \showarticletitle{In search of online deliberation: Towards a new method for examining the quality of online discussions}.
\newblock  (\bibinfo{year}{2003}).
\newblock


\bibitem[Greis et~al\mbox{.}(2014)]%
        {greis2014can}
\bibfield{author}{\bibinfo{person}{Miriam Greis}, \bibinfo{person}{Florian Alt}, \bibinfo{person}{Niels Henze}, {and} \bibinfo{person}{Nemanja Memarovic}.} \bibinfo{year}{2014}\natexlab{}.
\newblock \showarticletitle{I can wait a minute: uncovering the optimal delay time for pre-moderated user-generated content on public displays}. In \bibinfo{booktitle}{\emph{Proceedings of the SIGCHI Conference on Human Factors in Computing Systems}}. \bibinfo{pages}{1435--1438}.
\newblock


\bibitem[Habermas(1984)]%
        {Habermas1984}
\bibfield{author}{\bibinfo{person}{J\"urgen Habermas}.} \bibinfo{year}{1984}\natexlab{}.
\newblock \bibinfo{booktitle}{\emph{The theory of communicative action}}.
\newblock \bibinfo{publisher}{Beacon Press}.
\newblock


\bibitem[Hamblin and Crisp(2022)]%
        {hamblin2022negative}
\bibfield{author}{\bibinfo{person}{May Hamblin} {and} \bibinfo{person}{Philippe Crisp}.} \bibinfo{year}{2022}\natexlab{}.
\newblock \showarticletitle{Negative Focus, Self-Doubt, and Issues of ‘Tool Proficiency’: Beginner-Coaches’ Reflections on Reflective Practice}.
\newblock \bibinfo{journal}{\emph{Physical Culture and Sport. Studies and Research}} (\bibinfo{year}{2022}).
\newblock


\bibitem[Hansen and Jespersen(2013)]%
        {hansen2013nudge}
\bibfield{author}{\bibinfo{person}{Pelle~Guldborg Hansen} {and} \bibinfo{person}{Andreas~Maal{\o}e Jespersen}.} \bibinfo{year}{2013}\natexlab{}.
\newblock \showarticletitle{Nudge and the manipulation of choice: A framework for the responsible use of the nudge approach to behaviour change in public policy}.
\newblock \bibinfo{journal}{\emph{European Journal of Risk Regulation}} \bibinfo{volume}{4}, \bibinfo{number}{1} (\bibinfo{year}{2013}), \bibinfo{pages}{3--28}.
\newblock


\bibitem[Harbach et~al\mbox{.}(2014)]%
        {harbach2014using}
\bibfield{author}{\bibinfo{person}{Marian Harbach}, \bibinfo{person}{Markus Hettig}, \bibinfo{person}{Susanne Weber}, {and} \bibinfo{person}{Matthew Smith}.} \bibinfo{year}{2014}\natexlab{}.
\newblock \showarticletitle{Using personal examples to improve risk communication for security \& privacy decisions}. In \bibinfo{booktitle}{\emph{Proceedings of the SIGCHI conference on human factors in computing systems}}. \bibinfo{pages}{2647--2656}.
\newblock


\bibitem[Hinsz et~al\mbox{.}(1997)]%
        {hinsz1997using}
\bibfield{author}{\bibinfo{person}{Verlin~B Hinsz}, \bibinfo{person}{Lynn~R Kalnbach}, {and} \bibinfo{person}{Nichole~R Lorentz}.} \bibinfo{year}{1997}\natexlab{}.
\newblock \showarticletitle{Using judgmental anchors to establish challenging self-set goals without jeopardizing commitment}.
\newblock \bibinfo{journal}{\emph{Organizational Behavior and Human Decision Processes}} \bibinfo{volume}{71}, \bibinfo{number}{3} (\bibinfo{year}{1997}), \bibinfo{pages}{287--308}.
\newblock


\bibitem[Hruschka et~al\mbox{.}(2004)]%
        {hruschka2004reliability}
\bibfield{author}{\bibinfo{person}{Daniel~J Hruschka}, \bibinfo{person}{Deborah Schwartz}, \bibinfo{person}{Daphne~Cobb St.~John}, \bibinfo{person}{Erin Picone-Decaro}, \bibinfo{person}{Richard~A Jenkins}, {and} \bibinfo{person}{James~W Carey}.} \bibinfo{year}{2004}\natexlab{}.
\newblock \showarticletitle{Reliability in coding open-ended data: Lessons learned from HIV behavioral research}.
\newblock \bibinfo{journal}{\emph{Field methods}} \bibinfo{volume}{16}, \bibinfo{number}{3} (\bibinfo{year}{2004}), \bibinfo{pages}{307--331}.
\newblock


\bibitem[Jacobs et~al\mbox{.}(2009)]%
        {Jacobs2009}
\bibfield{author}{\bibinfo{person}{Lawrence~R. Jacobs}, \bibinfo{person}{Fay~Lomax Cook}, {and} \bibinfo{person}{Michael X.~Delli Carpini}.} \bibinfo{year}{2009}\natexlab{}.
\newblock \bibinfo{booktitle}{\emph{Talking Together: Public Deliberation and Political Participation in America}}.
\newblock \bibinfo{publisher}{The University of Chicago Press}.
\newblock


\bibitem[Jarvenpaa and Majchrzak(2010)]%
        {jarvenpaa2010research}
\bibfield{author}{\bibinfo{person}{Sirkka~L Jarvenpaa} {and} \bibinfo{person}{Ann Majchrzak}.} \bibinfo{year}{2010}\natexlab{}.
\newblock \showarticletitle{Research commentary—Vigilant interaction in knowledge collaboration: Challenges of online user participation under ambivalence}.
\newblock \bibinfo{journal}{\emph{Information Systems Research}} \bibinfo{volume}{21}, \bibinfo{number}{4} (\bibinfo{year}{2010}), \bibinfo{pages}{773--784}.
\newblock


\bibitem[Joinson(2003)]%
        {joinson2003understanding}
\bibfield{author}{\bibinfo{person}{Adam~N Joinson}.} \bibinfo{year}{2003}\natexlab{}.
\newblock \showarticletitle{Understanding the psychology of Internet behaviour: Virtual worlds, real lives}.
\newblock \bibinfo{journal}{\emph{Revista iberoamericana de educaci{\'o}n a distancia}} \bibinfo{volume}{6}, \bibinfo{number}{2} (\bibinfo{year}{2003}), \bibinfo{pages}{190}.
\newblock


\bibitem[Joinson(2007)]%
        {joinson2007disinhibition}
\bibfield{author}{\bibinfo{person}{Adam~N Joinson}.} \bibinfo{year}{2007}\natexlab{}.
\newblock \showarticletitle{Disinhibition and the Internet}.
\newblock In \bibinfo{booktitle}{\emph{Psychology and the Internet}}. \bibinfo{publisher}{Elsevier}, \bibinfo{pages}{75--92}.
\newblock


\bibitem[Keane(2000)]%
        {Keane2000}
\bibfield{author}{\bibinfo{person}{John Keane}.} \bibinfo{year}{2000}\natexlab{}.
\newblock \showarticletitle{Structural Transformations of the Public Sphere}.
\newblock In \bibinfo{booktitle}{\emph{Digital Democracy: Issues of Theory and Practice}}, \bibfield{editor}{\bibinfo{person}{Kenneth~L. Hacker} {and} \bibinfo{person}{Jan van Dijk}} (Eds.). \bibinfo{publisher}{SAGE Publications Ltd}, \bibinfo{address}{London}, \bibinfo{pages}{70--89}.
\newblock
\urldef\tempurl%
\url{https://doi.org/10.4135/9781446218891}
\showDOI{\tempurl}


\bibitem[Kim et~al\mbox{.}(2021)]%
        {kim2021starrythoughts}
\bibfield{author}{\bibinfo{person}{Hyunwoo Kim}, \bibinfo{person}{Haesoo Kim}, \bibinfo{person}{Kyung~Je Jo}, {and} \bibinfo{person}{Juho Kim}.} \bibinfo{year}{2021}\natexlab{}.
\newblock \showarticletitle{StarryThoughts: Facilitating Diverse Opinion Exploration on Social Issues}.
\newblock \bibinfo{journal}{\emph{Proceedings of the ACM on Human-Computer Interaction}} \bibinfo{volume}{5}, \bibinfo{number}{CSCW1} (\bibinfo{year}{2021}), \bibinfo{pages}{1--29}.
\newblock


\bibitem[Kim et~al\mbox{.}(2019)]%
        {kim2019crowdsourcing}
\bibfield{author}{\bibinfo{person}{Hyunwoo Kim}, \bibinfo{person}{Eun-Young Ko}, \bibinfo{person}{Donghoon Han}, \bibinfo{person}{Sung-Chul Lee}, \bibinfo{person}{Simon~T Perrault}, \bibinfo{person}{Jihee Kim}, {and} \bibinfo{person}{Juho Kim}.} \bibinfo{year}{2019}\natexlab{}.
\newblock \showarticletitle{Crowdsourcing perspectives on public policy from stakeholders}. In \bibinfo{booktitle}{\emph{Extended Abstracts of the 2019 CHI Conference on Human Factors in Computing Systems}}. \bibinfo{pages}{1--6}.
\newblock


\bibitem[Kim et~al\mbox{.}(2015)]%
        {kim2015factful}
\bibfield{author}{\bibinfo{person}{Juho Kim}, \bibinfo{person}{Eun-Young Ko}, \bibinfo{person}{Jonghyuk Jung}, \bibinfo{person}{Chang~Won Lee}, \bibinfo{person}{Nam~Wook Kim}, {and} \bibinfo{person}{Jihee Kim}.} \bibinfo{year}{2015}\natexlab{}.
\newblock \showarticletitle{Factful: Engaging taxpayers in the public discussion of a government budget}. In \bibinfo{booktitle}{\emph{Proceedings of the 33rd Annual ACM Conference on Human Factors in Computing Systems}}. \bibinfo{pages}{2843--2852}.
\newblock


\bibitem[Kriplean et~al\mbox{.}(2012)]%
        {kriplean2012supporting}
\bibfield{author}{\bibinfo{person}{Travis Kriplean}, \bibinfo{person}{Jonathan Morgan}, \bibinfo{person}{Deen Freelon}, \bibinfo{person}{Alan Borning}, {and} \bibinfo{person}{Lance Bennett}.} \bibinfo{year}{2012}\natexlab{}.
\newblock \showarticletitle{Supporting reflective public thought with considerit}. In \bibinfo{booktitle}{\emph{Proceedings of the ACM 2012 conference on Computer Supported Cooperative Work}}. \bibinfo{pages}{265--274}.
\newblock


\bibitem[Landis and Koch(1977)]%
        {10.2307/2529310}
\bibfield{author}{\bibinfo{person}{J.~Richard Landis} {and} \bibinfo{person}{Gary~G. Koch}.} \bibinfo{year}{1977}\natexlab{}.
\newblock \showarticletitle{The Measurement of Observer Agreement for Categorical Data}.
\newblock \bibinfo{journal}{\emph{Biometrics}} \bibinfo{volume}{33}, \bibinfo{number}{1} (\bibinfo{year}{1977}), \bibinfo{pages}{159--174}.
\newblock
\showISSN{0006341X, 15410420}
\urldef\tempurl%
\url{http://www.jstor.org/stable/2529310}
\showURL{%
\tempurl}


\bibitem[Lapidot-Lefler and Barak(2012)]%
        {lapidot2012effects}
\bibfield{author}{\bibinfo{person}{Noam Lapidot-Lefler} {and} \bibinfo{person}{Azy Barak}.} \bibinfo{year}{2012}\natexlab{}.
\newblock \showarticletitle{Effects of anonymity, invisibility, and lack of eye-contact on toxic online disinhibition}.
\newblock \bibinfo{journal}{\emph{Computers in human behavior}} \bibinfo{volume}{28}, \bibinfo{number}{2} (\bibinfo{year}{2012}), \bibinfo{pages}{434--443}.
\newblock


\bibitem[Li et~al\mbox{.}(2010)]%
        {li2010stage}
\bibfield{author}{\bibinfo{person}{Ian Li}, \bibinfo{person}{Anind Dey}, {and} \bibinfo{person}{Jodi Forlizzi}.} \bibinfo{year}{2010}\natexlab{}.
\newblock \showarticletitle{A stage-based model of personal informatics systems}. In \bibinfo{booktitle}{\emph{Proceedings of the SIGCHI conference on human factors in computing systems}}. \bibinfo{pages}{557--566}.
\newblock


\bibitem[Lindley et~al\mbox{.}(2009)]%
        {lindley2009desiring}
\bibfield{author}{\bibinfo{person}{Si{\^a}n~E Lindley}, \bibinfo{person}{Richard Harper}, {and} \bibinfo{person}{Abigail Sellen}.} \bibinfo{year}{2009}\natexlab{}.
\newblock \showarticletitle{Desiring to be in touch in a changing communications landscape: attitudes of older adults}. In \bibinfo{booktitle}{\emph{Proceedings of the SIGCHI Conference on Human Factors in Computing Systems}}. \bibinfo{pages}{1693--1702}.
\newblock


\bibitem[Lischetzke and Eid(2003)]%
        {lischetzke2003attention}
\bibfield{author}{\bibinfo{person}{Tanja Lischetzke} {and} \bibinfo{person}{Michael Eid}.} \bibinfo{year}{2003}\natexlab{}.
\newblock \showarticletitle{Is attention to feelings beneficial or detrimental to affective well-being? Mood regulation as a moderator variable.}
\newblock \bibinfo{journal}{\emph{Emotion}} \bibinfo{volume}{3}, \bibinfo{number}{4} (\bibinfo{year}{2003}), \bibinfo{pages}{361}.
\newblock


\bibitem[MacQueen et~al\mbox{.}(1998)]%
        {macqueen1998codebook}
\bibfield{author}{\bibinfo{person}{Kathleen~M MacQueen}, \bibinfo{person}{Eleanor McLellan}, \bibinfo{person}{Kelly Kay}, {and} \bibinfo{person}{Bobby Milstein}.} \bibinfo{year}{1998}\natexlab{}.
\newblock \showarticletitle{Codebook development for team-based qualitative analysis}.
\newblock \bibinfo{journal}{\emph{Cam Journal}} \bibinfo{volume}{10}, \bibinfo{number}{2} (\bibinfo{year}{1998}), \bibinfo{pages}{31--36}.
\newblock


\bibitem[Manin(2011)]%
        {manin2011comment}
\bibfield{author}{\bibinfo{person}{Bernard Manin}.} \bibinfo{year}{2011}\natexlab{}.
\newblock \showarticletitle{Comment promouvoir la d{\'e}lib{\'e}ration d{\'e}mocratique? Priorit{\'e} du d{\'e}bat contradictoire sur la discussion}.
\newblock \bibinfo{journal}{\emph{Raisons politiques}} \bibinfo{volume}{42}, \bibinfo{number}{02} (\bibinfo{year}{2011}), \bibinfo{pages}{83--113}.
\newblock


\bibitem[Marathe and Toyama(2018)]%
        {marathe2018semi}
\bibfield{author}{\bibinfo{person}{Megh Marathe} {and} \bibinfo{person}{Kentaro Toyama}.} \bibinfo{year}{2018}\natexlab{}.
\newblock \showarticletitle{Semi-automated coding for qualitative research: A user-centered inquiry and initial prototypes}. In \bibinfo{booktitle}{\emph{Proceedings of the 2018 CHI conference on human factors in computing systems}}. \bibinfo{pages}{1--12}.
\newblock


\bibitem[Menon et~al\mbox{.}(2020)]%
        {menon2020nudge}
\bibfield{author}{\bibinfo{person}{Sanju Menon}, \bibinfo{person}{Weiyu Zhang}, {and} \bibinfo{person}{Simon~T Perrault}.} \bibinfo{year}{2020}\natexlab{}.
\newblock \showarticletitle{Nudge for deliberativeness: How interface features influence online discourse}. In \bibinfo{booktitle}{\emph{Proceedings of the 2020 CHI Conference on Human Factors in Computing Systems}}. \bibinfo{pages}{1--13}.
\newblock


\bibitem[Moon(2013)]%
        {moon2013reflection}
\bibfield{author}{\bibinfo{person}{Jennifer~A Moon}.} \bibinfo{year}{2013}\natexlab{}.
\newblock \bibinfo{booktitle}{\emph{Reflection in learning and professional development: Theory and practice}}.
\newblock \bibinfo{publisher}{Routledge}.
\newblock


\bibitem[Moritz et~al\mbox{.}(2014)]%
        {moritz2014judgmental}
\bibfield{author}{\bibinfo{person}{Brent Moritz}, \bibinfo{person}{Enno Siemsen}, {and} \bibinfo{person}{Mirko Kremer}.} \bibinfo{year}{2014}\natexlab{}.
\newblock \showarticletitle{Judgmental forecasting: Cognitive reflection and decision speed}.
\newblock \bibinfo{journal}{\emph{Production and Operations Management}} \bibinfo{volume}{23}, \bibinfo{number}{7} (\bibinfo{year}{2014}), \bibinfo{pages}{1146--1160}.
\newblock


\bibitem[Morse(1997)]%
        {morse1997perfectly}
\bibfield{author}{\bibinfo{person}{Janice~M Morse}.} \bibinfo{year}{1997}\natexlab{}.
\newblock \bibinfo{title}{" Perfectly healthy, but dead": the myth of inter-rater reliability}.
\newblock , \bibinfo{numpages}{445--447}~pages.
\newblock


\bibitem[Moyer and Bayer(1976)]%
        {moyer1976mental}
\bibfield{author}{\bibinfo{person}{Robert~S Moyer} {and} \bibinfo{person}{Richard~H Bayer}.} \bibinfo{year}{1976}\natexlab{}.
\newblock \showarticletitle{Mental comparison and the symbolic distance effect}.
\newblock \bibinfo{journal}{\emph{Cognitive Psychology}} \bibinfo{volume}{8}, \bibinfo{number}{2} (\bibinfo{year}{1976}), \bibinfo{pages}{228--246}.
\newblock


\bibitem[Muradova(2021)]%
        {muradova2021seeing}
\bibfield{author}{\bibinfo{person}{Lala Muradova}.} \bibinfo{year}{2021}\natexlab{}.
\newblock \showarticletitle{Seeing the other side? Perspective-taking and reflective political judgements in interpersonal deliberation}.
\newblock \bibinfo{journal}{\emph{Political Studies}} \bibinfo{volume}{69}, \bibinfo{number}{3} (\bibinfo{year}{2021}), \bibinfo{pages}{644--664}.
\newblock


\bibitem[Neuman(1986)]%
        {Neuman1986}
\bibfield{author}{\bibinfo{person}{W.~Russell Neuman}.} \bibinfo{year}{1986}\natexlab{}.
\newblock \bibinfo{booktitle}{\emph{The Paradox of Mass Politics: Knowledge and Opinion in the American Electorate}}.
\newblock \bibinfo{publisher}{Harvard University Press}.
\newblock


\bibitem[Norman(2010)]%
        {norman2010likert}
\bibfield{author}{\bibinfo{person}{Geoff Norman}.} \bibinfo{year}{2010}\natexlab{}.
\newblock \showarticletitle{Likert scales, levels of measurement and the “laws” of statistics}.
\newblock \bibinfo{journal}{\emph{Advances in health sciences education}} \bibinfo{volume}{15}, \bibinfo{number}{5} (\bibinfo{year}{2010}), \bibinfo{pages}{625--632}.
\newblock


\bibitem[Park et~al\mbox{.}(2009)]%
        {park2009newscube}
\bibfield{author}{\bibinfo{person}{Souneil Park}, \bibinfo{person}{Seungwoo Kang}, \bibinfo{person}{Sangyoung Chung}, {and} \bibinfo{person}{Junehwa Song}.} \bibinfo{year}{2009}\natexlab{}.
\newblock \showarticletitle{NewsCube: delivering multiple aspects of news to mitigate media bias}. In \bibinfo{booktitle}{\emph{Proceedings of the SIGCHI conference on human factors in computing systems}}. \bibinfo{pages}{443--452}.
\newblock


\bibitem[Price et~al\mbox{.}(2002)]%
        {price2002does}
\bibfield{author}{\bibinfo{person}{Vincent Price}, \bibinfo{person}{Joseph~N Cappella}, {and} \bibinfo{person}{Lilach Nir}.} \bibinfo{year}{2002}\natexlab{}.
\newblock \showarticletitle{Does disagreement contribute to more deliberative opinion?}
\newblock \bibinfo{journal}{\emph{Political communication}} \bibinfo{volume}{19}, \bibinfo{number}{1} (\bibinfo{year}{2002}), \bibinfo{pages}{95--112}.
\newblock


\bibitem[Rendon-Velez et~al\mbox{.}(2012)]%
        {rendon2012pilot}
\bibfield{author}{\bibinfo{person}{Elizabeth Rendon-Velez}, \bibinfo{person}{Imre Horv{\'a}th}, {and} \bibinfo{person}{W Van~der Vegte}.} \bibinfo{year}{2012}\natexlab{}.
\newblock \showarticletitle{A pilot study to investigate time pressure as a surrogate of being in haste}. In \bibinfo{booktitle}{\emph{Proceedings of the ninth international symposium on tools and methods of competitive engineering}}. \bibinfo{pages}{393--406}.
\newblock


\bibitem[Richards and Hemphill(2018)]%
        {richards2018practical}
\bibfield{author}{\bibinfo{person}{K~Andrew~R Richards} {and} \bibinfo{person}{Michael~A Hemphill}.} \bibinfo{year}{2018}\natexlab{}.
\newblock \showarticletitle{A practical guide to collaborative qualitative data analysis}.
\newblock \bibinfo{journal}{\emph{Journal of Teaching in Physical education}} \bibinfo{volume}{37}, \bibinfo{number}{2} (\bibinfo{year}{2018}), \bibinfo{pages}{225--231}.
\newblock


\bibitem[Rietz and Maedche(2021)]%
        {rietz2021cody}
\bibfield{author}{\bibinfo{person}{Tim Rietz} {and} \bibinfo{person}{Alexander Maedche}.} \bibinfo{year}{2021}\natexlab{}.
\newblock \showarticletitle{Cody: An AI-based system to semi-automate coding for qualitative research}. In \bibinfo{booktitle}{\emph{Proceedings of the 2021 CHI Conference on Human Factors in Computing Systems}}. \bibinfo{pages}{1--14}.
\newblock


\bibitem[Rustichini(2009)]%
        {rustichini2009neuroeconomics}
\bibfield{author}{\bibinfo{person}{Aldo Rustichini}.} \bibinfo{year}{2009}\natexlab{}.
\newblock \showarticletitle{Neuroeconomics:: Formal Models of Decision Making and Cognitive Neuroscience}.
\newblock In \bibinfo{booktitle}{\emph{Neuroeconomics}}. \bibinfo{publisher}{Elsevier}, \bibinfo{pages}{33--46}.
\newblock


\bibitem[Saldivar et~al\mbox{.}(2019)]%
        {saldivar2019civic}
\bibfield{author}{\bibinfo{person}{Jorge Saldivar}, \bibinfo{person}{Cristhian Parra}, \bibinfo{person}{Marcelo Alcaraz}, \bibinfo{person}{Rebeca Arteta}, {and} \bibinfo{person}{Luca Cernuzzi}.} \bibinfo{year}{2019}\natexlab{}.
\newblock \showarticletitle{Civic technology for social innovation: A systematic literature review}.
\newblock \bibinfo{journal}{\emph{Computer Supported Cooperative Work (CSCW)}}  \bibinfo{volume}{28} (\bibinfo{year}{2019}), \bibinfo{pages}{169--207}.
\newblock


\bibitem[Sartori(1987)]%
        {Sartori1987}
\bibfield{author}{\bibinfo{person}{Giovanni Sartori}.} \bibinfo{year}{1987}\natexlab{}.
\newblock \bibinfo{booktitle}{\emph{The Theory of Democracy Revisited}}.
\newblock \bibinfo{publisher}{Steven Bridges Press}.
\newblock


\bibitem[Schwertman et~al\mbox{.}(2004)]%
        {schwertman2004simple}
\bibfield{author}{\bibinfo{person}{Neil~C Schwertman}, \bibinfo{person}{Margaret~Ann Owens}, {and} \bibinfo{person}{Robiah Adnan}.} \bibinfo{year}{2004}\natexlab{}.
\newblock \showarticletitle{A simple more general boxplot method for identifying outliers}.
\newblock \bibinfo{journal}{\emph{Computational statistics \& data analysis}} \bibinfo{volume}{47}, \bibinfo{number}{1} (\bibinfo{year}{2004}), \bibinfo{pages}{165--174}.
\newblock


\bibitem[Slack and Draugalis~Jr(2001)]%
        {slack2001establishing}
\bibfield{author}{\bibinfo{person}{Marion~K Slack} {and} \bibinfo{person}{Jolaine~R Draugalis~Jr}.} \bibinfo{year}{2001}\natexlab{}.
\newblock \showarticletitle{Establishing the internal and external validity of experimental studies}.
\newblock \bibinfo{journal}{\emph{American journal of health-system pharmacy}} \bibinfo{volume}{58}, \bibinfo{number}{22} (\bibinfo{year}{2001}), \bibinfo{pages}{2173--2181}.
\newblock


\bibitem[Spatariu et~al\mbox{.}(2004)]%
        {Spatariu2004}
\bibfield{author}{\bibinfo{person}{Alexandru Spatariu}, \bibinfo{person}{Kendall Hartley}, {and} \bibinfo{person}{Lisa~D. Bendixen}.} \bibinfo{year}{2004}\natexlab{}.
\newblock \showarticletitle{Defining and Measuring Quality in Online Discussions}.
\newblock \bibinfo{journal}{\emph{Journal of Interactive Online Learning}} \bibinfo{volume}{2}, \bibinfo{number}{4} (\bibinfo{year}{2004}).
\newblock
\showISSN{1541-4914}
\urldef\tempurl%
\url{https://www.ncolr.org/issues/jiol/v2/n4/defining-and-measuring-quality-in-online-discussions.html}
\showURL{%
\tempurl}


\bibitem[Steenbergen et~al\mbox{.}(2003)]%
        {steenbergen2003measuring}
\bibfield{author}{\bibinfo{person}{Marco~R Steenbergen}, \bibinfo{person}{Andr{\'e} B{\"a}chtiger}, \bibinfo{person}{Markus Sp{\"o}rndli}, {and} \bibinfo{person}{J{\"u}rg Steiner}.} \bibinfo{year}{2003}\natexlab{}.
\newblock \showarticletitle{Measuring political deliberation: A discourse quality index}.
\newblock \bibinfo{journal}{\emph{Comparative European Politics}}  \bibinfo{volume}{1} (\bibinfo{year}{2003}), \bibinfo{pages}{21--48}.
\newblock


\bibitem[Stromer-Galley(2007)]%
        {stromer2007measuring}
\bibfield{author}{\bibinfo{person}{Jennifer Stromer-Galley}.} \bibinfo{year}{2007}\natexlab{}.
\newblock \showarticletitle{Measuring deliberation’s content: A coding scheme}.
\newblock \bibinfo{journal}{\emph{Journal of Deliberative Democracy}} \bibinfo{volume}{3}, \bibinfo{number}{1} (\bibinfo{year}{2007}).
\newblock


\bibitem[Suler(2004)]%
        {suler2004online}
\bibfield{author}{\bibinfo{person}{John Suler}.} \bibinfo{year}{2004}\natexlab{}.
\newblock \showarticletitle{The online disinhibition effect}.
\newblock \bibinfo{journal}{\emph{Cyberpsychology \& behavior}} \bibinfo{volume}{7}, \bibinfo{number}{3} (\bibinfo{year}{2004}), \bibinfo{pages}{321--326}.
\newblock


\bibitem[Switzer~III and Sniezek(1991)]%
        {switzer1991judgment}
\bibfield{author}{\bibinfo{person}{Fred~S Switzer~III} {and} \bibinfo{person}{Janet~A Sniezek}.} \bibinfo{year}{1991}\natexlab{}.
\newblock \showarticletitle{Judgment processes in motivation: Anchoring and adjustment effects on judgment and behavior}.
\newblock \bibinfo{journal}{\emph{Organizational Behavior and Human Decision Processes}} \bibinfo{volume}{49}, \bibinfo{number}{2} (\bibinfo{year}{1991}), \bibinfo{pages}{208--229}.
\newblock


\bibitem[Takayoshi(2015)]%
        {takayoshi2015short}
\bibfield{author}{\bibinfo{person}{Pamela Takayoshi}.} \bibinfo{year}{2015}\natexlab{}.
\newblock \showarticletitle{Short-form writing: Studying process in the context of contemporary composing technologies}.
\newblock \bibinfo{journal}{\emph{Computers and Composition}}  \bibinfo{volume}{37} (\bibinfo{year}{2015}), \bibinfo{pages}{1--13}.
\newblock


\bibitem[Thaler and Sunstein(2008)]%
        {thaler2008nudge}
\bibfield{author}{\bibinfo{person}{Richard Thaler} {and} \bibinfo{person}{Cass Sunstein}.} \bibinfo{year}{2008}\natexlab{}.
\newblock \showarticletitle{Nudge: The gentle power of choice architecture}.
\newblock \bibinfo{journal}{\emph{New Haven, Conn.: Yale}} (\bibinfo{year}{2008}).
\newblock


\bibitem[Thomas(1946)]%
        {hobbes1946}
\bibfield{author}{\bibinfo{person}{Hobbes Thomas}.} \bibinfo{year}{1946}\natexlab{}.
\newblock \bibinfo{booktitle}{\emph{Leviathan}}.
\newblock \bibinfo{publisher}{London: Basil Blackwell}.
\newblock


\bibitem[Trénel(2004)]%
        {Trenel2004}
\bibfield{author}{\bibinfo{person}{Matthias Trénel}.} \bibinfo{year}{2004}\natexlab{}.
\newblock \bibinfo{title}{Measuring the quality of online deliberation. Coding scheme 2.0.}
\newblock
\newblock
\urldef\tempurl%
\url{http://www.wz-berlin.de/~trenel/tools/qod_2_0.pdf}
\showURL{%
Retrieved December 1, 2023 from \tempurl}


\bibitem[Umansky and Pipal(2023)]%
        {umansky2023dances}
\bibfield{author}{\bibinfo{person}{Natalia Umansky} {and} \bibinfo{person}{Christian Pipal}.} \bibinfo{year}{2023}\natexlab{}.
\newblock \showarticletitle{Dances, Duets, and Debates: Analysing political communication and viewer engagement on TikTok}.
\newblock \bibinfo{journal}{\emph{OSF Preprints. September}}  \bibinfo{volume}{21} (\bibinfo{year}{2023}).
\newblock


\bibitem[Voggeser et~al\mbox{.}(2018)]%
        {voggeser2018self}
\bibfield{author}{\bibinfo{person}{Birgit~J Voggeser}, \bibinfo{person}{Ranjit~K Singh}, {and} \bibinfo{person}{Anja~S G{\"o}ritz}.} \bibinfo{year}{2018}\natexlab{}.
\newblock \showarticletitle{Self-control in online discussions: Disinhibited online behavior as a failure to recognize social cues}.
\newblock \bibinfo{journal}{\emph{Frontiers in psychology}}  \bibinfo{volume}{8} (\bibinfo{year}{2018}), \bibinfo{pages}{316647}.
\newblock


\bibitem[Wang et~al\mbox{.}(2014)]%
        {wang2014field}
\bibfield{author}{\bibinfo{person}{Yang Wang}, \bibinfo{person}{Pedro~Giovanni Leon}, \bibinfo{person}{Alessandro Acquisti}, \bibinfo{person}{Lorrie~Faith Cranor}, \bibinfo{person}{Alain Forget}, {and} \bibinfo{person}{Norman Sadeh}.} \bibinfo{year}{2014}\natexlab{}.
\newblock \showarticletitle{A field trial of privacy nudges for facebook}. In \bibinfo{booktitle}{\emph{Proceedings of the SIGCHI conference on human factors in computing systems}}. \bibinfo{pages}{2367--2376}.
\newblock


\bibitem[Wang et~al\mbox{.}(2011)]%
        {wang2011regretted}
\bibfield{author}{\bibinfo{person}{Yang Wang}, \bibinfo{person}{Gregory Norcie}, \bibinfo{person}{Saranga Komanduri}, \bibinfo{person}{Alessandro Acquisti}, \bibinfo{person}{Pedro~Giovanni Leon}, {and} \bibinfo{person}{Lorrie~Faith Cranor}.} \bibinfo{year}{2011}\natexlab{}.
\newblock \showarticletitle{" I regretted the minute I pressed share" a qualitative study of regrets on Facebook}. In \bibinfo{booktitle}{\emph{Proceedings of the seventh symposium on usable privacy and security}}. \bibinfo{pages}{1--16}.
\newblock


\bibitem[Xiao et~al\mbox{.}(2015)]%
        {xiao2015design}
\bibfield{author}{\bibinfo{person}{Lu Xiao}, \bibinfo{person}{Weiyu Zhang}, \bibinfo{person}{Anna Przybylska}, \bibinfo{person}{Anna De~Liddo}, \bibinfo{person}{Gregorio Convertino}, \bibinfo{person}{Todd Davies}, {and} \bibinfo{person}{Mark Klein}.} \bibinfo{year}{2015}\natexlab{}.
\newblock \showarticletitle{Design for online deliberative processes and technologies: Towards a multidisciplinary research agenda}. In \bibinfo{booktitle}{\emph{Proceedings of the 33rd Annual ACM Conference Extended Abstracts on Human Factors in Computing Systems}}. \bibinfo{pages}{865--868}.
\newblock


\bibitem[Yeo et~al\mbox{.}(2024)]%
        {10.1145/3613904.3642530}
\bibfield{author}{\bibinfo{person}{ShunYi Yeo}, \bibinfo{person}{Gionnieve Lim}, \bibinfo{person}{Jie Gao}, \bibinfo{person}{Weiyu Zhang}, {and} \bibinfo{person}{Simon~Tangi Perrault}.} \bibinfo{year}{2024}\natexlab{}.
\newblock \showarticletitle{Help Me Reflect: Leveraging Self-Reflection Interface Nudges to Enhance Deliberativeness on Online Deliberation Platforms}. In \bibinfo{booktitle}{\emph{Proceedings of the CHI Conference on Human Factors in Computing Systems}} \emph{(\bibinfo{series}{CHI '24})}. \bibinfo{publisher}{Association for Computing Machinery}, \bibinfo{address}{New York, NY, USA}, Article \bibinfo{articleno}{806}, \bibinfo{numpages}{32}~pages.
\newblock
\showISBNx{9798400703300}
\urldef\tempurl%
\url{https://doi.org/10.1145/3613904.3642530}
\showDOI{\tempurl}


\bibitem[Zhang et~al\mbox{.}(2024)]%
        {zhang2024form}
\bibfield{author}{\bibinfo{person}{Amy~X Zhang}, \bibinfo{person}{Michael~S Bernstein}, \bibinfo{person}{David~R Karger}, {and} \bibinfo{person}{Mark~S Ackerman}.} \bibinfo{year}{2024}\natexlab{}.
\newblock \showarticletitle{Form-From: A Design Space of Social Media Systems}.
\newblock \bibinfo{journal}{\emph{arXiv preprint arXiv:2402.05388}} (\bibinfo{year}{2024}).
\newblock


\bibitem[Zhang et~al\mbox{.}(2022)]%
        {zhang2022driving}
\bibfield{author}{\bibinfo{person}{Cevin Zhang}, \bibinfo{person}{Hemingxi Zheng}, {and} \bibinfo{person}{Qing Wang}.} \bibinfo{year}{2022}\natexlab{}.
\newblock \showarticletitle{Driving factors and moderating effects behind citizen engagement with mobile short-form videos}.
\newblock \bibinfo{journal}{\emph{IEEE Access}}  \bibinfo{volume}{10} (\bibinfo{year}{2022}), \bibinfo{pages}{40999--41009}.
\newblock


\bibitem[Zhang(2005)]%
        {zhang2005online}
\bibfield{author}{\bibinfo{person}{Weiyu Zhang}.} \bibinfo{year}{2005}\natexlab{}.
\newblock \showarticletitle{Are online discussions deliberate? A case study of a Chinese online discussion board}.
\newblock \bibinfo{journal}{\emph{Tr{\'\i}podos}} \bibinfo{volume}{1}, \bibinfo{number}{1} (\bibinfo{year}{2005}), \bibinfo{pages}{119--134}.
\newblock


\bibitem[Zhang et~al\mbox{.}(2013)]%
        {zhang2013structural}
\bibfield{author}{\bibinfo{person}{Weiyu Zhang}, \bibinfo{person}{Xiaoxia Cao}, {and} \bibinfo{person}{Minh~Ngoc Tran}.} \bibinfo{year}{2013}\natexlab{}.
\newblock \showarticletitle{The structural features and the deliberative quality of online discussions}.
\newblock \bibinfo{journal}{\emph{Telematics and informatics}} \bibinfo{volume}{30}, \bibinfo{number}{2} (\bibinfo{year}{2013}), \bibinfo{pages}{74--86}.
\newblock


\bibitem[Zhang and Chang(2014)]%
        {zhang2014perceived}
\bibfield{author}{\bibinfo{person}{Weiyu Zhang} {and} \bibinfo{person}{Leanne Chang}.} \bibinfo{year}{2014}\natexlab{}.
\newblock \showarticletitle{Perceived speech conditions and disagreement of everyday talk: A proceduralist perspective of citizen deliberation}.
\newblock \bibinfo{journal}{\emph{Communication Theory}} \bibinfo{volume}{24}, \bibinfo{number}{2} (\bibinfo{year}{2014}), \bibinfo{pages}{124--145}.
\newblock


\bibitem[Zhang et~al\mbox{.}(2021)]%
        {zhang2021nudge}
\bibfield{author}{\bibinfo{person}{Weiyu Zhang}, \bibinfo{person}{Tian Yang}, {and} \bibinfo{person}{Simon Tangi~Perrault}.} \bibinfo{year}{2021}\natexlab{}.
\newblock \showarticletitle{Nudge for Reflection: More Than Just a Channel to Political Knowledge}. In \bibinfo{booktitle}{\emph{Proceedings of the 2021 CHI Conference on Human Factors in Computing Systems}}. \bibinfo{pages}{1--10}.
\newblock


\bibitem[Zhu et~al\mbox{.}(2017)]%
        {zhu2017exploring}
\bibfield{author}{\bibinfo{person}{Fengyuan Zhu}, \bibinfo{person}{Ke Fang}, {and} \bibinfo{person}{Xiaojuan Ma}.} \bibinfo{year}{2017}\natexlab{}.
\newblock \showarticletitle{Exploring the effects of strategy and arousal of cueing in computer-human persuasion}. In \bibinfo{booktitle}{\emph{Proceedings of the 2017 CHI Conference Extended Abstracts on Human Factors in Computing Systems}}. \bibinfo{pages}{2276--2283}.
\newblock


\end{thebibliography}

\newpage
\section*{Appendix}
The appendix contains supplementary data, which, while not part of the analysis, may provide additional details that can help in replicating the study. 

\subsection*{Methodology Appendix A1: Content Analysis for Analyzing Qualitative Data}
\label{sec:methodologyappendixA1}

\subsubsection*{\textbf{Data Preparation.}}
\label{sec:data preparation}
Deliberativeness was operationalized through three measurements: argument repertoire, argument diversity and word count as discussed in section~\ref{sec:quality}. Of the three, argument repertoire and argument diversity were derived from a content analysis of participants’ responses, which we elucidate as follows:

Collected textual data to the four-day workweek task was subjected to qualitative research techniques following the Grounded Theory method~\cite{charmaz2014constructing,corbin2014basics}. This involved open coding of participants' responses by two coders. Both coders were PhD students with respectively two and five years of experience in content analysis. A single independent coder was used for reliability checks and to review the coding process. 

To ensure the consistency and clarity of the coding process~\cite{chen2016challenges, rietz2021cody, marathe2018semi}, coding was done sentence-by-sentence where one or more codes from the codebook could be applied to each sentence.

To generate the codebook, a subset of the data (10\% of the data from study 1 and 30\% of the data from study 2) were distributed to the coders. Both coders independently examined the participants' responses and proposed a set of codes. They then met to compare and discuss similar codes and to resolve code conflicts while focusing on (1) how relevant the codes were to the dependent variable of deliberativeness and (2) whether the code appeared in the sentence. After agreeing on an initial master list of codes, the coders followed the practice by Hruschka et al.~\cite{hruschka2004reliability} and MacQueen et al.~\cite{macqueen1998codebook} to derive a set of rules. These rules were meant to ensure efficient coding between the two coders in which they decide on the criteria of whether a specific sentence fell under a particular code. More specifically, an inclusion list (see Appendix Table \ref{tab:inclusionlist}) was generated to clearly define the code and to clarify which sentence should or should not constitute an instance of a code.

After the initial draft of the codebook was developed, the coders checked for inter-coder reliability rates in an iterative process. A total of three coding rounds were conducted and Cohen's Kappa was computed: Argument Repertoire ($\kappa = 0.808$); Argument Diversity ($\kappa = 0.808$), both above the satisfactory threshold of 0.80~\cite{10.2307/2529310,fleiss1979large, morse1997perfectly}.

Afterwards, the entire set of responses was coded according to the finalized codebook which comprised of 24 codes. As Hruschka et al.~\cite{hruschka2004reliability} found that a large number of codes consequentially decreases the likelihood of any real agreement between the coders as there are just too many codes to choose from, we restricted the number of possible codes to less than 30.

\subsubsection*{\textbf{Content Analysis.}}
\label{sec: qualitative analysis}
\begin{figure*}[!htbp]
   \centering
   \includegraphics[scale=0.5]{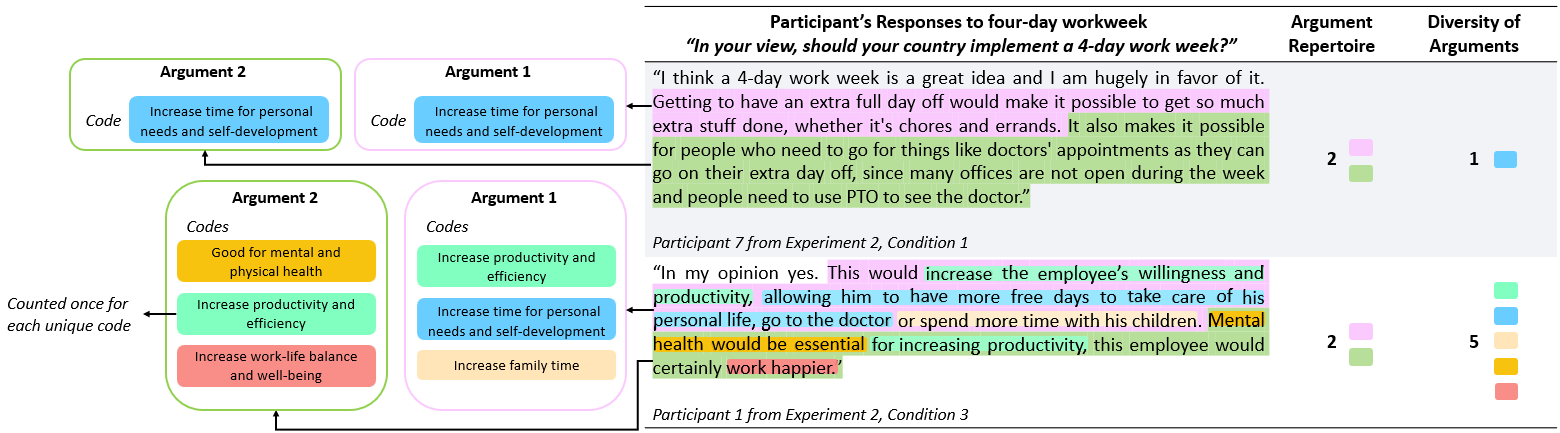}
   \caption{Illustration of two participants' opinions. Both participants have the same number of argument repertoire but participant 1 has a higher diversity count. Note that even though P1 and P7 were subjected to different conditions in study 2, this does not impact the diversity of the arguments as participants were given the same task.}
   \label{diversityofarguments}
\end{figure*}

For argument repertoire, a relevant sentence is counted if it has at least one code. Sentences stating one's personal preference (e.g., \textit{"I am hugely in favour of four-day workweek."}) or emotional response (e.g., \textit{"I felt very happy for the idea on four-day workweek."}) were not counted. For argument diversity, a theme is a unique code that is present in the participant's opinion. Codes that were repeated were only counted once. An illustration of the computation for argument repertoire and argument diversity is depicted in Figure~\ref{diversityofarguments}.

Open coding was conducted in two instances. In the first instance, open coding was conducted in both studies to prepare the data for analyzing deliberativeness quantitatively. The second instance was conducted in study 2 in which open coding was conducted by the primary author on participants' feedback towards the interface-based time nudges.

\subsection*{Demographics Appendix B1: Demographics Profile}
We found no significant differences in terms of the demographics profile for any of the experimental conditions in study 1. Similarly, we found no significant differences in terms of the demographics profile for any of the experimental conditions in study 2. 

\subsection*{Supplementary Statistical Details C1: Additional Statistics for Study 2}
\begin{figure*}[!htbp]
   \centering
   \includegraphics[scale=0.5]{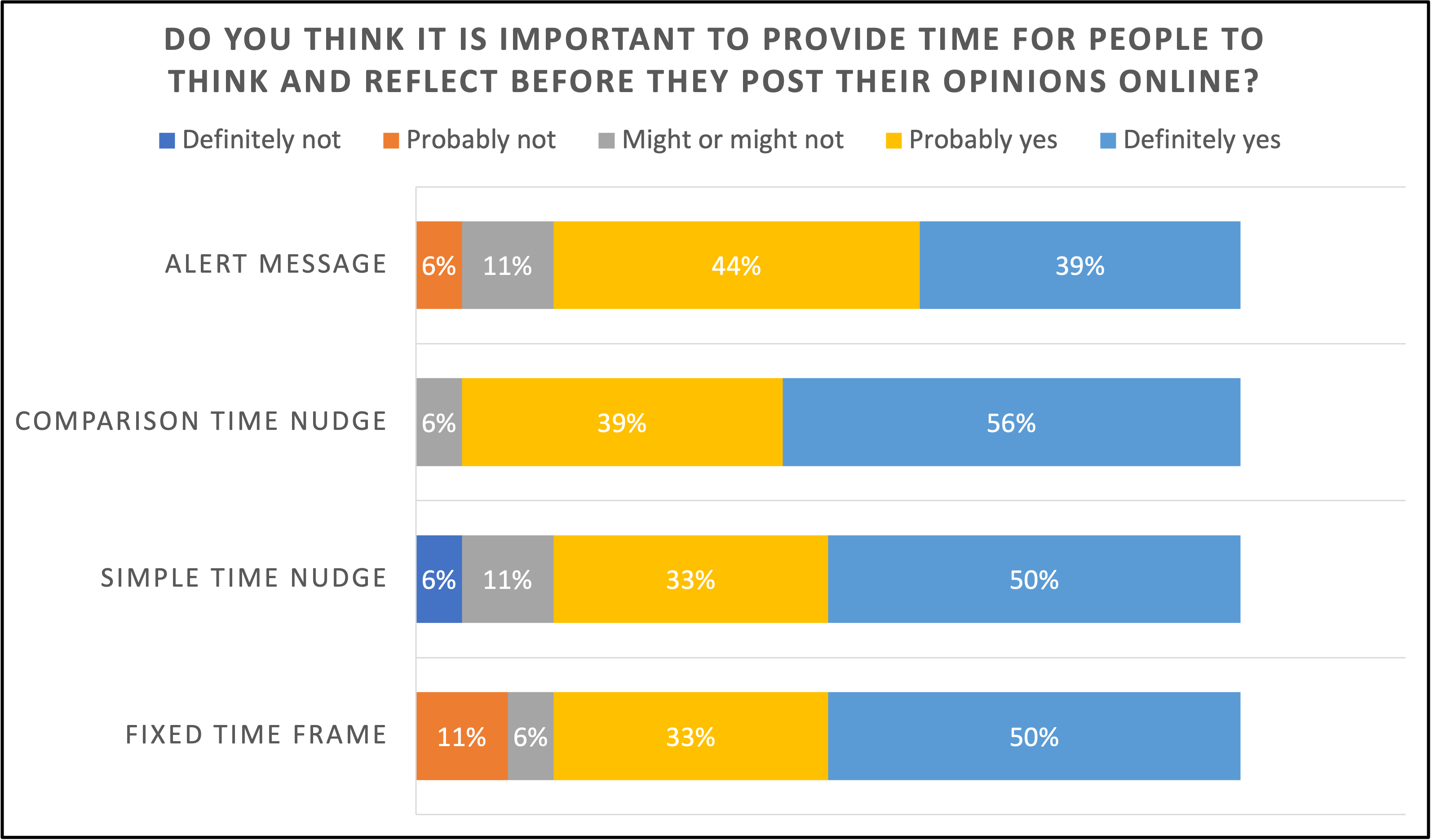}
   \caption{Participants' perspectives on the importance of allowing time for reflection before posting. A significant majority 86.1\% (62 out of 72 participants), expressed agreement by selecting 'yes' (probably yes and definitely yes).}
   \label{importance of time}
\end{figure*}

Although not included in the primary analysis, we collated participants' perspectives of reflection time in the post-task survey of study 2. We asked participants to rate their perceptions to a question that reads \textit{"Do you think it is important to provide time for people to think and reflect before they post their opinions online?"}. They have to rate this question on a five-point Likert scale (1 = definitely not, 2 = probably not, 3 = might or might not, 4 = probably yes and 5 = definitely yes). Majority (86.1\%) of the participants indicated `yes' to providing time for people to reflect during the deliberation process (see Figure~\ref{importance of time}).

\subsection*{Supplementary Statistical Details C2: Regression Analysis for Study 1}
Regression analysis of both linear and logarithmic model for the three deliberative measures (word count, argument repertoire and argument diversity) are shown in Table~\ref{tab:regression analysis}. 

\subsection*{Benefits \& Drawbacks for Study 2}
\begin{figure*}[!htbp]
  \centering
  \includegraphics[scale=0.26]{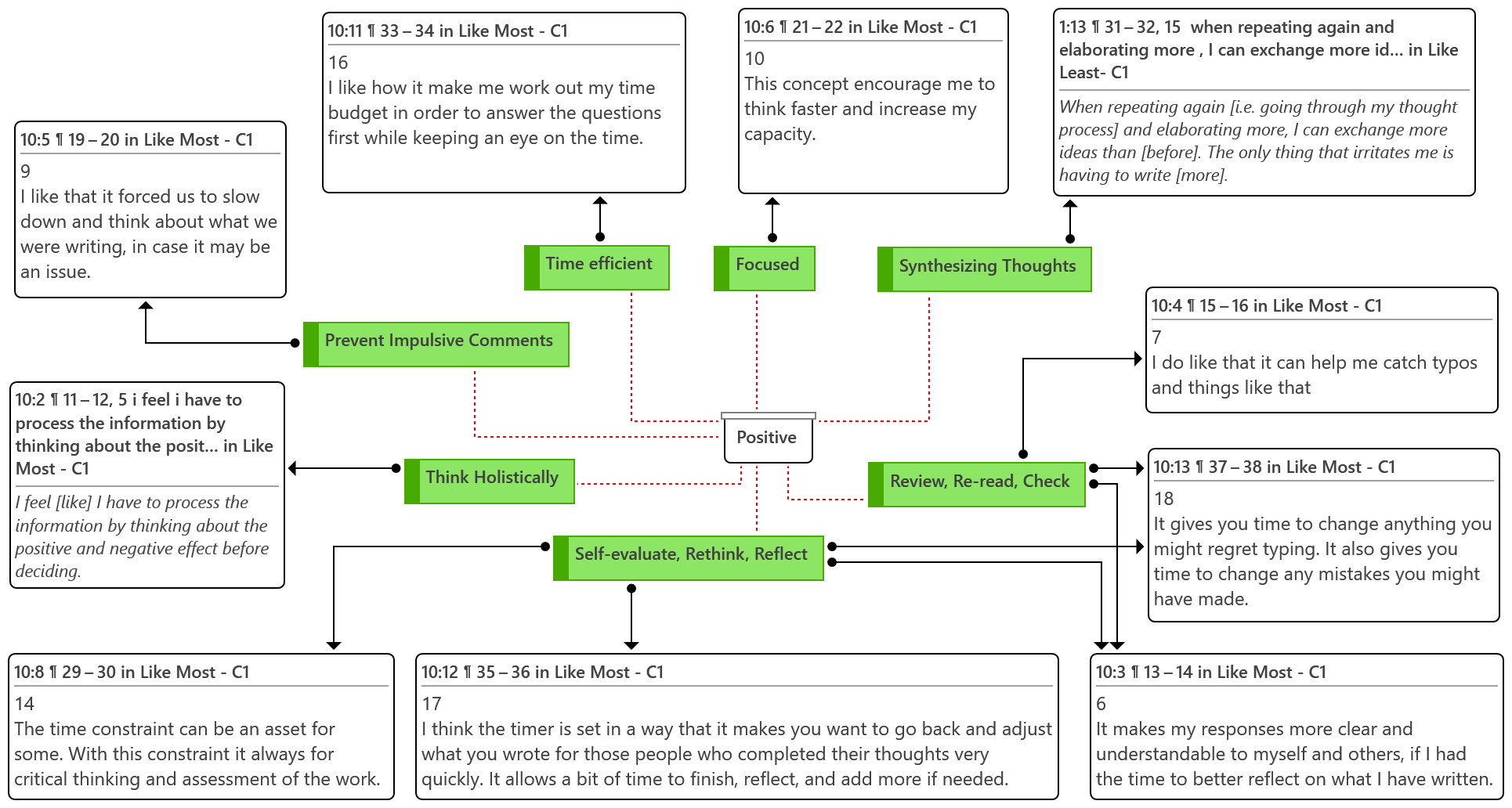}
  \caption{Benefits: Fixed Time Frame}
  \label{fig:likes fixedtimeframe}
\end{figure*}

\begin{figure*}[!htbp]
  \centering
  \includegraphics[scale=0.26]{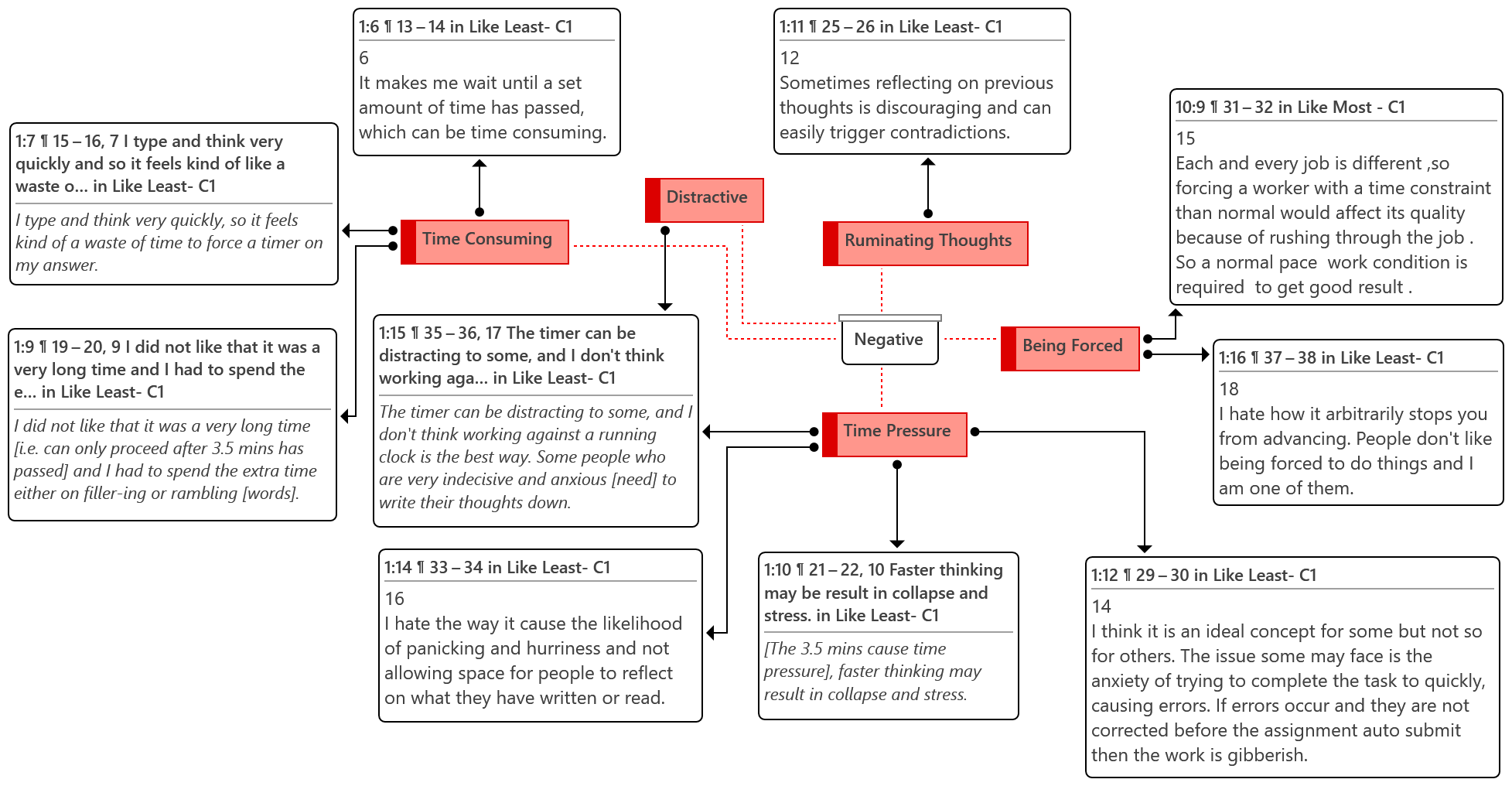}
  \caption{Drawbacks: Fixed Time Frame}
  \label{fig:dislikes fixedtimeframe}
\end{figure*}

\begin{figure*}[!htbp]
  \centering
  \includegraphics[scale=0.23]{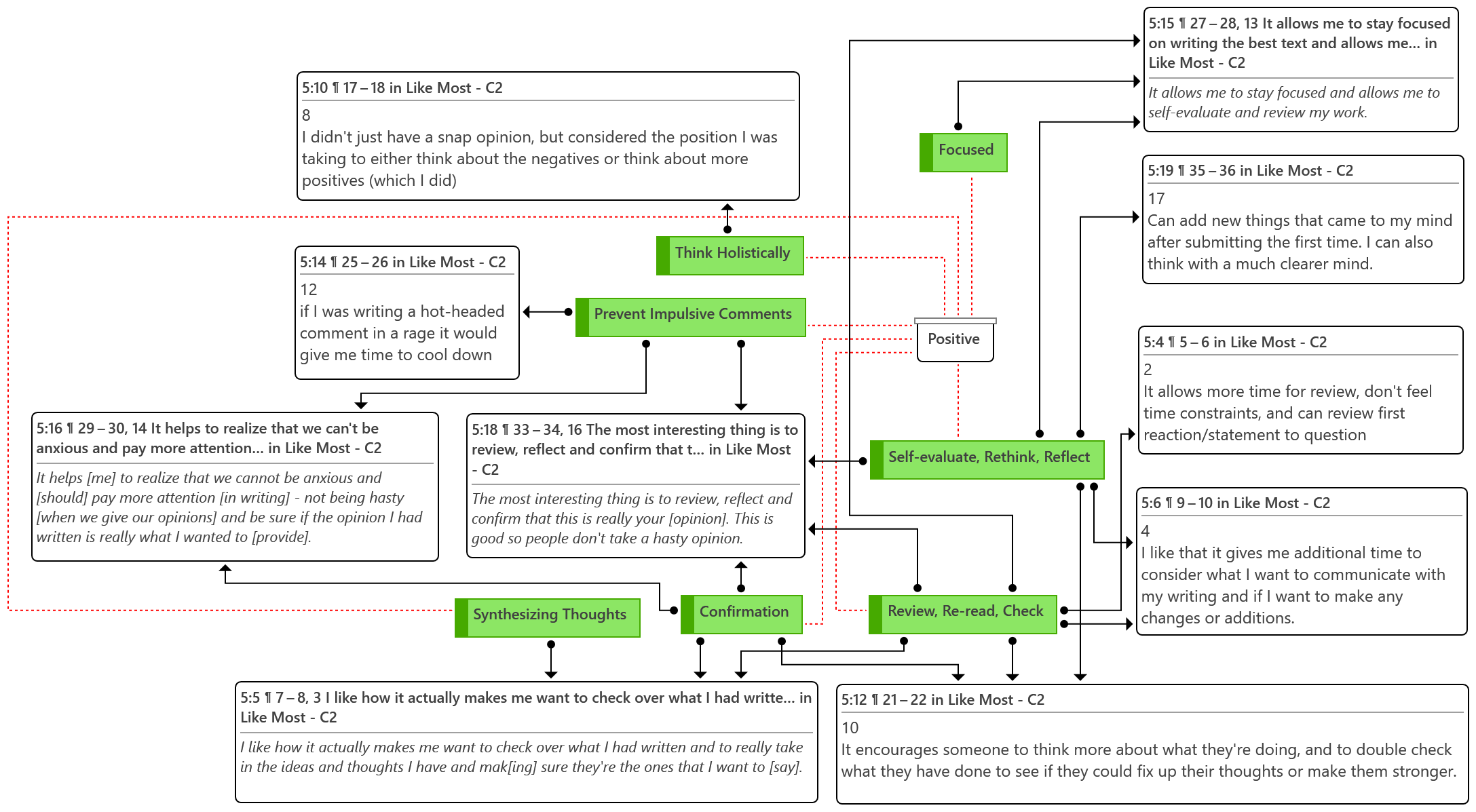}
  \caption{Benefits: Simple Time Nudge}
  \label{fig:likes simpletimenudge}
\end{figure*}

\begin{figure*}[!htbp]
  \centering
  \includegraphics[scale=0.30]{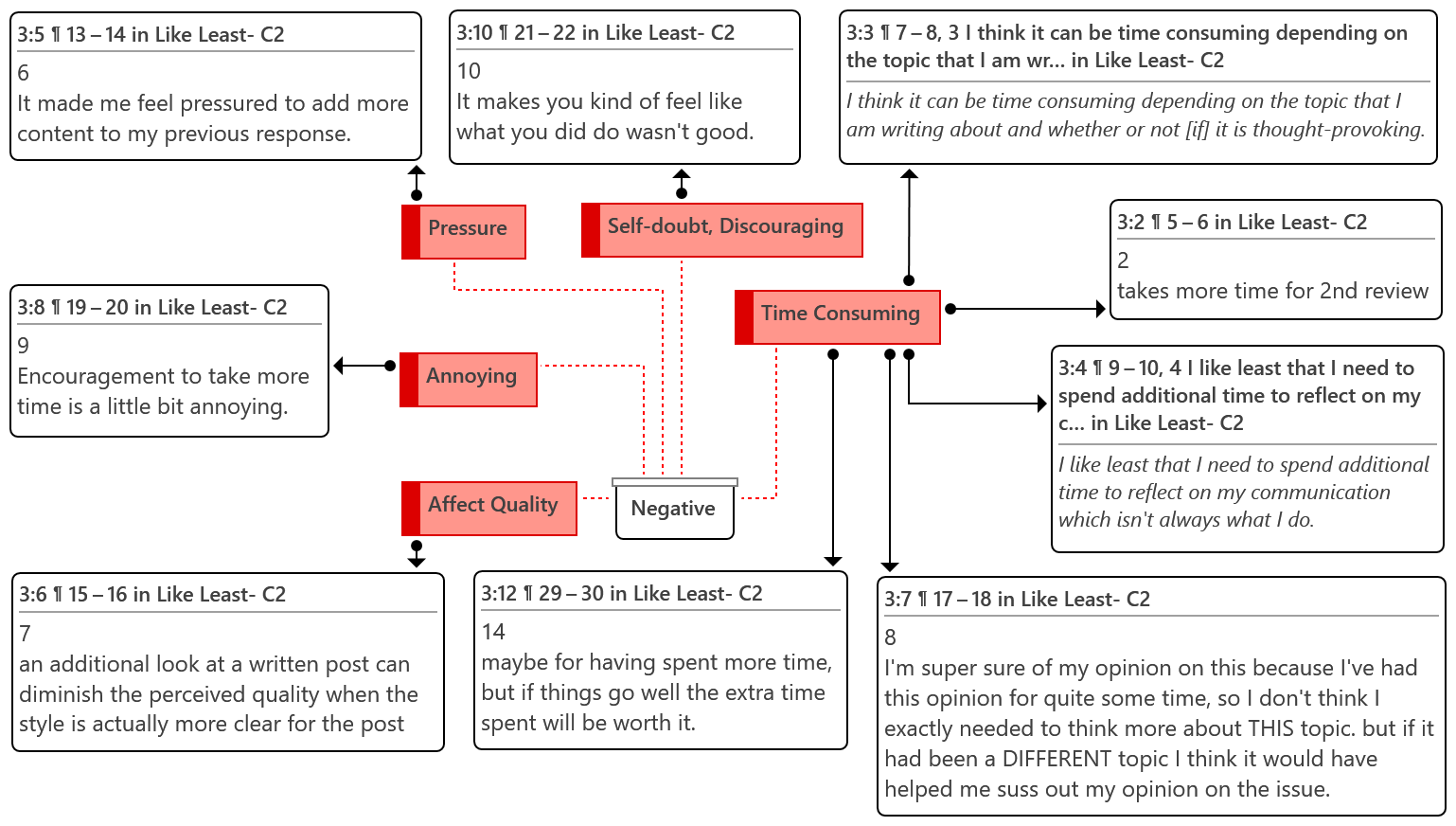}
  \caption{Drawbacks: Simple Time Nudge}
  \label{fig:dislikes simpletimenudge}
\end{figure*}

\begin{figure*}[!htbp]
  \centering
  \includegraphics[scale=0.29]{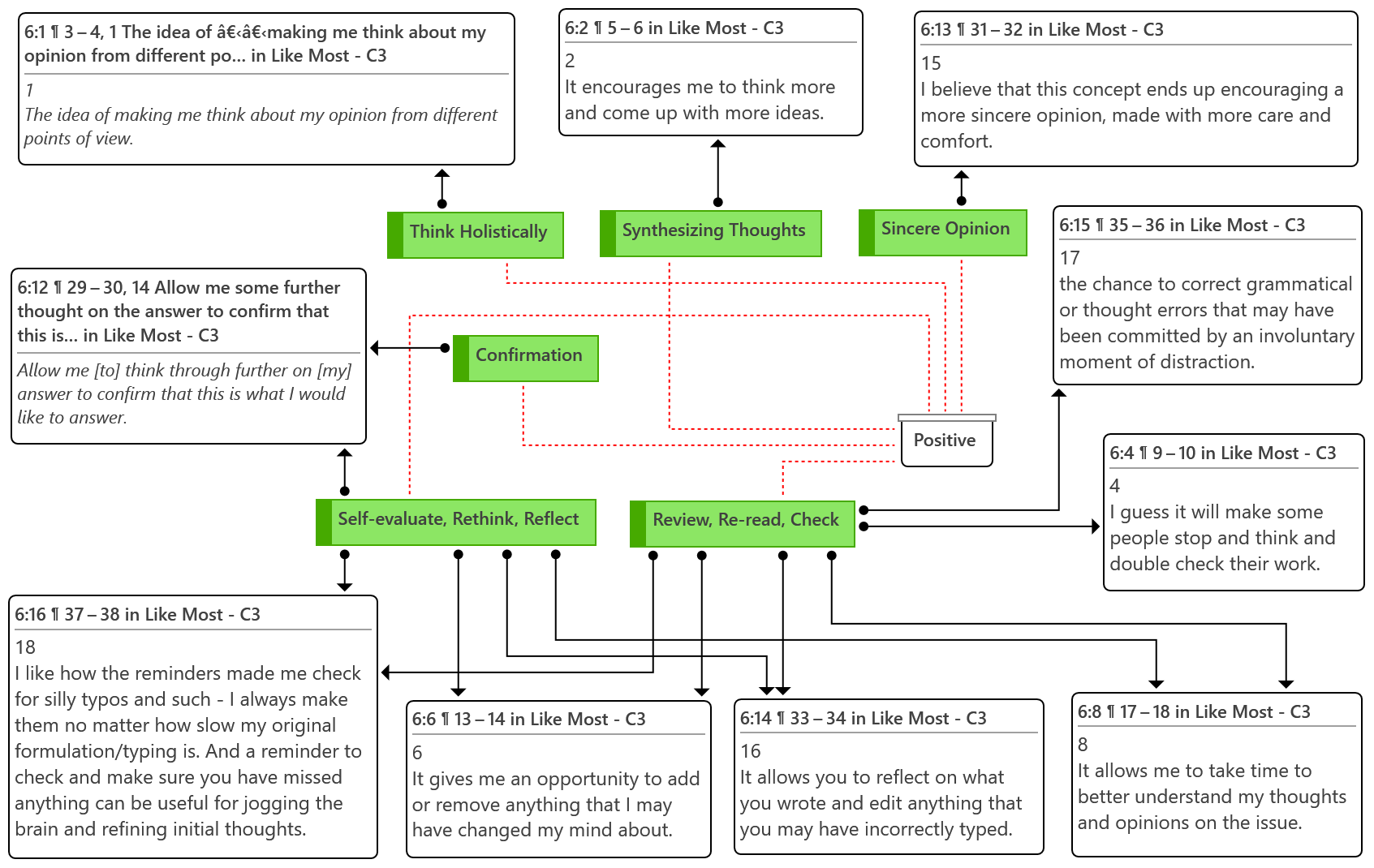}
  \caption{Benefits: Comparison Time Nudge}
  \label{fig:likes comparisontimenudge}
\end{figure*}

\begin{figure*}[!htbp]
  \centering
  \includegraphics[scale=0.25]{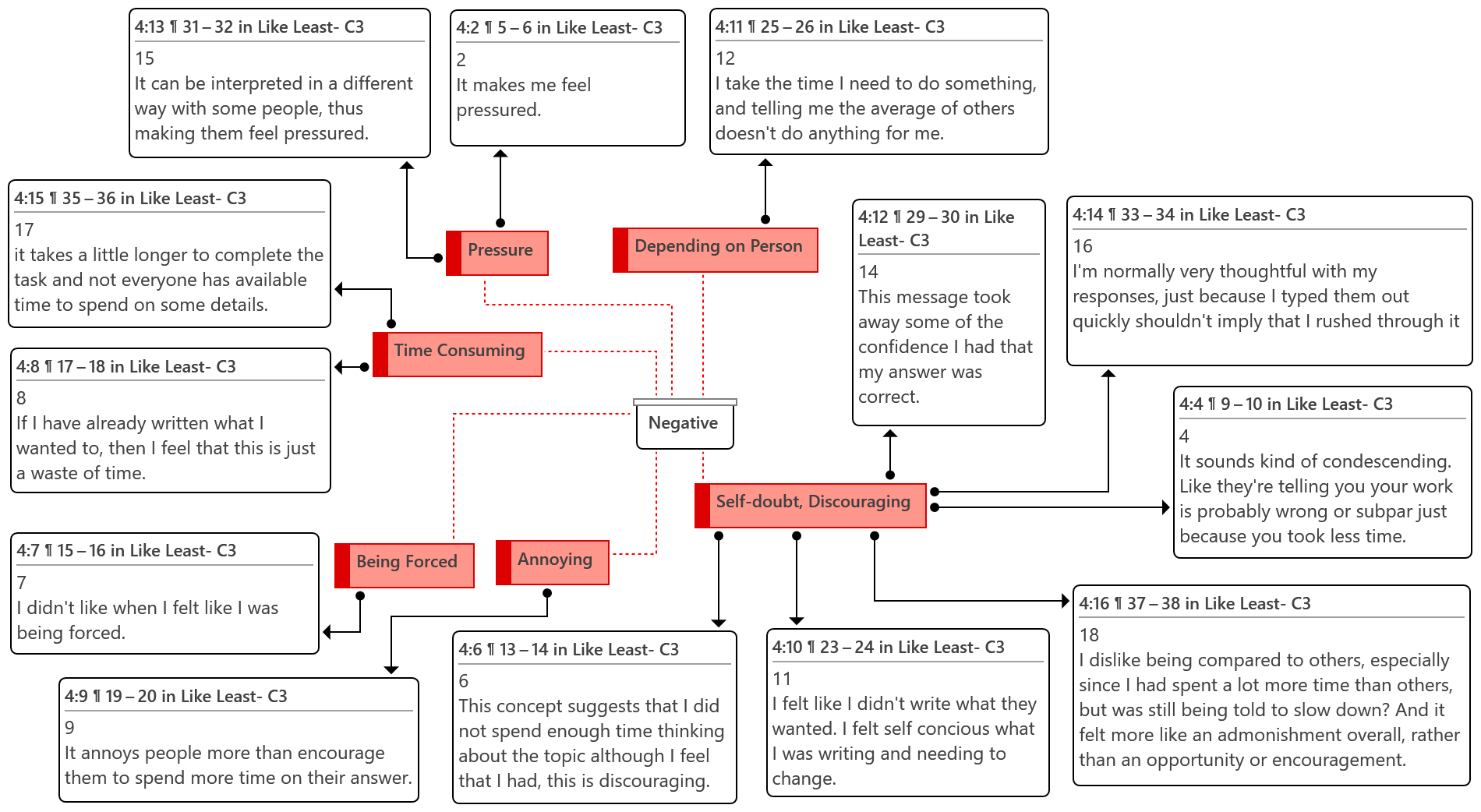}
  \caption{Drawbacks: Comparison Time Nudge}
  \label{fig:dislikes comparisontimenudge}
\end{figure*}

\begin{figure*}[!htbp]
  \centering
  \includegraphics[scale=0.25]{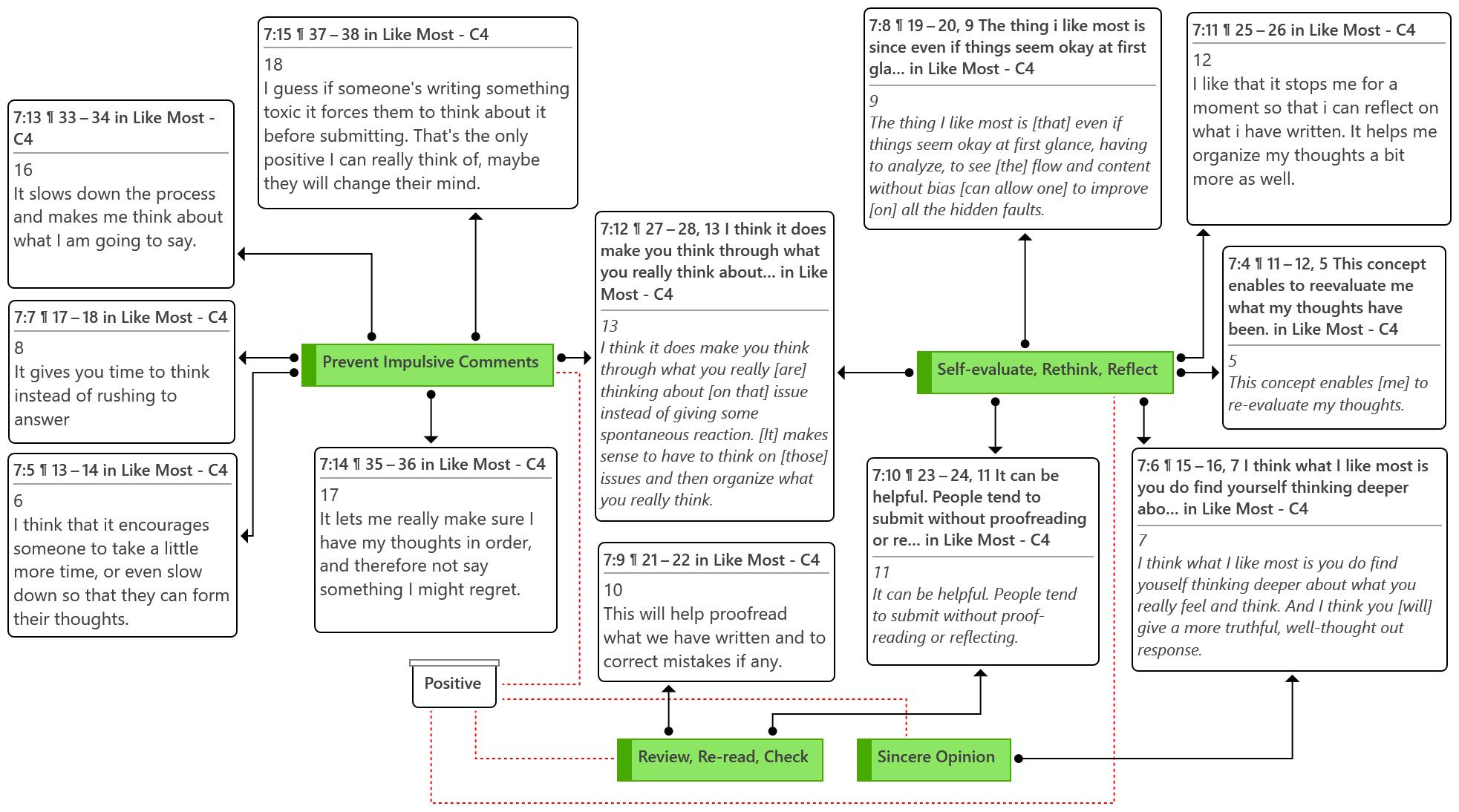}
  \caption{Benefits: Alert Message}
  \label{fig:likes alertmessage}
\end{figure*}

\begin{figure*}[!htbp]
  \centering
  \includegraphics[scale=0.23]{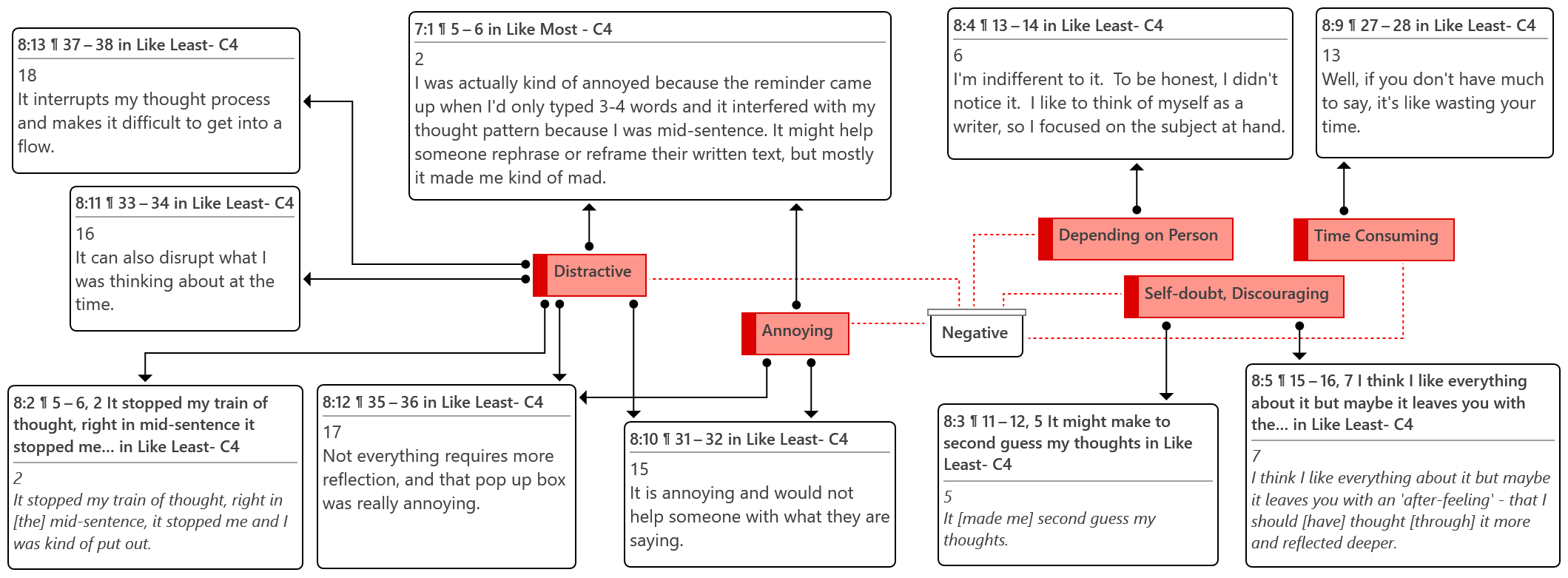}
  \caption{Drawbacks: Alert Message}
  \label{fig:dislikes alertmessage}
\end{figure*}

Figures~\ref{fig:likes fixedtimeframe} -~\ref{fig:dislikes alertmessage} show a snapshot of the benefits and drawbacks (both key and non-key benefits/drawbacks) for each time nudge. During the coding process, one or more codes could be applied to each participant's feedback.

\newpage

\begin{table*}[!htbp]
\caption{A section of the finalized codebook for Study 1. The inclusion list is not exhaustive.}
\label{tab:inclusionlist}
\scalebox{0.8}{\begin{tabular}{@{}cl@{}}
\toprule
\textbf{Codes} & \textbf{Inclusion List}\\
\midrule
\multirow{5}{*}{\begin{tabular}[c]{@{}c@{}}4-day workweek\\ is not generalizable\end{tabular}} & Depends on different types of work\\ & Depends on different industries\\ & Depends on cultures\\ & Depends on the country's customs\\ & Don't trust Japan's study \\[0.3cm]
\multirow{5}{*}{Good deal despite trade-offs} & Good deal\\ & Gain and loss\\ & Comparison of trade-offs\\ & Merits of 4-day workweek outweighs longer working hours\\ & Additional 2 working hours is not a big deal compared to a whole day off \\[0.3cm] 
\multirow{5}{*}{\begin{tabular}[c]{@{}c@{}}Increase time for personal\\ needs and self development\end{tabular}} & More quality time for self\\ & Increase time for skills upgrading and/or learning\\ & More time for hobbies and/or pursuing other interests\\ & More time for personal business\\ & More time to attend to personal needs \\[0.3cm]
\multirow{6}{*}{\begin{tabular}[c]{@{}c@{}}Increase productivity\\ and efficiency\end{tabular}} & Increase productivity/productive*\\ & Increase efficiency/efficient*\\ & Increase work output/able to accomplish more\\ & Good for working\\ & Increase concentration\\ & Increase motivation/able to focus better \\[0.3cm]
\multicolumn{2}{l}{\begin{tabular}[c]{@{}l@{}}\textit{* broadens the inclusion list to include various word endings and spellings}\end{tabular}}\\
\bottomrule
\end{tabular}}
\end{table*}  

\begin{table*}[!htbp]
\caption{Summary Statistics of the Demographic Profiles in Study 1}
\label{tab:demographic-experiment1}
\centering
\scalebox{0.85}{
\begin{tblr}{
  width = \linewidth,
  colspec = {Q[200]Q[350]Q[67]Q[77]Q[77]Q[77]Q[77]},
  row{1} = {c},
  row{2} = {c},
  cell{1}{1} = {c=2,r=2}{0.45\linewidth},
  cell{1}{3} = {c=5}{0.3\linewidth},
  cell{3}{3} = {c},
  cell{3}{4} = {c},
  cell{3}{5} = {c},
  cell{3}{6} = {c},
  cell{3}{7} = {c},
  cell{4}{1} = {r=2}{},
  cell{4}{3} = {c},
  cell{4}{4} = {c},
  cell{4}{5} = {c},
  cell{4}{6} = {c},
  cell{4}{7} = {c},
  cell{5}{3} = {c},
  cell{5}{4} = {c},
  cell{5}{5} = {c},
  cell{5}{6} = {c},
  cell{5}{7} = {c},
  cell{6}{1} = {r=6}{},
  cell{6}{3} = {c},
  cell{6}{4} = {c},
  cell{6}{5} = {c},
  cell{6}{6} = {c},
  cell{6}{7} = {c},
  cell{7}{3} = {c},
  cell{7}{4} = {c},
  cell{7}{5} = {c},
  cell{7}{6} = {c},
  cell{7}{7} = {c},
  cell{8}{3} = {c},
  cell{8}{4} = {c},
  cell{8}{5} = {c},
  cell{8}{6} = {c},
  cell{8}{7} = {c},
  cell{9}{3} = {c},
  cell{9}{4} = {c},
  cell{9}{5} = {c},
  cell{9}{6} = {c},
  cell{9}{7} = {c},
  cell{10}{3} = {c},
  cell{10}{4} = {c},
  cell{10}{5} = {c},
  cell{10}{6} = {c},
  cell{10}{7} = {c},
  cell{11}{3} = {c},
  cell{11}{4} = {c},
  cell{11}{5} = {c},
  cell{11}{6} = {c},
  cell{11}{7} = {c},
  cell{12}{3} = {c},
  cell{12}{4} = {c},
  cell{12}{5} = {c},
  cell{12}{6} = {c},
  cell{12}{7} = {c},
  cell{13}{3} = {c},
  cell{13}{4} = {c},
  cell{13}{5} = {c},
  cell{13}{6} = {c},
  cell{13}{7} = {c},
  hline{1,3-4,6,12-14} = {-}{},
  hline{2} = {3-7}{},}
\textbf{Demographics Profile} & & \textbf{Reflection Time} & & & & \\
& & \textbf{1 min} & \textbf{2 mins} & \textbf{3 mins} & \textbf{4 mins} & \textbf{5 mins} \\
\textbf{Participants} & \begin{tabular}[c]{@{}l@{}} Total number of participants \\ (after excluding outliers) \end{tabular} & 20 & 20 & 20 & 18 & 20 \\
\textbf{Gender} & Number of Males & 13 & 11 & 13 & 10 & 13 \\ 
& Number of Females & 7 & 9 & 7 & 8 & 7 \\
\textbf{Country} & USA & 10 & 14 & 16 & 11 & 17 \\
& India & 5 & 4 & 3 & 6 & 1 \\
& \begin{tabular}[c]{@{}l@{}} Europe (UK, Germany, \\ Bulgaria, Italy) \end{tabular} & 2 & 2 & 0 & 0 & 0 \\
& Canada & 1 & 0 & 0 & 0 & 0 \\
& Brazil & 1 & 0 & 1 & 1 & 2 \\
& Iran & 1 & 0 & 0 & 0 & 0\\
\textbf{Age} & Mean Age & 39.0 & 37.3 & 38.2 & 37.6 & 37.7 \\
\textbf{Familiarity} & Mean Familiarity & 60.6 & 65.0 & 61.5 & 61.6 & 53.5            
\end{tblr}}
\end{table*}

\begin{table*}[!htbp]
\caption{Summary Statistics of the Demographic Profiles in Study 2}
\label{tab:demographic-experiment2}
\centering
\scalebox{0.85}{
\begin{tblr}{
  width = \linewidth,
  colspec = {Q[180]Q[250]Q[146]Q[156]Q[160]Q[119]},
  row{1} = {c},
  row{2} = {c},
  cell{1}{1} = {c=2,r=2}{0.3\linewidth},
  cell{1}{3} = {c=4}{0.2\linewidth},
  cell{3}{3} = {c},
  cell{3}{4} = {c},
  cell{3}{5} = {c},
  cell{3}{6} = {c},
  cell{4}{1} = {r=3}{},
  cell{4}{3} = {c},
  cell{4}{4} = {c},
  cell{4}{5} = {c},
  cell{4}{6} = {c},
  cell{5}{3} = {c},
  cell{5}{4} = {c},
  cell{5}{5} = {c},
  cell{5}{6} = {c},
  cell{6}{3} = {c},
  cell{6}{4} = {c},
  cell{6}{5} = {c},
  cell{6}{6} = {c},
  cell{7}{1} = {r=3}{},
  cell{7}{3} = {c},
  cell{7}{4} = {c},
  cell{7}{5} = {c},
  cell{7}{6} = {c},
  cell{8}{3} = {c},
  cell{8}{4} = {c},
  cell{8}{5} = {c},
  cell{8}{6} = {c},
  cell{9}{3} = {c},
  cell{9}{4} = {c},
  cell{9}{5} = {c},
  cell{9}{6} = {c},
  cell{10}{3} = {c},
  cell{10}{4} = {c},
  cell{10}{5} = {c},
  cell{10}{6} = {c},
  cell{11}{3} = {c},
  cell{11}{4} = {c},
  cell{11}{5} = {c},
  cell{11}{6} = {c},
  hline{1,3-4,7,10-12} = {-}{},
  hline{2} = {3-6}{},}
\textbf{Demographics Profile } &  & \textbf{Time Nudges } &  &  & \\
 &  & \textbf{Fixed Time Frame} & \textbf{Simple Time Nudge} & \textbf{Comparison Time Nudge} & \textbf{Alert Message}\\
\textbf{Participants} & \begin{tabular}[c]{@{}l@{}} Total number of \\ participants \end{tabular} & 18 & 18 & 18 & 18\\
\textbf{Gender } & Number of Males & 10 & 9 & 11 & 10\\
 & Number of Females & 8 & 9 & 7 & 7\\
 & Prefer not to say & 0 & 0 & 0 & 1\\
\textbf{Country } & USA & 15 & 14 & 15 & 15\\
 & India & 1 & 1 & 0 & 3\\
 & \begin{tabular}[c]{@{}l@{}} Europe \\ (Italy, Brazil) \end{tabular} & 2 & 3 & 3 & 0\\
\textbf{Age} & Mean Age & 36.2 & 39.7 & 35.6 & 44.3\\
\textbf{Familiarity} & Mean Familiarity & 58.8 & 58.8 & 65.1 & 71.4
\end{tblr}}
\end{table*}

\begin{table*}[!htbp] \centering 
  \caption{Regression analysis of linear and logarithmic model for deliberativeness. Note: $^{*}$p$<$0.1; $^{**}$p$<$0.05; $^{***}$p$<$0.01} 
  \label{tab:regression analysis} 
\begin{subtable}[t]{0.48\textwidth}
\scalebox{0.8}{\begin{tabular}{@{\extracolsep{5pt}}lc} 
\\[-1.8ex]\hline 
\hline \\[-2.3ex] 
 & \multicolumn{1}{c}{\textit{Dependent variable:}} \\ 
\cline{2-2} 
\\[-1.8ex] & Word Count \\ 
\hline \\[-1.8ex] 
 Time & 15.559$^{***}$ \\ 
  & (2.979) \\ 
  & \\ [-2.5ex]
 Constant & 10.520 \\ 
  & (9.024) \\ 
  & \\ [-3ex]
\hline \\[-3.0ex]  
Observations & 6 \\ 
R$^{2}$ & 0.872 \\ 
Adjusted R$^{2}$ & 0.840 \\ 
Residual Std. Error & 12.079 (df = 4) \\ 
F Statistic & 27.277$^{***}$ (df = 1; 4) \\ 
\hline 
\hline \\[-1.8ex]  
\end{tabular}}
\caption{Linear model for word count}
\label{tab:table1_a}
\end{subtable}
\hspace{0.5em}
\begin{subtable}[t]{0.48\textwidth}
\flushright
\scalebox{0.8}{\begin{tabular}{@{\extracolsep{5pt}}lc} 
\\[-1.8ex]\hline 
\hline \\[-2.3ex] 
 & \multicolumn{1}{c}{\textit{Dependent variable:}} \\ 
\cline{2-2} 
\\[-1.8ex] & Word Count \\ 
\hline \\[-1.8ex] 
 Time & 25.727$^{***}$ \\ 
  & (2.156) \\ 
  & \\ [-2.5ex]
 Constant & 35.952$^{***}$ \\ 
  & (2.564) \\ 
  & \\ [-3ex]
\hline \\[-3.0ex]  
Observations & 6 \\ 
R$^{2}$ & 0.973 \\ 
Adjusted R$^{2}$ & 0.966 \\ 
Residual Std. Error & 5.582 (df = 4) \\ 
F Statistic & 142.453$^{***}$ (df = 1; 4) \\ 
\hline 
\hline \\[-1.8ex]  
\end{tabular}}
\caption{Logarithmic model for word count}
\label{tab:table1_b}
\end{subtable}

\smallskip 

\begin{subtable}[t]{0.48\textwidth} 
\scalebox{0.8}{\begin{tabular}{@{\extracolsep{5pt}}lc} 
\\[-1.8ex]\hline 
\hline \\[-2.3ex]
 & \multicolumn{1}{c}{\textit{Dependent variable:}} \\ 
\cline{2-2} 
\\[-1.8ex] & Argument Repertoire \\ 
\hline \\[-1.8ex] 
 Time & 0.645$^{**}$ \\ 
  & (0.170) \\ 
  & \\ [-2.5ex]
 Constant & 0.744 \\ 
  & (0.515) \\ 
  & \\ [-3ex]
\hline \\[-3.0ex] 
Observations & 6 \\ 
R$^{2}$ & 0.782 \\ 
Adjusted R$^{2}$ & 0.728 \\ 
Residual Std. Error & 0.690 (df = 4) \\ 
F Statistic & 14.381$^{**}$ (df = 1; 4) \\ 
\hline 
\hline \\[-1.8ex]
\end{tabular}} 
\caption{Linear model for argument repertoire}
\label{tab:table1_c}
\end{subtable}
\hspace{0.5em}
\begin{subtable}[t]{0.48\textwidth}
\flushright
\scalebox{0.8}{\begin{tabular}{@{\extracolsep{5pt}}lc} 
\\[-1.8ex]\hline 
\hline \\[-2.3ex] 
 & \multicolumn{1}{c}{\textit{Dependent variable:}} \\ 
\cline{2-2} 
\\[-1.8ex] & Argument Repertoire \\ 
\hline \\[-1.8ex] 
 Time & 1.131$^{***}$ \\ 
  & (0.081) \\ 
  & \\ [-2.5ex]
 Constant & 1.765$^{***}$ \\ 
  & (0.096) \\ 
  & \\ [-3ex]
\hline \\[-3.0ex]  
Observations & 6 \\ 
R$^{2}$ & 0.980 \\ 
Adjusted R$^{2}$ & 0.975 \\ 
Residual Std. Error & 0.210 (df = 4) \\
F Statistic & 194.663$^{***}$ (df = 1; 4) \\ 
\hline 
\hline \\[-1.8ex] 
\end{tabular}} 
\caption{Logarithmic model for argument repertoire}
\label{tab:table1_d}
\end{subtable}

\smallskip

\begin{subtable}[t]{0.48\textwidth} 
\scalebox{0.8}{\begin{tabular}{@{\extracolsep{5pt}}lc} 
\\[-1.8ex]\hline 
\hline \\[-2.3ex] 
 & \multicolumn{1}{c}{\textit{Dependent variable:}} \\ 
\cline{2-2} 
\\[-1.8ex] & Argument Diversity \\ 
\hline \\[-1.8ex] 
 Time & 0.633$^{**}$ \\ 
  & (0.159) \\ 
  & \\ [-2.5ex]
 Constant & 0.684 \\ 
  & (0.483) \\ 
  & \\ [-3ex]
\hline \\[-3.0ex]  
Observations & 6 \\
R$^{2}$ & 0.798 \\ 
Adjusted R$^{2}$ & 0.748 \\ 
Residual Std. Error & 0.646 (df = 4) \\ 
F Statistic & 15.804$^{**}$ (df = 1; 4) \\ 
\hline 
\hline \\[-1.8ex]  
\end{tabular}} 
\caption{Linear model for argument diversity}
\label{tab:table1_e}
\end{subtable}
\hspace{0.5em}
\begin{subtable}[t]{0.48\textwidth}
\flushright
\scalebox{0.8}{\begin{tabular}{@{\extracolsep{5pt}}lc} 
\\[-1.8ex]\hline 
\hline \\[-2.3ex] 
 & \multicolumn{1}{c}{\textit{Dependent variable:}} \\ 
\cline{2-2} 
\\[-1.8ex] & Argument Diversity \\ 
\hline \\[-1.8ex] 
 Time & 1.101$^{***}$ \\ 
  & (0.070) \\ 
  & \\ [-2.5ex]
 Constant & 1.690$^{***}$ \\ 
  & (0.083) \\ 
  & \\ [-3ex]
\hline \\[-3.0ex]  
Observations & 6 \\ 
R$^{2}$ & 0.984 \\ 
Adjusted R$^{2}$ & 0.980 \\ 
Residual Std. Error & 0.181 (df = 4) \\ 
F Statistic & 249.136$^{***}$ (df = 1; 4) \\ 
\hline 
\hline \\[-1.8ex] 
\end{tabular}}
\caption{Logarithmic model for argument diversity}
\label{tab:table1_f}
\end{subtable}
\end{table*} 

\clearpage

\end{document}